\documentclass[review]{elsarticle}
\usepackage{}

\usepackage{lineno,hyperref}
\usepackage{amssymb}
\usepackage{color}
\usepackage{amsmath}
\usepackage{tikz}
\usepackage{amsfonts}
\usepackage{mathrsfs}
 \usepackage{amsthm}

\usepackage{bbding}
\usepackage{mathrsfs}
\usepackage{amsmath, amsfonts, amssymb, mathrsfs, txfonts}
\usepackage{graphicx, subfigure}
\usepackage{wasysym}
\usepackage{color}
\usepackage{bm}
\usepackage{enumerate}

\usepackage{pifont}
\usepackage{txfonts}
\usepackage{amsmath}
\usepackage{graphicx}
\usepackage{amsfonts}
\usepackage{amssymb}
\usepackage{mathrsfs,psfrag,eepic,epsfig}
\usepackage{epstopdf}

\modulolinenumbers[5]

\journal{Journal of \LaTeX\ Templates}

\makeatletter \@addtoreset{equation}{section}
\renewcommand{\theequation}{\arabic{section}.\arabic{equation}}

\newtheorem{thm}{Theorem}[section]

\newtheorem{prop}[thm]{Proposition}
\theoremstyle{definition}

\renewcommand{\baselinestretch}{1.25}
\begin{document}

\begin{frontmatter}

\title{Inverse scattering transform and dynamics of soliton solutions for nonlocal focusing modified Korteweg-de Vries equation \tnoteref{mytitlenote}}
\tnotetext[mytitlenote]{%Project supported by the Fundamental Research Fund for the Central Universities under the grant No. 2017XKQY101.\\
%\hspace*{3ex}$^{*}$
Corresponding author.\\
\hspace*{3ex}\emph{E-mail addresses}: sftian@cumt.edu.cn,
shoufu2006@126.com (S. F. Tian) }

%% Group authors per affiliation:
\author{Xiao-Fan Zhang, Shou-Fu Tian$^{*}$ and Jin-Jie Yang}
\address{School of Mathematics, China University of Mining and Technology,  Xuzhou 221116, People's Republic of China}

\begin{abstract}
In this work, we mainly study the general $N$-soliton solutions of the nonlocal  modified Korteweg-de Vries (mKdV) equation by utilizing the Riemann-Hilbert (RH) method. For the initial value  belonging to Schwarz space, we firstly obtain the corresponding eigenfunctions and scattering data in the direct scattering process. Then we successfully  establish a suitable RH problem of the nonlocal mKdV equation. The exact expression of the solution for the equation is derived via solving the RH problem. Using the symmetry of scattering data, the phenomena corresponding to different eigenvalues are analyzed, including bounded solutions, singular solutions, position solutions and kink solutions. Finally, the propagation path of the solution is observed, and the characteristic line is further used to analyze the continuity or other phenomena of the solution. The new dynamic behavior of the solution is observed by rotating the characteristic line at a certain angle.
\end{abstract}

\begin{keyword} Nonlocal mKdV equation \sep
Initial value problem  \sep Riemann-Hilbert method \sep Soliton solutions \sep Characteristic line analysis.
\end{keyword}

\end{frontmatter}

%\linenumbers
%\tableofcontents
%\newpage

\section{Introduction}
The scattering problem of Ablowitz-Kaup-Newell-Segur (AKNS) type are often discussed in integrable systems
\begin{align}
\psi_{x}(x,t)=U\psi(x,t),\\
\psi_{t}(x,t)=V\psi(x,t),
\end{align}
where
\begin{align}
U=\left(
            \begin{array}{cc}
              -ik & q(x,t) \\
              r(x,t) & ik \\
            \end{array}
          \right),~~
V=\left(
            \begin{array}{cc}
              A & B \\
              C & -A \\
            \end{array}
          \right),
\end{align}
and the eigenfunction $\psi$ is a two-component vector, $q(x,t)$, $r(x,t)$ are complex valued functions of the real variables $x$, and $t$, $A$, $B$ and $C$ are scalar functions. Taking
\begin{align*}
A=-4ik^{3}-2iq(x,t)r(x,t)k+r(x,t)q_{x}(x,t)-r_{x}(x,t)q(x,t),\\
B=4k^{2}q(x,t)+2iq_{x}(x,t)k+2q^{2}(x,t)r(x,t)-q_{xx}(x,t),\\
C=4k^{2}r(x,t)-2ir_{x}(x,t)k+2q(x,t)r^{2}(x,t)-r_{xx}(x,t),
\end{align*}
we obtain coupled mKdV equations by the compatibility condition of Lax pair
\addtocounter{equation}{1}
\begin{align}
q_{t}(x, t)+q_{xxx}(x, t)-6q(x,t)r(x,t)q_{x}(x,t)=0,\tag{\theequation a}\label{Q1}\\
r_{t}(x, t)+r_{xxx}(x, t)-6q(x,t)r(x,t)r_{x}(x,t)=0.\tag{\theequation b}\label{Q2}
\end{align}
Under the symmetry reduction
\begin{align}
r(x,t)=\sigma q^{*}(-x,-t),~~\sigma=\pm1,
\end{align}
since \eqref{Q1} and \eqref{Q2} are compatible, the nonlocal complex mKdV equation \cite{Ablowitz-2017} with reverse space-time is derived by
\begin{align}
q_{t}(x,t)+q_{xxx}(x,t)-6\sigma q(x,t)q^{*}(-x,-t)q_{x}(x,t)=0,
\end{align}
where ${*}$ denotes the complex conjugation, and $q(x,t)$ is a complex value function. Using the symmetry property, that is
\begin{align}
r(x,t)=\sigma q(-x,-t),~~\sigma=\pm1,
\end{align}
the nonlocal real mKdV equation with reverse space-time proposed by Zhu and Ji in \cite{nmKdV-CNSNS-2017} can be derived
\begin{align}\label{mkdv}
q_{t}(x,t)+q_{xxx}(x,t)-6\sigma q(x,t)q(-x,-t)q_{x}(x,t)=0,
\end{align}
in addition, they reported the soliton, kink, anti-kink and rogue-wave solutions of equation \eqref{mkdv} by the Darboux transformation. The soliton solution and breathe solution of equation \eqref{mkdv} have been studied by using inverse scattering transformation in \cite{mKdV-JMAA-2017}.
In this work, we mainly study the dynamic behavior of the nonlocal real focusing modified Korteweg-de Vries (mKdV) equation of reverse space-time
\begin{align}\label{Q3}
q_{t}(x,t)+6q(-x,-t)q(x,t)q_{x}(x,t)+q_{xxx}(x,t)=0,~~~~q(x,0)=q_{0}(x),
\end{align}
 where $q(x,t)$ is a real function. Unlike the case where the equation is local, the nonlinear induced ``potential" is parity-time $(\mathcal {P}\mathcal {T})$ symmetry \cite{Ablowitz-2017}. Since Bender et al. introduced $\mathcal {P}\mathcal {T}$ symmetry into generalized Hamiltonian \cite{Bender-1998} in 1998, which plays an important role in many fields \cite{Bender-2007}-\cite{Konotop-2016}. In 2013, $\mathcal {P}\mathcal {T}$ symmetry was introduced into the first AKNS type system and a nonlocal nonlinear Sch\"{o}rdinger equation was proposed with $q^{*}(-x,-t)$, which is a nonlocal term \cite{AM-PRL-2013,Ablowitz-2016}.

In 1967, the inverse scattering transformation (IST) method was proposed  by Gardner, Greene, Kruskal and Miura (GGKM), which was used to solve the initial value problem of the KdV equation with Lax pair \cite{Gardner-1967}. Subsequently Zakharov and Shabat employed the  IST method to solve the nonlinear Schr\"{o}dinger equation in 1972 \cite{Shabat-1972}. After that, AKNS further proposed a new integrable system, called AKNS system, and constructed a general framework of IST \cite{Ablowitz-1973,Ablowitz-1974}. So far, it is still a very effective method to solve the Cauchy problem of nonlinear partial differential equations (PDEs). The main idea of IST is to establish the Gelfand-Levitan-Marchenko (GLM) integral equation from the scattering data obtained in the direct scattering process,  from which the potential function can be recovered. But it is worth noting that the calculation of this process is very complicated, so a modern version of IST method, Riemann-Hilbert (RH) problem \cite{Beal-Confiman-1984}-\cite{Zhou-1989}, is gradually widely used in integrable systems.

Since the 1970s, people have developed the RH method for studying integrable systems, which is a method of transforming a given problem into an analytic function based on the jump conditions. The core idea of the RH method is that establishing the RH problem of nonlinear evolution integrable equation with Lax pair,  from which we can obtain the relationship between the solution of the RH problem and the initial value problem of the equation. In recent years, some remarkable results have been obtained by using RH method, including  coupled nonlinear
Schr\"{o}dinger equations \cite{cNLS-JDE-2017}-\cite{CNLS-PRSL-2016}, the Sasa-Satsuma equation \cite{SS-WM-2016, SS-NA-2019}, and the coupled modified Korteweg-de Vries equation \cite{cmKdV-JPA-2017,cmKdV-JGP-2018}, other equations \cite{Yang-JMP-2003}-\cite{NLS-TMP-2020}.
Inspired by Yang's work \cite{NNLS-PLA-2019}, in this work, we mainly discuss the nonlocal real focusing mKdV equation via RH method. In addition, the nonlocal mKdV equation has also been reported in many works \cite{NNmKdV-PDNP-2019,nonlocal mKdV-JGP-2020}. Especially, we also find the relationship between nonlocal and local equations, the results show that many nonlocal equations can be transformed into local equations by transformation \cite{Yangjianke-2018}. For example, first, when $q(-x,-t)=q(x,t)$, the equation \eqref{Q3} becomes the normal mKdV equation; second, when $q(-x,-t)=-q(x,t)$, the equation \eqref{Q3} becomes the defocusing  mKdV equation; third, when $q(-x,-t)=-q(-x,-t)$, the equation \eqref{Q3} becomes the nonlocal defocusing  mKdV equation.

The exact solution of the nonlocal mKdV equation with the finite density type initial value  condition  is studied in \cite{NNmKdV-PDNP-2019}, and one-soliton solution and two-soliton solutions are discussed. However, nonlocal equations have $\mathcal {P}\mathcal {T}$ symmetry, and actually have more abundant phenomena worth exploring. It is worth noting that when the potential function $q_{\pm}(x,t)=0$ in the nonzero boundary conditions (NZBCs), it will reduce to the zero boundary condition (ZBC). The basic difference between ZBC and the finite density type initial value condition is the construction of Jost function, which further affects the follow-up work. There is not much essential difference in the final solution. Therefore, our work is mainly to further study the more abundant phenomena of the nonlocal mKdV equation with the ZBCs. The main highlights of this work can be described as follows. The solution of the non-regular RH problem is derived through the regular RH problem, and the $N$-soliton solution expression of the nonlocal mKdV equation is obtained. Based on the nonlocal symmetry reduction, in order to discuss the symmetry of eigenvalues, additional constraints are further added. As a result,  many abundant phenomena are obtained, including degenerate solutions, kink solutions, position solutions, and solution collapse phenomena. Eventually, utilizing the characteristic line to further study the propagation phenomenon of the solution after rotating a certain angle.
%\begin{enumerate}[(i)]
%\item Based on the nonlocal symmetry reduction, the symmetry of the eigenvalue problem is proved.
%\item The solution of the non-regular RH problem is derived through the regular RH problem, and the
%$N$-soliton solution expression of the nonlocal mKdV equation is given.
%\item In order to discuss the symmetry of eigenvalues, additional constraints are further added. As a result,  many abundant phenomena are obtained, including degenerate solutions, kink solutions, position solutions, and solution collapse phenomena.
%\item Utilizing the characteristic line to further study the propagation phenomenon of the solution after rotating a certain angle.
%\end{enumerate}
%The main motivation of this work is that although the nonlocal mKdV equation has been studied, we analyze the phenomenon of the reverse space-time mKdV equation more comprehensively. By contrast, we observed more abundant phenomena. In this work, we combine the image of soliton solution with characteristic line analysis, which has not appeared before. We find out the relationship between $x$ and $t$ in the characteristic line, keep the angle of the characteristic line unchanged, and change the slope so that the image is rotated, then  the  new propagation behaviors of the soliton solution are obtained. These propagation behaviors may only change in amplitude or position, or may change from singular solution to bounded solution with time.

The concrete structure of this work is as follows. In section 2, we mainly use the RH method to derive the general form of $N$-soliton solutions of the nonlocal mKdV equation. In section 3, we further analyze the dynamic behavior of soliton solutions, including one-soliton, two-soliton and three-soliton. By selecting different eigenvalues and comparing different phenomena, we find that some soliton solutions always collapse repeatedly, but still maintain the original velocity and shape after collision and  remain bounded. In addition, we find that the two solitons will collapse repeatedly with different directions, which is a nonlinear superposition of two one-soliton, but the dynamic behavior of one-soliton is very stable.

\section{The spectral analysis}
In this section, we focus on the spectral analysis of the nonlocal mKdV equation, that is, the analyticity, symmetry and asymptotic property of the nonlocal mKdV equation.

Taking the compatibility condition into account, $A_{t}-B_{x}+[X,T]=0$, we obtain the scattering problem of Lax pair for the nonlocal mKdV equation at infinity, that is
\addtocounter{equation}{1}
\begin{align}
\Phi_{x}&=A\Phi=(ik\sigma_{3}+Q)\Phi,\tag{\theequation a}\label{Q4}\\
\Phi_{t}&=B\Phi=[(4k^{2}-2q(x,t)q(-x,-t))X-2ik\sigma_{3}Q_{x}+[Q_{x},Q]-Q_{xx}]\Phi,\tag{\theequation b}\label{Q5}
\end{align}
where
\begin{align}
Q=Q(x,t)=\left(
           \begin{array}{cc}
             0 & q(x,t) \\
             -q(-x,-t) & 0 \\
           \end{array}
         \right).
\end{align}
The three Pauli matrices are introduced as
\begin{align*}
\begin{split}
\sigma_{1}=\left(
             \begin{array}{cc}
               0 & 1 \\
               1 & 0 \\
             \end{array}
           \right),~~~~
\sigma_{2}=\left(
           \begin{array}{cc}
             0 & -i \\
             i & 0 \\
           \end{array}
         \right),~~~~
\sigma_{3}=\left(
             \begin{array}{cc}
               1 & 0 \\
               0 & -1 \\
             \end{array}
           \right),
\end{split}
\end{align*}
and $e^{\beta\hat{\sigma_{3}}}A=e^{\beta\sigma_{3}}Ae^{-\beta\sigma_{3}}$.
At the same time, we will seek for the simultaneous solutions $\Phi_{\pm}(x,t,k)$, the so-called Jost solutions
\begin{align*}
\Phi_{\pm}(x,t,k)=e^{ik\sigma_{3}x+4ik^{3}t\sigma_{3}}+o(1), as ~~x\rightarrow\pm\infty.
\end{align*}
The modified Jost solutions are obtained by removing the asymptotic exponential oscillations
\begin{align}\label{1}
\Phi(x,t,k)=\varphi(x,t,k)e^{ik\sigma_{3}x+4ik^{3}t\sigma_{3}},
\end{align}
with the asymptotic properties
\begin{align}\label{8}
\lim_{x\rightarrow\pm\infty}\varphi_{\pm}(x,t,k)=\mathbb{I},~~~x\rightarrow\pm\infty.
\end{align}
As a result, when all the potentials rapidly vanish as $x\rightarrow\pm\infty$, $\varphi_{\pm}(x,t,k)$ acquire the equivalent pair of matrix spectral problems to \eqref{Q4}, \eqref{Q5}:
\addtocounter{equation}{1}
\begin{align}
\varphi_{x}(x,t,k)&=ik[\sigma_{3},\varphi]+Q\varphi(x,t,k),\tag{\theequation a}\label{Q6}\\
\varphi_{t}(x,t,k)&=4ik^{3}[\sigma_{3},\varphi]+\tilde{Q}\varphi(x,t,k),\tag{\theequation b}\label{Q7}
\end{align}
where
\begin{align*}
\tilde{Q}=4k^{2}Q-2ik\sigma_{3}q(x,t)q(-x,-t)-2Qq(x,t)q(-x,-t)-2ik\sigma_{3}Q_{x}+[Q_{x},Q]-Q_{xx}.
\end{align*}
Then we can turn the $x$-part of the equation \eqref{Q6} into the Jost integral equations:
\begin{align}\label{2}
\varphi_{\pm}(x,t,k)=I+\int_{\pm\infty}^{x}e^{ik\widehat{\sigma_{3}}(x-y)}Q(k,y)\varphi_{\pm}(k,y)dy,
\end{align}
where the asymptotic conditions \eqref{8} have been used.
\begin{prop}\label{2.2}
The Jost integral equation \eqref{2} has unique solutions $\varphi_{\pm}(x,t,k)$ determined by \eqref{1} in $\mathbb{R}$. In addition, the modified eigenfunctions $\varphi_{+,1}(x,t,k)$ and $\varphi_{-,2}(x,t,k)$ are analytic in $\mathbb{D}^{+}=\{k|Im~k>0\}$, the modified eigenfunctions $\varphi_{-,1}(x,t,k)$ and $\varphi_{+,2}(x,t,k)$ are analytic in $\mathbb{D}^{-}=\{k|Im~k<0\}$, where $\varphi_{\pm,i}(x,t,k)$ $(i=1,2)$ denote the i-th column of the modified eigenfunctions $\varphi_{\pm}(x,t,k)$.
\end{prop}
Furthermore, since $tr(A)=tr(B)=0$ and employing Abel's theorem, it follows that $(\det\Phi)_{x}=(\det\Phi)_{t}=0$, we have $\det\varphi_{\pm}(x,t,k)=1$.
Let $N(x,t,k)=e^{ik\sigma_{3}x+4ik^{3}\sigma_{3}t}$, we find the Lax pairs  existing two fundamental solutions $\varphi_{-}(x,t,k)N$ $(x,t,k)$ and $\varphi_{+}(x,t,k)N(x,t,k)$, so there is a matrix $S(k)$ that only depend on $k$, such that
\begin{align}
\varphi_{-}(x,t,k)N(x,t,k)=\varphi_{+}(x,t,k)N(x,t,k)S(k),
\end{align}
and one has $\det S(k)=1$. The Jost integral equations \eqref{2} also guarantee the normalization condition.

\section{Riemann-Hilbert problem and $N$-solitons for nonlocal mKdV equation}

In what follows, we will formulate the RH problem via using the analytic properties and asymptotic properties of the Jost functions $\varphi_{\pm}(x,t,k)$ for the nonlocal focusing reverse space-time mKdV equation.

Based on the above results, we  first introduce two matrices,
\begin{align}
H_{1}=\left(
  \begin{array}{cc}
    1 & 0 \\
    0 & 0 \\
  \end{array}
\right),\quad
H_{2}=\left(
  \begin{array}{cc}
    0 & 0 \\
    0 & 1 \\
  \end{array}
\right),
\end{align}
and define these two matrices
\begin{align}
P^{+}(x,t,k)&=\varphi_{-}(x,t,k)H_{1}+\varphi_{+}(x,t,k)H_{2},\\
P^{-}(x,t,k)&=H_{1}\varphi_{-}(x,t,k)^{-1}+H_{2}\varphi_{+}(x,t,k)^{-1},
\end{align}
due to the analytical and asymptotic properties of $\varphi_{\pm}(x,t,k)$. Consequently, it follows from \eqref{2.2} that $P^{+}(x,t,k)$ and $P^{-}(x,t,k)$ are analytic in $\mathbb{D}^{+}$ and $\mathbb{D}^{-}$ respectively.
\begin{prop}
Based on the definition of $P^{\pm}(x,t,k)$, the RH problem admit:

$\bullet$ Analysis:~$P^{\pm}(x,t,k)$ are analytic in $\mathbb{D}^{\pm}$  respectively.

$\bullet$ Jump condition:~$P^{-}(x,t,k)P^{+}(x,t,k)=J(x,t,k),~k\in\mathbb{R}$,
where
\begin{align}
J(x,t,k)=N\left(
            \begin{array}{cc}
              1 & s_{12}^{-1}(k) \\
              s_{21}(k) & 1 \\
            \end{array}
          \right)
N^{-1}.
\end{align}

$\bullet$ Normalization:~$P^{\pm}(x,t,k)\rightarrow\mathbb{I},~k\rightarrow\infty$.
\end{prop}
Under the condition of no reflection potential, the jump condition in the RH problem can be reduced to
\begin{align}
P^{-}(x,t,k)P^{+}(x,t,k)=\mathbb{I}.
\end{align}
Similar to \cite{Yangjianke-2010}, the exact solution of $P^{+}$ can be expressed as
\begin{align}
P^{+}=I+\sum_{j,\ell=1}^{N}\frac{(M^{-1})_{j,\ell} T_{j}\widetilde{T}_{\ell}}{k_{j}-\widetilde{k}_{\ell}},
\end{align}
where
\begin{align*}
T_{j}(x,t)=e^{ik_{j}\sigma_{3}x+4ik_{j}^{3}\sigma_{3}t}T_{j_{0}},~\qquad
\widetilde{T}_{j}(x,t)=\widetilde{T}_{j_{0}}e^{-i\widetilde{k}_{j}\sigma_{3}x-4i\widetilde{k}_{j}^{3}\sigma_{3}t}.
\end{align*}
Let
\begin{align*}
\theta_{j}=ik_{j}\sigma_{3}x+4ik_{j}^{3}\sigma_{3}t,~\qquad
\widetilde{\theta}_{j}=-i\widetilde{k}_{j}\sigma_{3}x-4i\widetilde{k}_{j}^{3}\sigma_{3}t,
\end{align*}
where $k_{j}\in D^{+}$, $\widetilde{k}_{j}\in D^{-}$, $j,\ell=1,2,\ldots,N$, $T_{j_{0}}=\left(
                                                                                    \begin{array}{cc}
                                                                                      a_{j} & b_{j} \\
                                                                                    \end{array}
                                                                                  \right)
^{T}$ and $\widetilde{T}_{j_{0}}=\left(
                                  \begin{array}{cc}
                                    \widetilde{a}_{j} & \widetilde{b}_{j} \\
                                  \end{array}
                                \right)$,
$M=(m_{j\ell})$ is a $N\times N$ matrix with elements
\begin{align}
m_{j\ell}=\frac{\widetilde{T}_{j}T_{\ell}}{\widetilde{k}_{j}-k_{\ell}},~1\leq j,\ell\leq N.
\end{align}
To obtain the expression of the potential $q(x,t)$, then making Taylor expansion for $P^{+}$,
taking \eqref{Q6} and \eqref{Q7} into account, %and matching the $O(k^{2})$, $O(k)$ and $O(1)$ term both in $x-$part and $t-$part
one can obtain the solution
%The general $N$-soliton solutions can be explictly expressed as
\begin{align}
q(x,t)=2i(P_{1})_{12}
=2i\left(\sum_{j,\ell=1}^{N}\frac{(M^{-1})_{j\ell}T_{j}\widetilde{T}_{\ell}}
{k_{j}-\widetilde{k}_{\ell}}\right)_{12},
\end{align}
which
 can be rewritten to the form of matrix
\begin{align}\label{7}
q(x,t)=-2i\frac{\det M^{\natural}}{\det M},
\end{align}
where $M^{\natural}$ is a $(N+1)\times(N+1)$ matrix, that is
\begin{align}
M^{\natural}=\left(
  \begin{array}{cccc}
    0 & a_{1}e^{\theta_{1}} & \cdots & a_{N}e^{\theta_{N}} \\
    \widetilde{b}_{1}e^{-\widetilde{\theta}_{1}} & m_{11} & \cdots & m_{1N} \\
    \vdots & \vdots & \ddots & \vdots \\
    \widetilde{b}_{N}e^{-\widetilde{\theta}_{N}} & m_{N1} & \cdots & m_{NN} \\
  \end{array}
\right).
\end{align}

Since symmetry affects eigenvalues and further affects eigenvectors, the study of symmetry is very important. To study the symmetry of scattering data for the reverse space-time nonlocal mKdV, we first give the following relation according to the reduction relation $r(x,t)=-q(-x,-t)$
%We study the symmetric relationship of scattering data for the reverse space-time nonlocal mKdV, in order to achieve this purpose, let us introduce two notations before we give the  theorem as follows.
%By the reduction $r(x,t)=-q(-x,-t)$, we have the relation
\begin{align}
Q(x,t)=-\sigma_{1}Q(-x,-t)\sigma_{1},\quad
{\Phi}(x,t,k)=\sigma_{1}\Phi(-x,-t,k).
\end{align}
Taking these conditions into account, we can obtain the theorem obviously.
\begin{thm}\label{4}
For the reverse space-time mKdV equation \eqref{Q3}, eigenvalues $k_{j}$ can be extended analytically to $\mathbb{D}^{+}$, and similarly eigenvalues $\widetilde{k}_{j}$ are analytic in $\mathbb{D}^{-}$. In addition, their eigenvectors can be expressed as the form
\begin{align}
T_{j_{0}}=\left(
            \begin{array}{cc}
              1 & \omega_{j} \\
            \end{array}
          \right)^{T},
~\widetilde{T}_{j_{0}}=\left(
            \begin{array}{cc}
              1 & \widetilde{\omega}_{j} \\
            \end{array}
          \right),
\end{align}
where $\omega_{j}=\pm1$ and $\widetilde{\omega}_{j}=\pm1$.
\end{thm}
Similar to the method in \cite{NNLS-PLA-2019}, by using the spectral problem and combining the asymptotic properties of the eigenfunction function
\begin{align}\label{9}
\begin{split}
&\Phi_{j}(x)\sim\left(
                 \begin{array}{c}
                   a_{j}e^{ik_{j}x} \\
                   0 \\
                 \end{array}
               \right),~~~~~~~x\rightarrow-\infty,\\
&\Phi_{j}(x)\sim\left(
                 \begin{array}{c}
                   0 \\
                   -b_{j}e^{-ik_{j}x} \\
                 \end{array}
               \right),~~~x\rightarrow+\infty,
               \end{split}
\end{align}
then Theorem \ref{4} can be derived.

It follows that if $k$ is an eigenvalue of \eqref{Q4}, then $-k$ will be another eigenvalue of \eqref{Q4}. If the eigenvalues and eigenvectors satisfy the condition given by Theorem \ref{4},  the nonlocal focusing mKdV equation has $N$-soliton solutions
\begin{align}\label{5}
q(x,t)=-2i\frac{\det M^{\natural}}{\det M},
\end{align}
where $M^{\natural}$, $M$ are the same as above.
\section{Dynamics of solutions in the reverse space-time nonlacal mKdV equation}

In this section,  let $N\in\mathbb{Z}$ be arbitrary. We will give the specific expression of the $N$-soliton solution for the reverse space-time mKdV equation, and select different eigenvalues according to the symmetry relation in Theorem \ref{4} to analyze the propagation of the solution in detail. Specifically, we let
\begin{align}
T_{j_{0}}=\left(
            \begin{array}{cc}
              1 & \omega_{j} \\
            \end{array}
          \right)^{T},
\widetilde{T}_{j_{0}}=\left(
            \begin{array}{cc}
              1 & \widetilde{\omega}_{j} \\
            \end{array}
          \right),
\end{align}
where $\omega_{j}=\pm1$ and $\widetilde{\omega}_{j}=\pm1$. It applies that $\omega_{j}$ and $\widetilde{\omega}_{j}$ have two choices between $\pm1$.
\subsection{Fundamental soliton solutions}
Taking $N$=1 for the one-soliton, the specific form of the solution of the equation \eqref{5} can be expressed as
\begin{align}\label{6}
q(x,t)=\frac{2i\widetilde{\omega}_{1}(\widetilde{k}_{1}-k_{1})e^{\theta_{1}-\widetilde{\theta}_{1}}}
{e^{\theta_{1}+\widetilde{\theta}_{1}}+\omega_{1}\widetilde{\omega}_{1}
e^{-(\theta_{1}+\widetilde{\theta}_{1})}},
\end{align}
where $\omega_{1}=\pm1$, $\widetilde{\omega}_{1}=\pm1$, $\theta_{1}\in \mathbb{D}^{+}$
and $\widetilde{\theta}_{1}\in \mathbb{D}^{-}$. For $k_{1}$ and $\widetilde{k}_{1}$, we select three sets of eigenvalues.   As a comparison, we discuss the following different situations, including non-pure imaginary eigenvalues and pure imaginary eigenvalues.

\subsubsection{Pure imaginary eigenvalues with $\omega_{1}=\widetilde{\omega}_{1}$}
\
\newline
For a soliton solution, we notice that there are six free parameters $\omega_{1}, \widetilde{\omega}_{1}, k_{1}, \widetilde{k}_{1},$ $\theta_{1}$ and $\widetilde{\theta}_{1}$ in the expression of the solution. Now considering the case that the eigenvalues are pure imaginary numbers, the propagation behavior of the solution can be characterized by the bright soliton solution shown in Fig. 1. In addition, another bright soliton solution Fig. 1(\textbf{b}) is obtained by rotating the characteristic line of the bright soliton solution Fig. 1(\textbf{a}). Note that since the characteristic line is related to the eigenvalue, different eigenvalues are selected to obtain the propagation behavior of rotation with $\tan\theta=-\frac{2}{9}$.

\noindent{\rotatebox{0}{\includegraphics[width=3.3cm,height=2.8cm,angle=0]{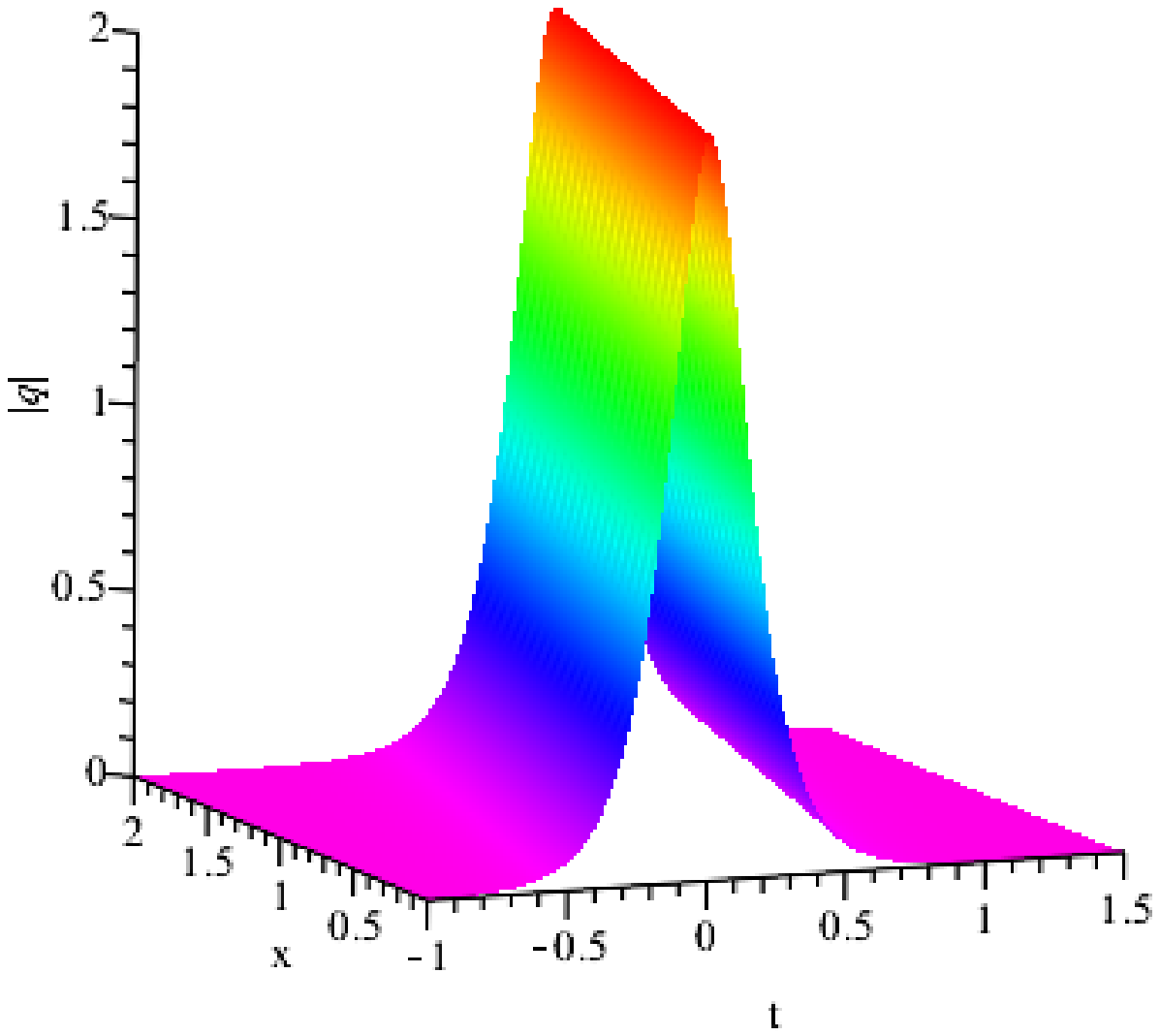}}
~\quad\rotatebox{0}{\includegraphics[width=2.5cm,height=2.4cm,angle=0]{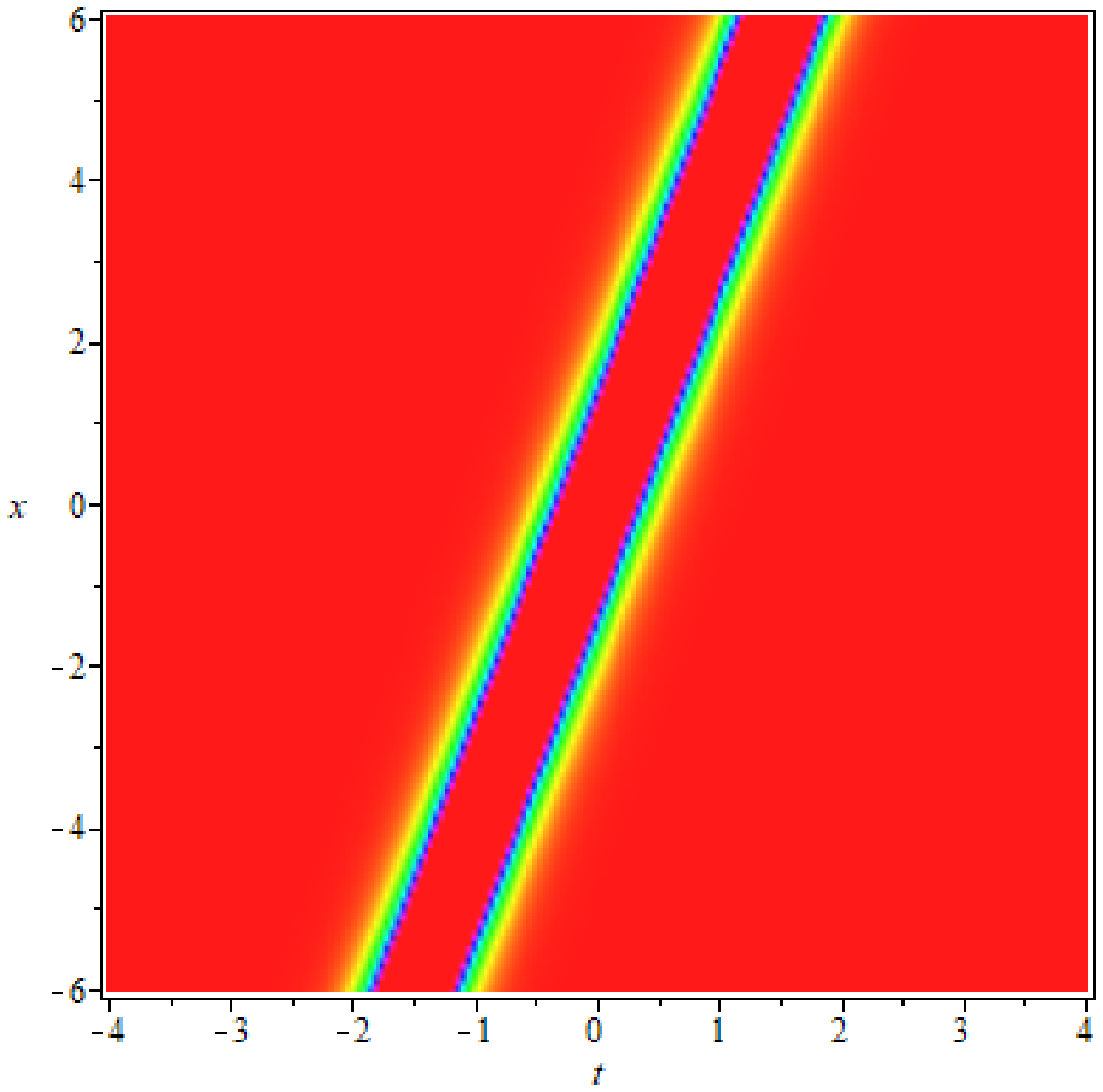}}
~\quad\rotatebox{0}{\includegraphics[width=2.5cm,height=2.4cm,angle=0]{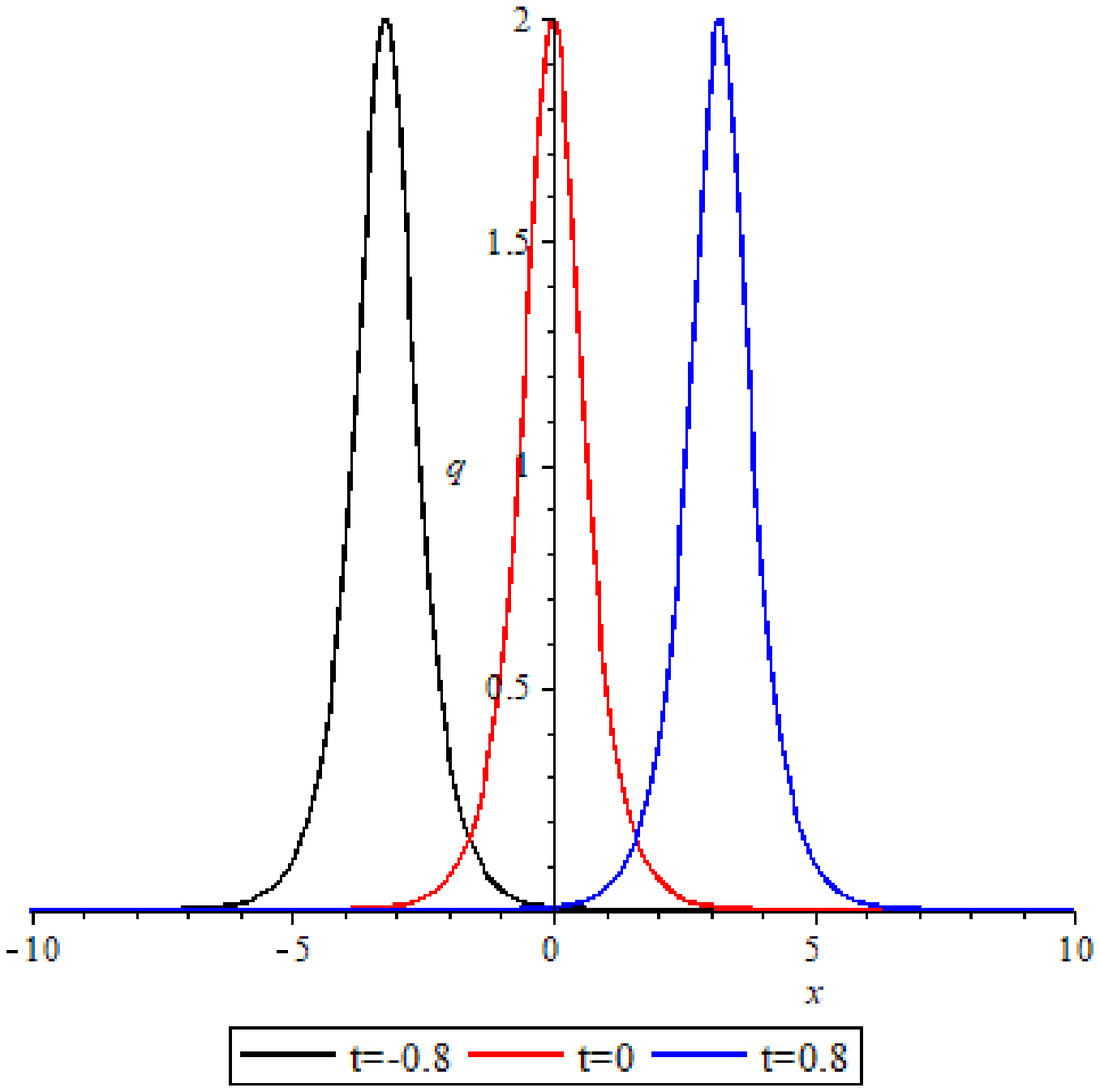}}
~\quad\rotatebox{0}{\includegraphics[width=2.5cm,height=2.4cm,angle=0]{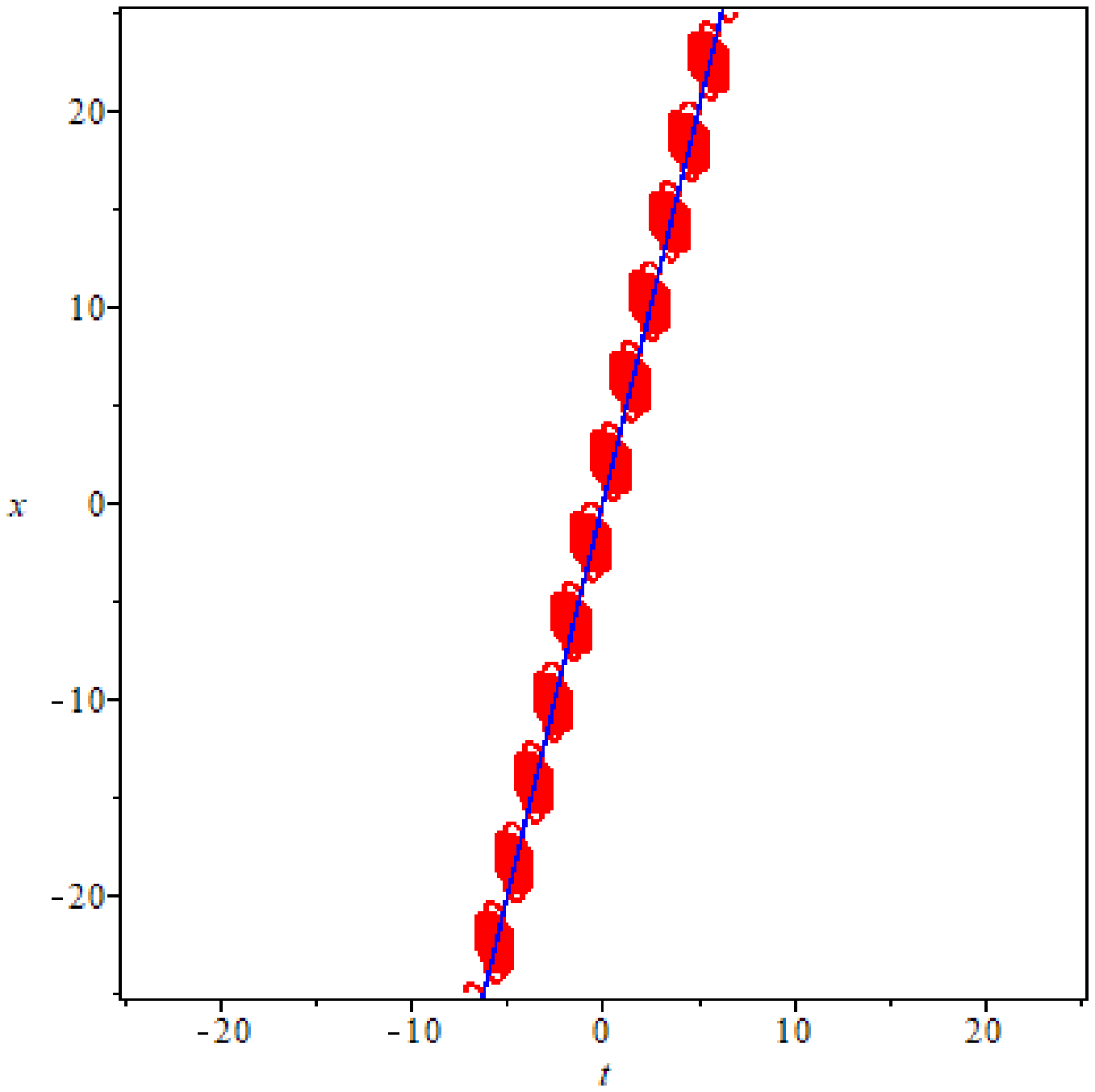}}}

$~~\quad\quad(\textbf{a})\qquad\qquad\qquad\qquad\quad(\textbf{b})
\qquad\qquad\qquad~~~(\textbf{c})\qquad\qquad~~~\qquad\quad(\textbf{d})$\\

\noindent{\rotatebox{0}{\includegraphics[width=3.3cm,height=2.8cm,angle=0]{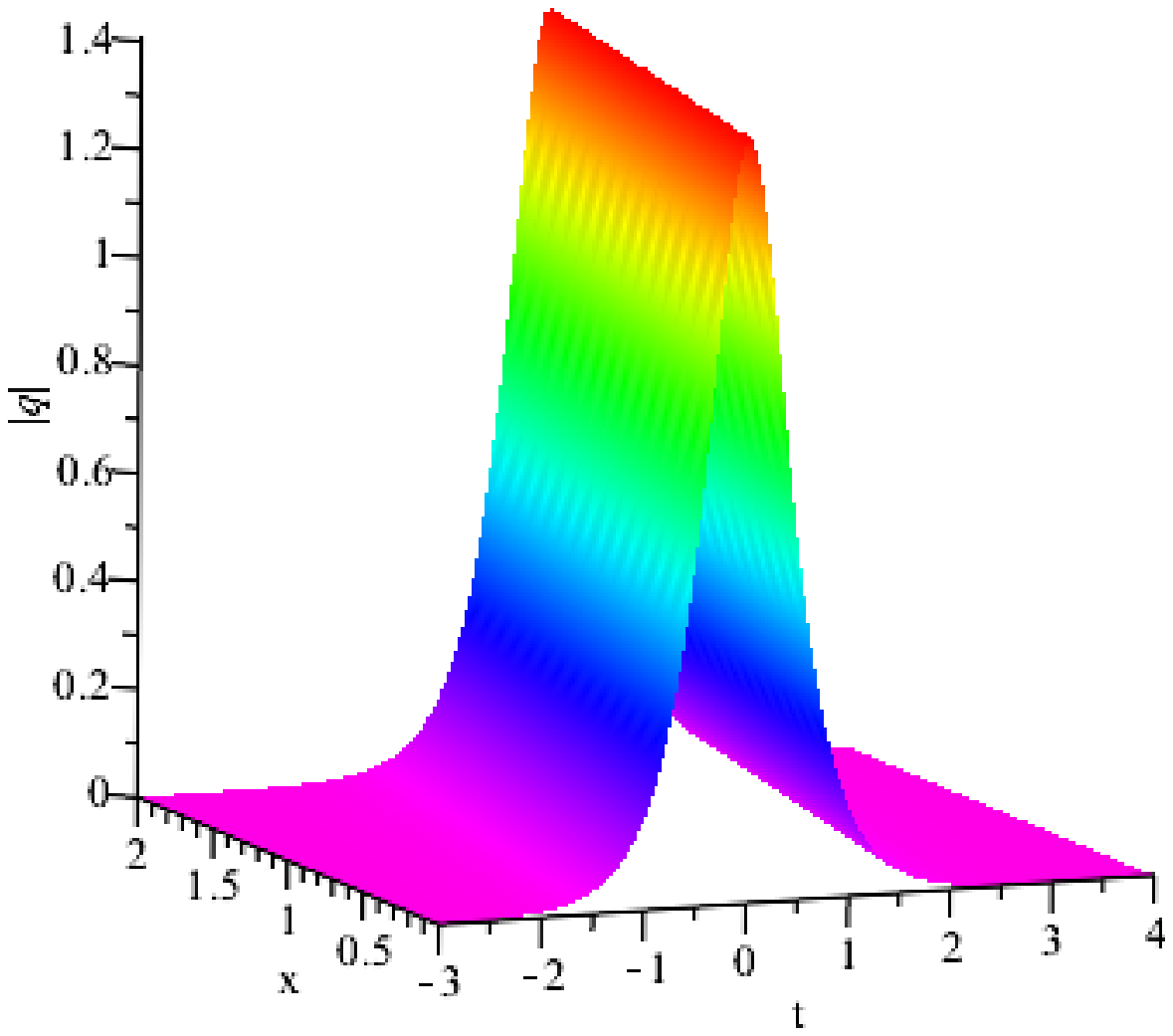}}
~\quad\rotatebox{0}{\includegraphics[width=2.5cm,height=2.4cm,angle=0]{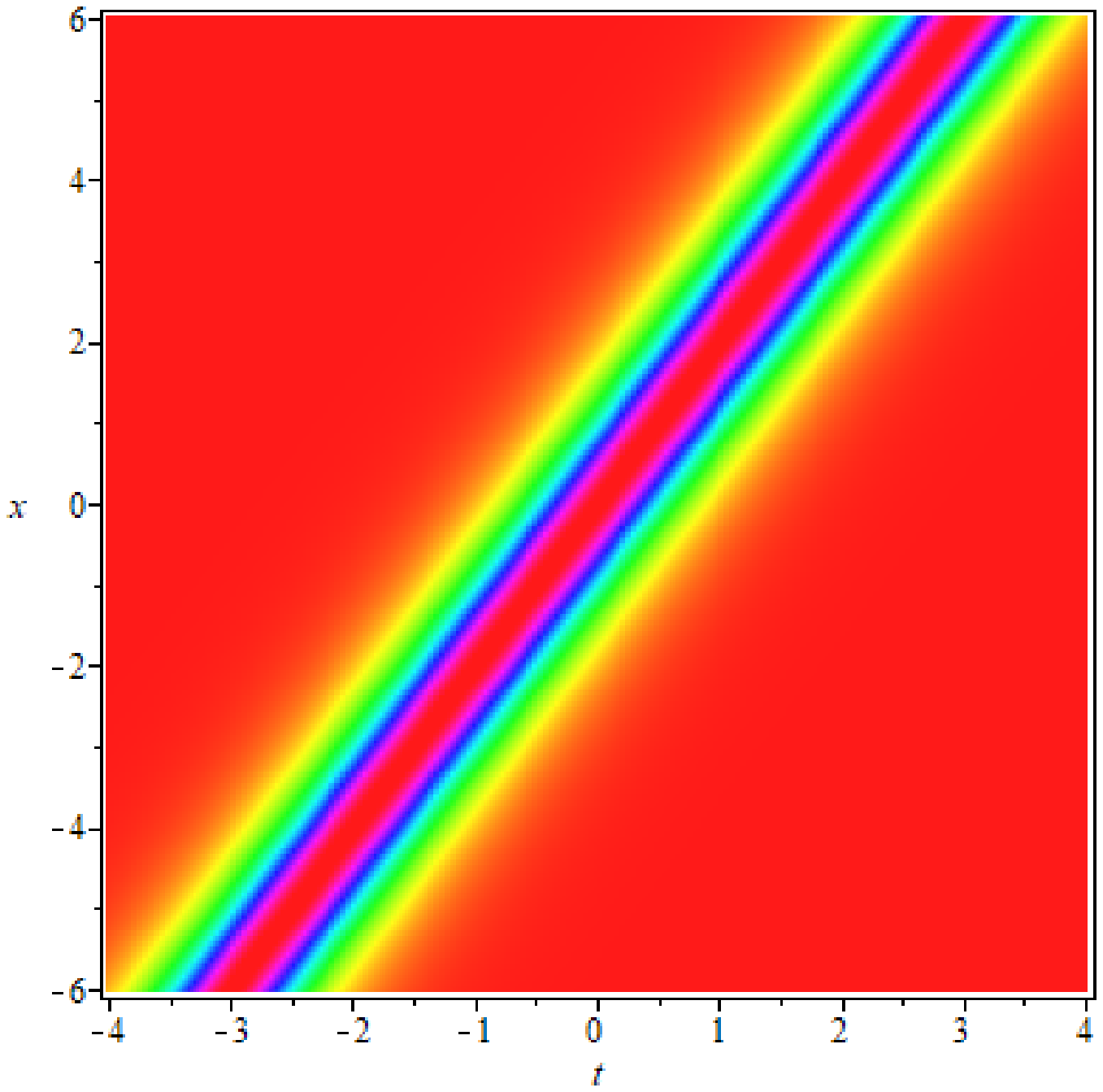}}
~\quad\rotatebox{0}{\includegraphics[width=2.5cm,height=2.4cm,angle=0]{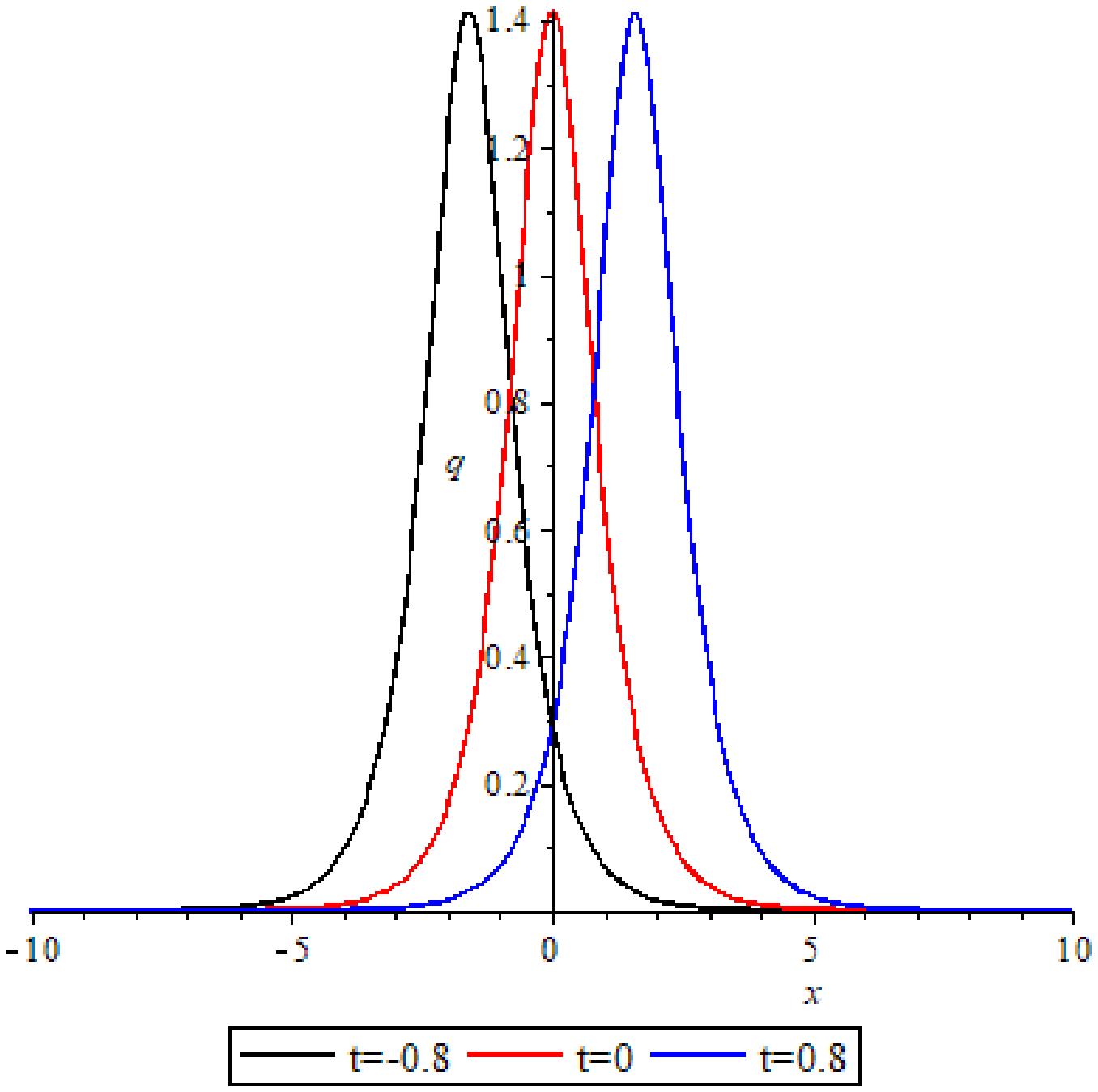}}
~\quad\rotatebox{0}{\includegraphics[width=2.5cm,height=2.4cm,angle=0]{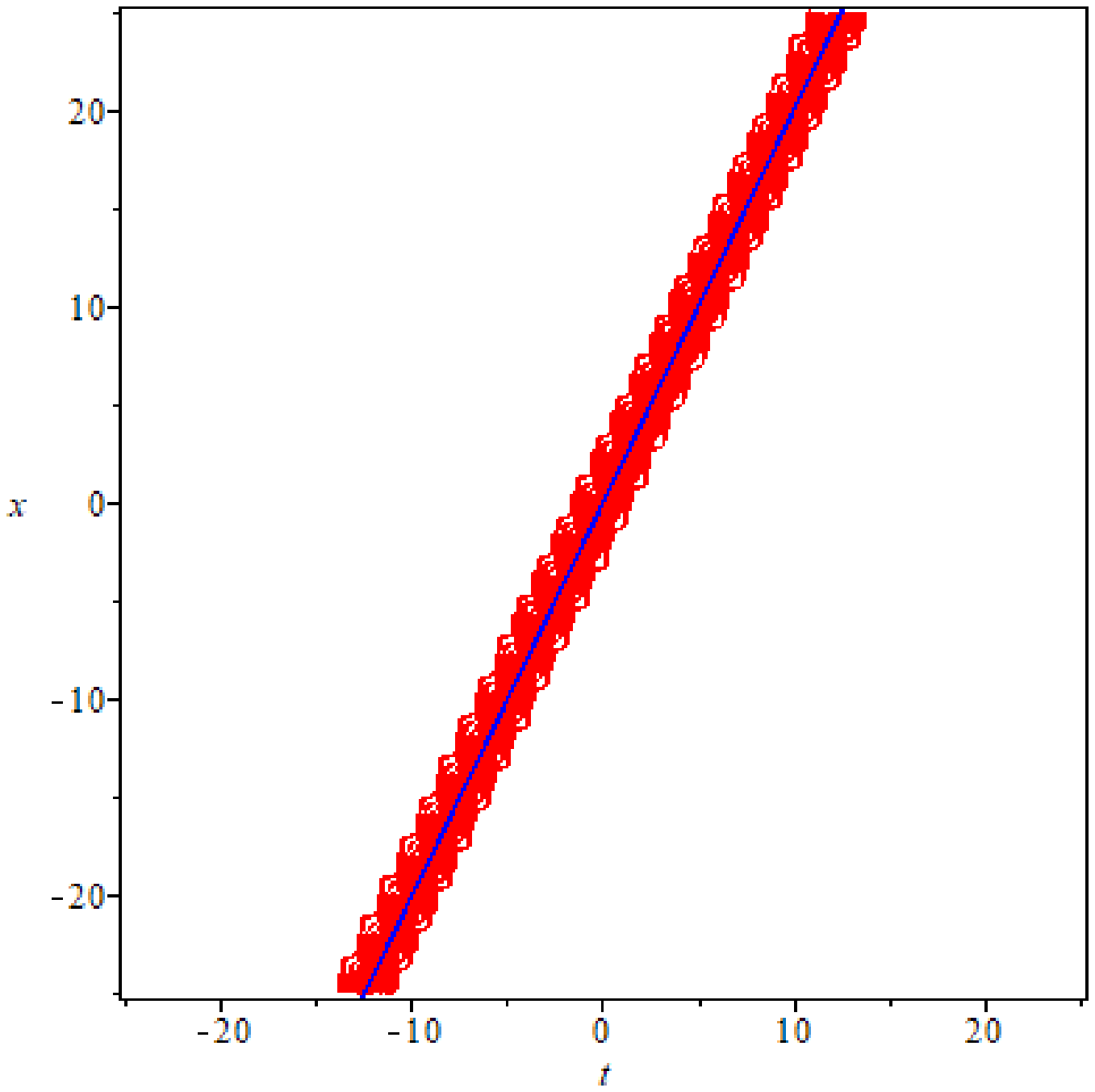}}}

$~~\quad\quad(\textbf{e})\qquad\qquad\qquad\qquad\quad(\textbf{f})
\qquad\qquad\qquad~~~~(\textbf{g})\qquad\qquad~~~\qquad\quad(\textbf{h})$\\
\\
\noindent { \small \textbf{Figure 1.}  The solution \eqref{6} of the equation \eqref{Q1} with the parameters $\textbf{(a)}$ $k_{1}=i$, $\widetilde{k}_{1}=-i$, $\omega_{1}=1$, and $\widetilde{\omega}_{1}=1$; $\textbf{(e)}$:  $k_{1}=\frac{\sqrt{2}}{2}i$, $\widetilde{k}_{1}=-\frac{\sqrt{2}}{2}i$, $\omega_{1}=1$, and $\widetilde{\omega}_{1}=1$; $\textbf{(b)}$ and $\textbf{(f)}$ denote the density of $(\textbf{a})$ and $\textbf{(e)}$, respectively; $\textbf{(c)}$ and $\textbf{(g)}$ represents the dynamic behavior of the one-soliton solutions at different times;  $\textbf{(d)}$ is the characteristic line graph (blue line $L_{1}:x-4t=0$) and contour map of $\textbf{(a)}$; $\textbf{(h)}$  is the characteristic line graph (blue line $L_{2}:x-2t=0$) and contour map of $\textbf{(e)}$.}

From the $\textbf{(a,e)}$, it is found that the situation is similar to the breathing on the top of the bright solitons. We can observe that the amplitude of the graph has changed after rotation. From the combination of characteristic line and contour line in Fig. $\textbf{(c)}$, it appears wave crest. The obvious difference of the Fig. $\textbf{(d)}$ chart is that its dynamic propagation behavior is continuous.

It is not hard to see that they are bright soliton structures, and they both have a peak and two symmetrical valleys. It also applies that these solutions are bounded, that is bounded solutions, which are stable under any parameter. For the characteristic line, we find the relationship between $x$ and $t$. In the process of rotation, the angle between the two lines must be fixed. We change the slope of the characteristic line, which leads to the change of the eigenvalue parameters, so as to construct the dynamic behavior of the soliton solution after rotation. We can also see from the graph that the graph has not changed, only the position or amplitude $|q(x,t)|$ has been changed.
\subsubsection{Pure imaginary eigenvalues with $\omega_{1}\neq\widetilde{\omega}_{1}$}
\
\newline
Considering the case that the eigenvalues are pure imaginary eigenvalues with $\omega_{1}\neq\widetilde{\omega}_{1}$, the propagation behavior of the solution can be characterized by the bright soliton solutions shown in Fig. 2. In addition, another singular soliton solution Fig. 2(\textbf{b}) is obtained by rotating the characteristic line of the singular soliton solution Fig. 2(\textbf{a}). Note that since the characteristic line is related to the eigenvalue, different eigenvalues are selected to obtain the propagation behavior of rotation with $\tan\theta=\frac{2}{9}$.

\qquad\quad\rotatebox{0}{\includegraphics[width=4.7cm,height=4.4cm,angle=0]{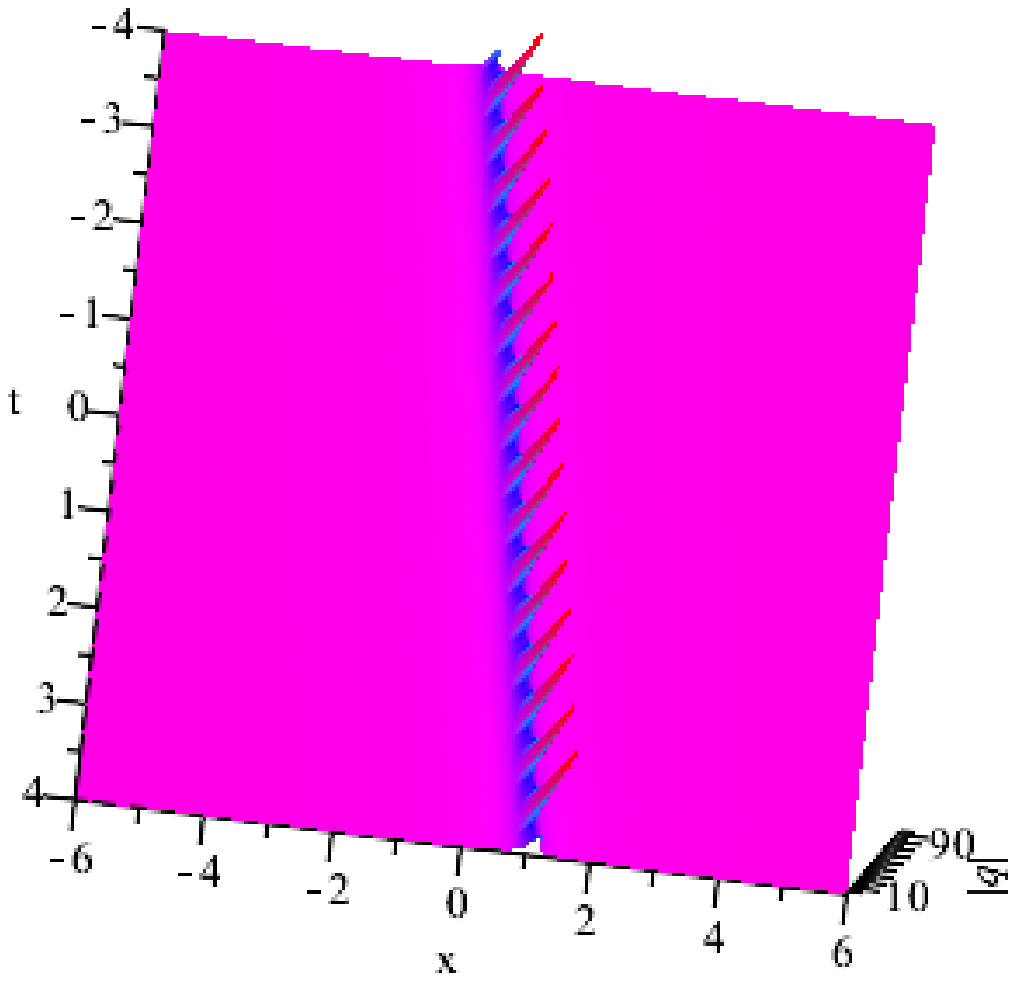}}
~\quad\rotatebox{0}{\includegraphics[width=4.7cm,height=4.4cm,angle=0]{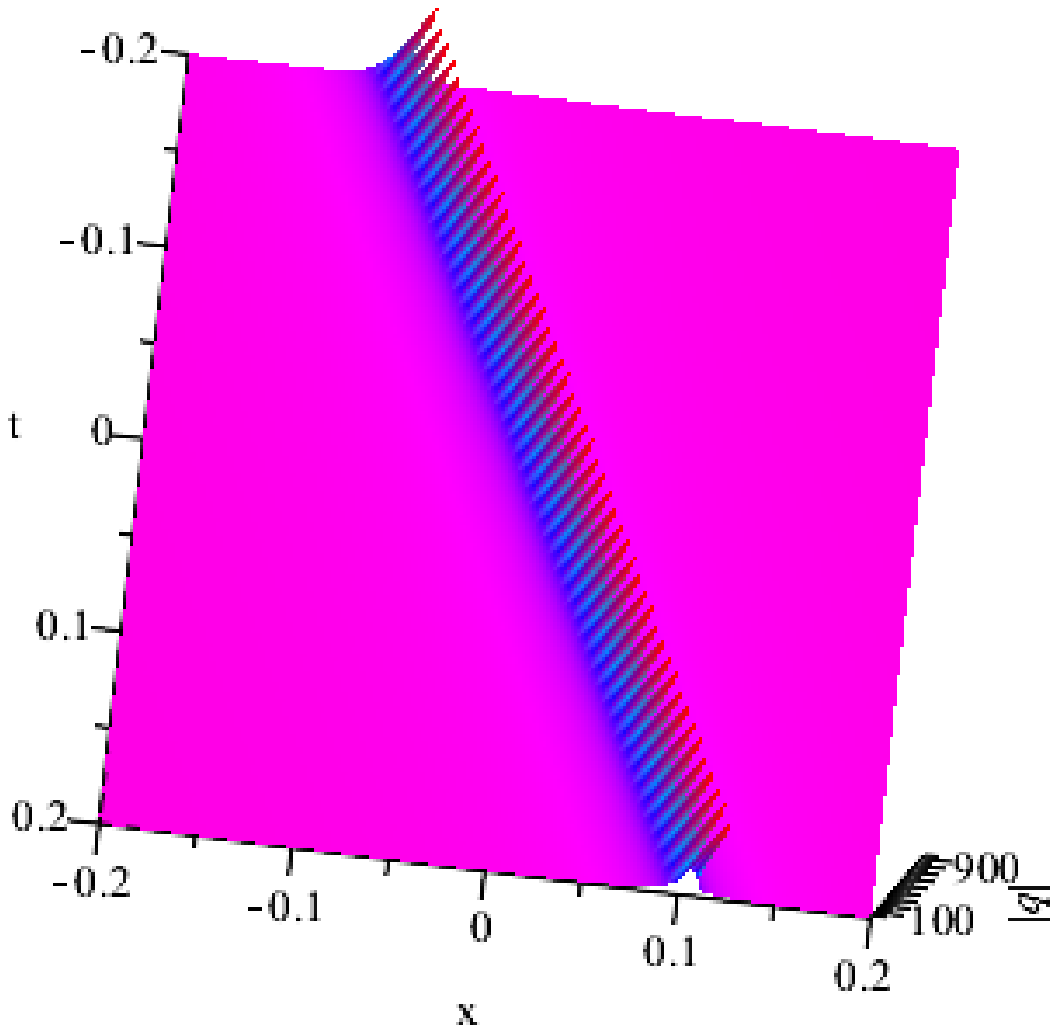}}

$\qquad\qquad\qquad\qquad(\textbf{a})\qquad\qquad\qquad\qquad\qquad\qquad\qquad(\textbf{b})$\\

\qquad~~\quad\rotatebox{0}{\includegraphics[width=3.2cm,height=2.7cm,angle=0]{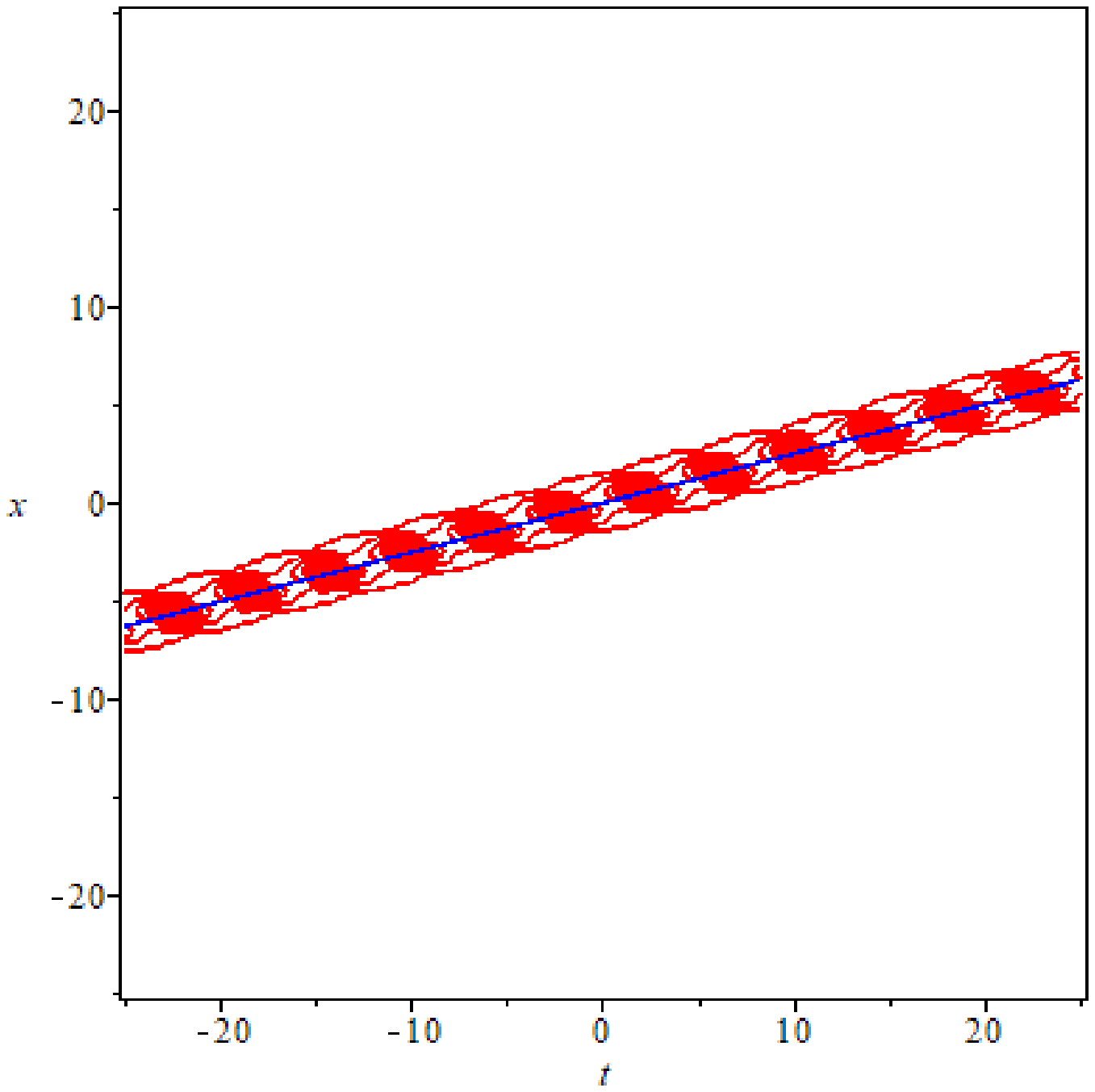}}
\qquad\qquad~~\quad\rotatebox{0}{\includegraphics[width=3.2cm,height=2.7cm,angle=0]{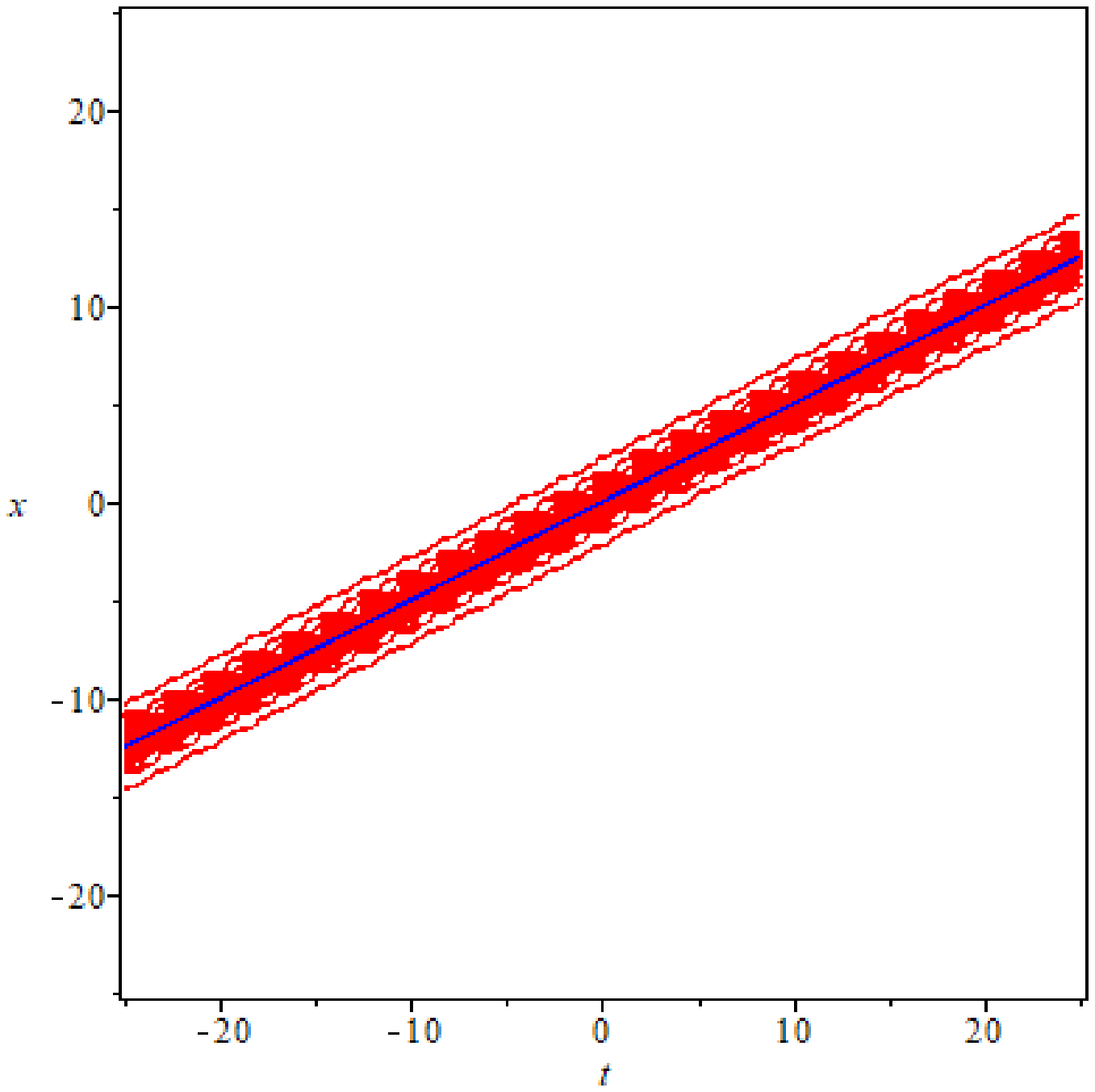}}

$\qquad\qquad\qquad\qquad(\textbf{c})~~~\qquad\qquad\quad\qquad\qquad\qquad\qquad(\textbf{d})$
\\
\noindent { \small \textbf{Figure 2.} The solution \eqref{6} of the equation \eqref{Q1} with the parameters $\textbf{(a)}$  $k_{1}=\frac{i}{4}$, $\widetilde{k}_{1}=-\frac{i}{4}$, $\omega_{1}=1$, and $\widetilde{\omega}_{1}=-1$; $\textbf{(b)}$  $k_{1}=\frac{\sqrt{2}}{4}i$, $\widetilde{k}_{1}=-\frac{\sqrt{2}}{4}i$, $\omega_{1}=1$, and $\widetilde{\omega}_{1}=-1$; $\textbf{(c)}$  is the characteristic line graph (blue line $L_{1}=x-\frac{1}{4}t$) and contour map of $\textbf{(a)}$; $\textbf{(d)}$  is the characteristic line graph (blue line $L_{2}:x-\frac{1}{2}t=0$) and contour map of $\textbf{(b)}$.}

In the case of these parameters, the propagation behavior of the soliton solution is shown in the upper half of the graph, and the corresponding characteristic lines and contour maps are shown in the lower half of the graph.

Comparing with the figures in Fig. 1, the eigenvalue parameters we choose are both pure imaginary numbers,  but the dynamic behavior is different. The reason for this phenomenon may be the different values of $\omega$ in the symmetry relationship. We can observe from the image that one-soliton solution is collapsing repeatedly, and this kind of solution is called singular solution. Finally, we can clearly see that one-soliton wave has a characteristic line on $(x, t)$. A new characteristic line is generated via rotating the angle. The change of the slope for the characteristic line leads to the change of the eigenvalues. In general, it does not affect the change of the graph, but the position of the graph.

\subsubsection{Non-pure imaginary eigenvalues}
\
\newline
For the third case, we take the  eigenvalues as follows,
\begin{align}
k_{1}=0.3+0.7i,~ \widetilde{k}_{1}=0.3-0.7i, ~\omega_{1}=\widetilde{\omega}_{1}=1,
\end{align}
then the solution of \eqref{6} can be characterized in Fig. 3.

\noindent{\rotatebox{0}{\includegraphics[width=3.3cm,height=2.8cm,angle=0]{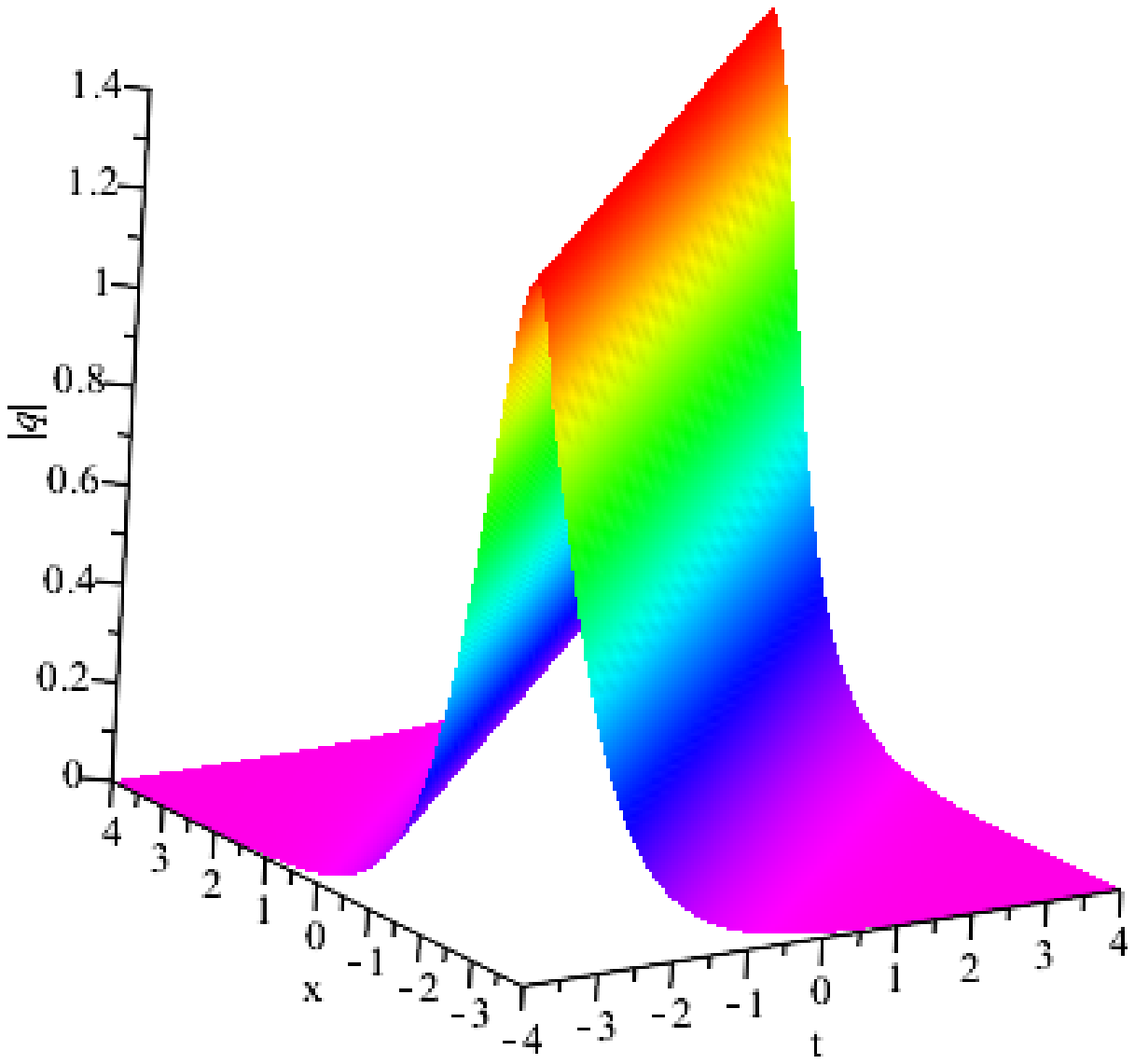}}
~\quad\rotatebox{0}{\includegraphics[width=2.5cm,height=2.5cm,angle=0]{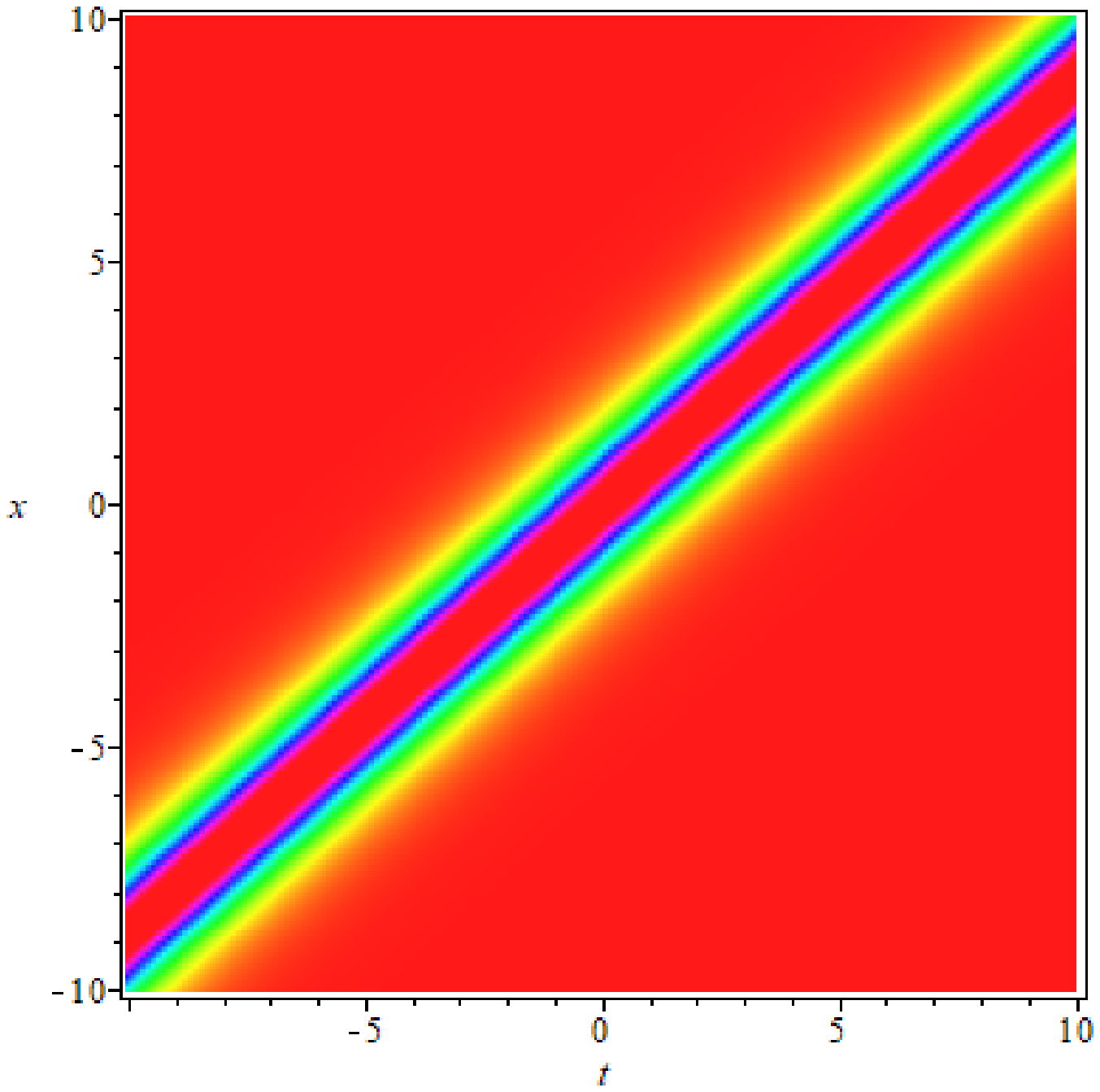}}
~\quad\rotatebox{0}{\includegraphics[width=2.5cm,height=2.5cm,angle=0]{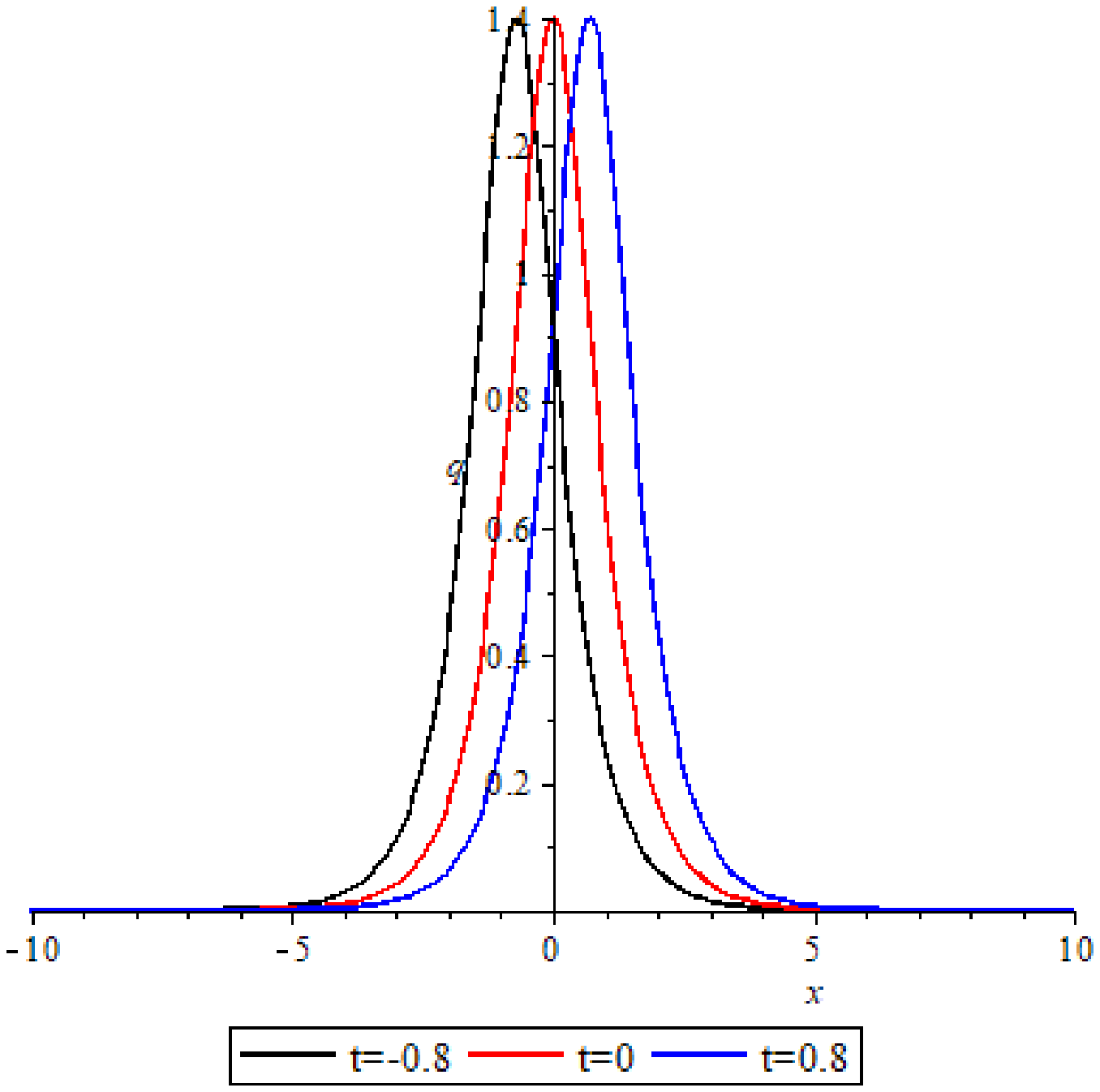}}
~\quad\rotatebox{0}{\includegraphics[width=2.5cm,height=2.5cm,angle=0]{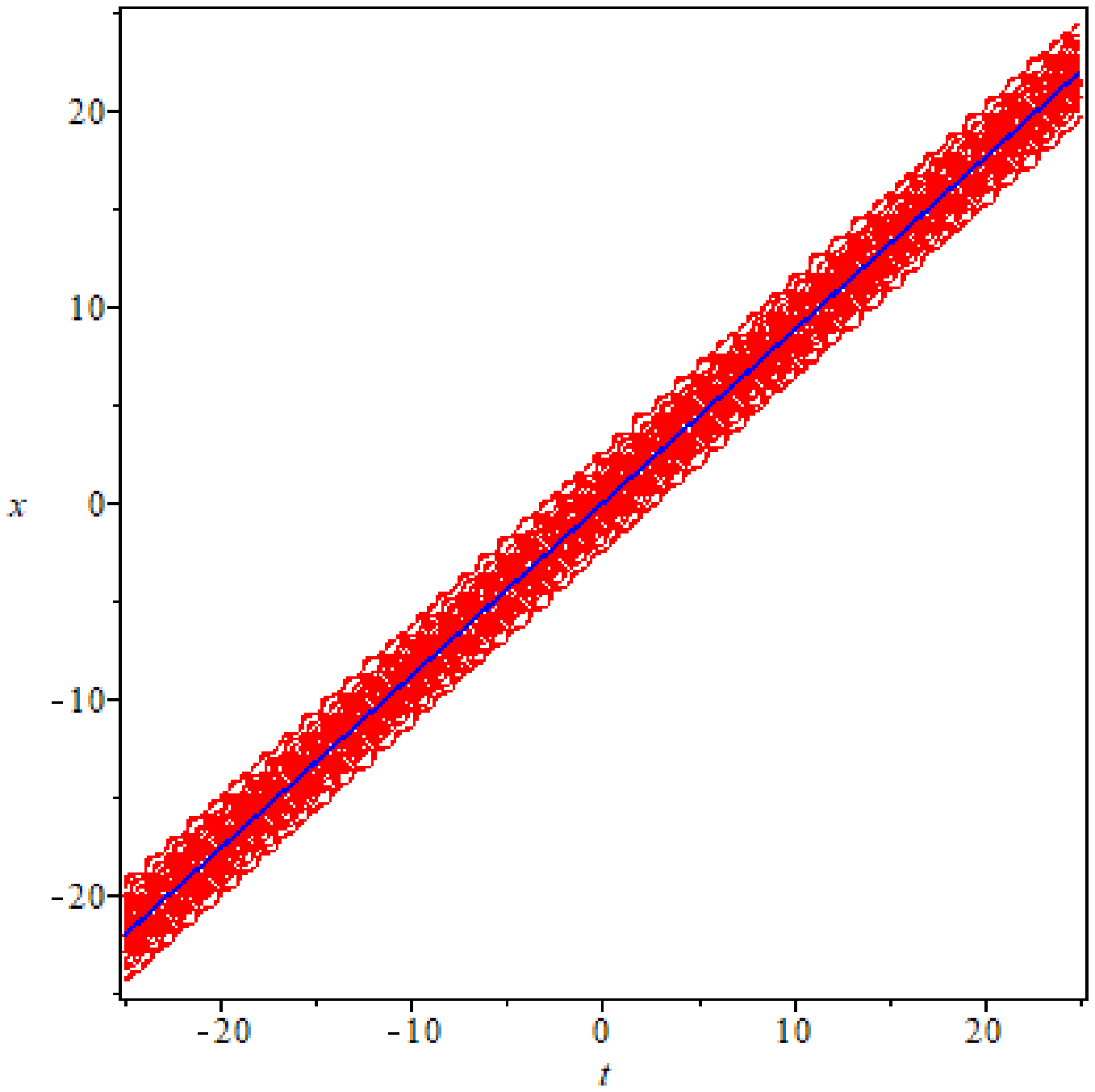}}}

$~~\quad\quad(\textbf{a})\qquad\qquad\qquad\qquad\quad(\textbf{b})
\qquad\qquad\qquad~~~(\textbf{c})\qquad\qquad~~~\qquad\quad(\textbf{d})$\\

\noindent{\rotatebox{0}{\includegraphics[width=3.3cm,height=2.8cm,angle=0]{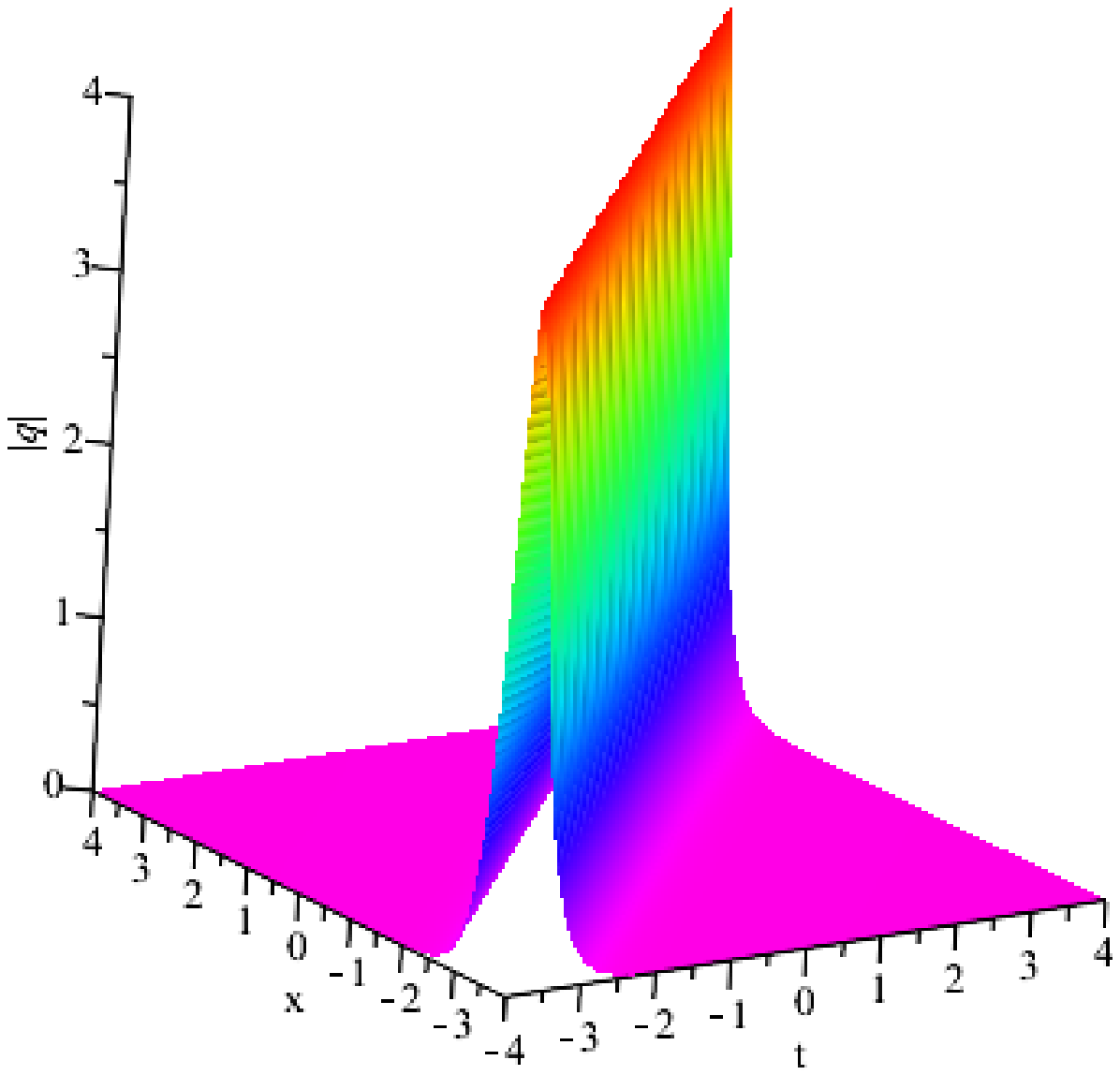}}
~\quad\rotatebox{0}{\includegraphics[width=2.5cm,height=2.5cm,angle=0]{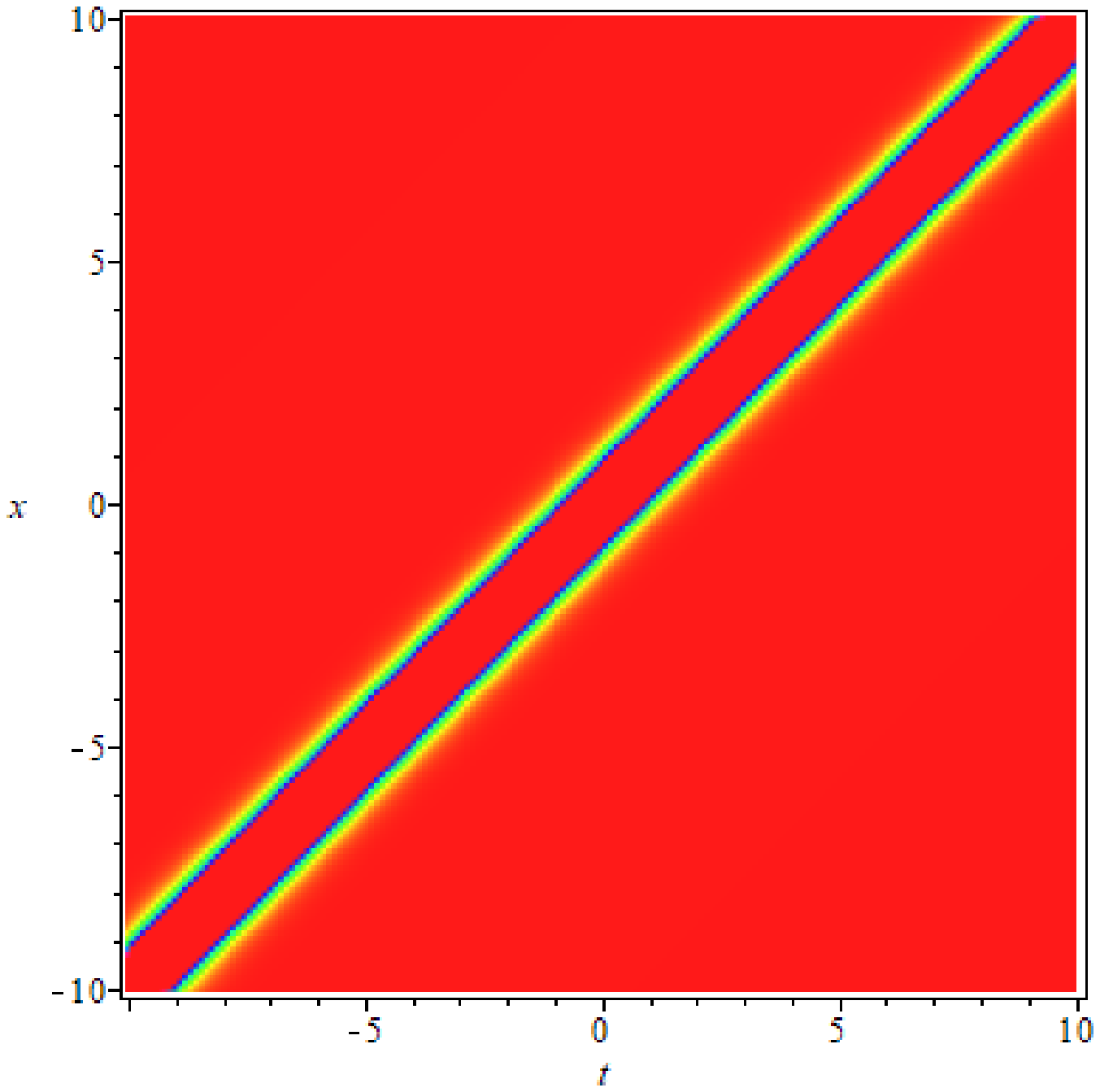}}
~\quad\rotatebox{0}{\includegraphics[width=2.5cm,height=2.5cm,angle=0]{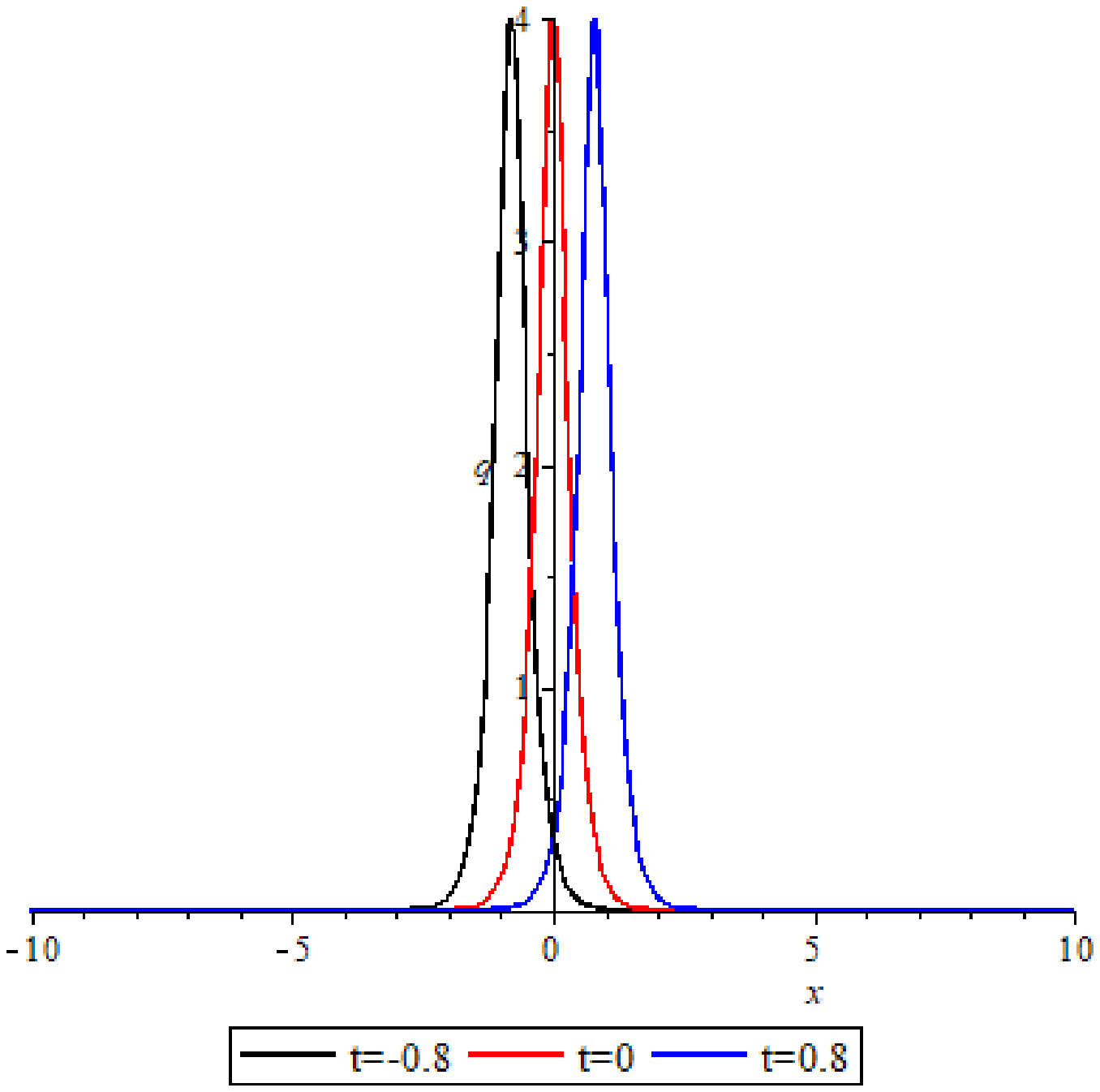}}
~\quad\rotatebox{0}{\includegraphics[width=2.5cm,height=2.5cm,angle=0]{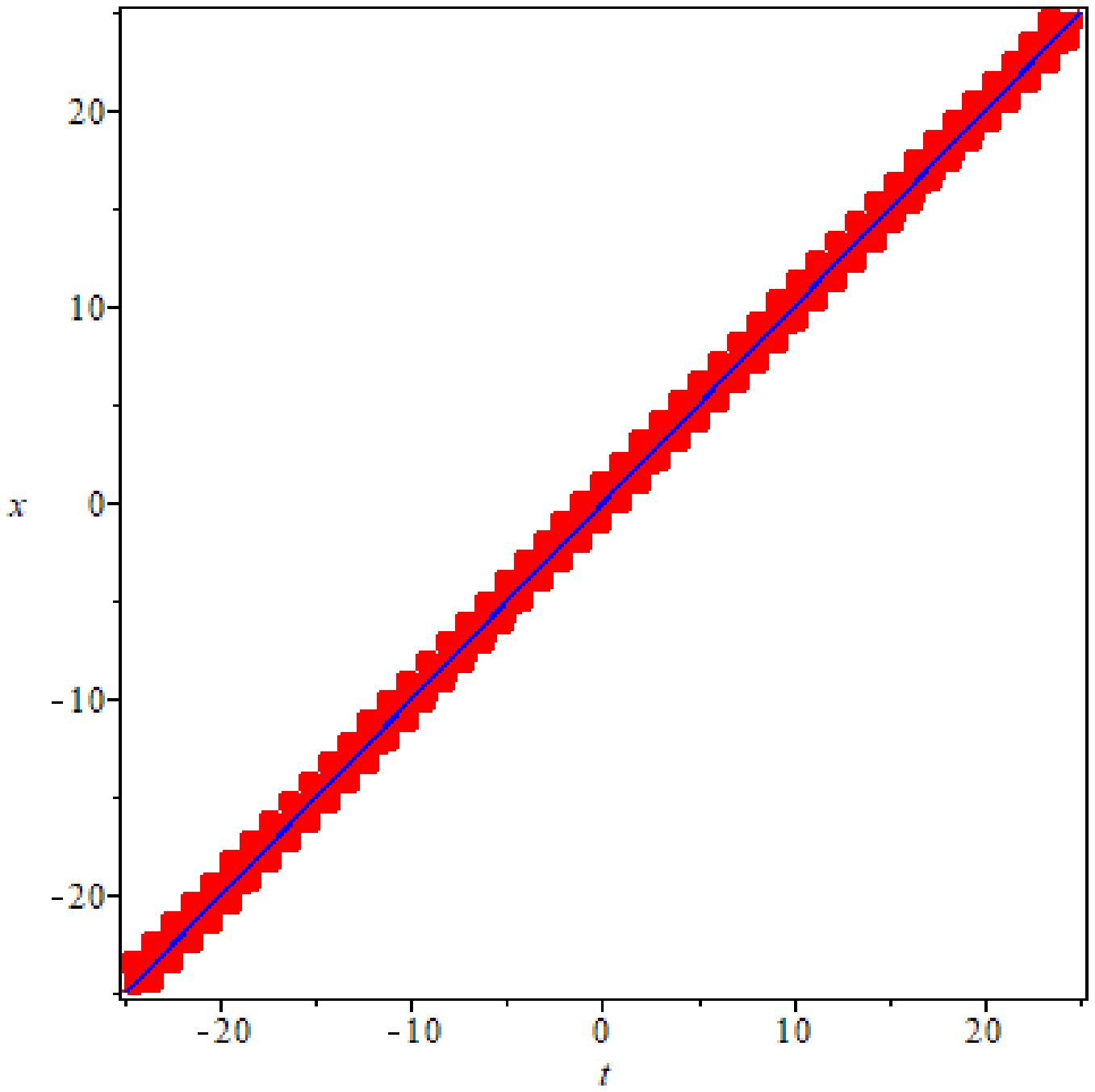}}}

$~~\quad\quad(\textbf{e})\qquad\qquad\qquad\qquad\quad(\textbf{f})
\qquad\qquad\qquad~~~~(\textbf{g})\qquad\qquad~~~\qquad\quad(\textbf{h})$\\

\noindent { \small \textbf{Figure 3.}  The solution \eqref{6} of the equation \eqref{Q1} with the parameters $\textbf{(a)}$ $k_{1}=0.3+0.7i$, $\widetilde{k}_{1}=0.3-0.7i$, $\omega_{1}=1$, and $\widetilde{\omega}_{1}=1$; $\textbf{(e)}$:  $k_{1}=\frac{\sqrt{5}}{2}+2i$, $\widetilde{k}_{1}=\frac{\sqrt{5}}{2}-2i$, $\omega_{1}=1$, and $\widetilde{\omega}_{1}=1$; $\textbf{(b)}$ and $\textbf{(f)}$ denote the density of $(\textbf{a})$ and $\textbf{(e)}$, respectively; $\textbf{(c)}$ and $\textbf{(g)}$ represents the dynamic behavior of the one-soliton solutions at different times;  $\textbf{(d)}$ is the characteristic line graph (blue line $L_{1}:x-0.88t=0$) and contour map of $\textbf{(a)}$; $\textbf{(h)}$  is the characteristic line graph (blue line $L_{2}:x-t=0$) and contour map of $\textbf{(e)}$.}

As seen in  Fig. 3, this is the third group of $N=1$, where Fig. 3(\textbf{b}) is the graph of Fig. 3(\textbf{a}) after rotation with $\tan\theta=\frac{3}{47}$. By comparing the dynamic behaviors of soliton solutions, we find that the factors that can affect the propagation behavior not only related to the selected eigenvalue parameters, but also related to the coefficients $\omega_{1}$ and $\widetilde{\omega}_{1}$ in the symmetry relationship.

Comparing with the pictures shown in Fig. 1, we find that the two groups have different parameter forms (one is pure imaginary number, the other is non-pure imaginary number), but the dynamic behaviors of their soliton solutions is very similar, which have one peak and two symmetrical valleys. In the process of propagation, the amplitude, velocity and width of single soliton are changed. It can be seen that the amplitude of the excited state is limited and this soliton solutions are bounded and stable.

\subsection{Two-soliton solutions}
\
\newline
We take $N$=2 in the formulae \eqref{5} in order to get two-soliton solutions, the specific form of the solution of the equation \eqref{5} can be expressed as
\begin{align}
q(x,t)=-2i\frac{\det\left(
                      \begin{array}{ccc}
                        0 & a_{1}e^{\theta_{1}} & a_{2}e^{\theta_{2}} \\
                        \widetilde{\omega}_{1}e^{-\widetilde{\theta}_{1}} & m_{11} & m_{12} \\
                        \widetilde{\omega}_{2}e^{-\widetilde{\theta}_{2}} & m_{21} & m_{22} \\
                      \end{array}
                    \right)
}{\det\left(
        \begin{array}{cc}
          m_{11} & m_{12} \\
          m_{21} & m_{22} \\
        \end{array}
      \right)
},
\end{align}
where
\begin{align*}
m_{11}&=\frac{e^{\theta_{1}+\widetilde{\theta}_{1}}+
\omega_{1}\widetilde{\omega}_{1}e^{-(\theta_{1}+\widetilde{\theta}_{1})}}{\widetilde{k}_{1}-k_{1}},~~
m_{12}=\frac{e^{\theta_{2}+\widetilde{\theta}_{1}}+
\omega_{1}\widetilde{\omega}_{1}e^{-(\theta_{2}+\widetilde{\theta}_{1})}}{\widetilde{k}_{1}-k_{2}},
\end{align*}
\begin{align*}
m_{21}&=\frac{e^{\theta_{1}+\widetilde{\theta}_{2}}+
\omega_{1}\widetilde{\omega}_{2}e^{-(\theta_{1}+\widetilde{\theta}_{2})}}{\widetilde{k}_{2}-k_{1}},~~
m_{22}=\frac{e^{\theta_{2}+\widetilde{\theta}_{2}}+
\omega_{2}\widetilde{\omega}_{2}e^{-(\theta_{2}+\widetilde{\theta}_{2})}}{\widetilde{k}_{2}-k_{2}},
\end{align*}
where $\omega_{1}=\pm1$, $\widetilde{\omega}_{1}=\pm1$, $\theta_{1}\in \mathbb{D}^{+}$
and $\widetilde{\theta}_{1}\in \mathbb{D}^{-}$.  As a comparison, we discuss the following two situations.
\subsubsection{Two-soliton solutions with pure imaginary eigenvalues}
\
\newline
Now we disscuss the case that $k_{1},k_{2}\in \mathbb{D}^{+}$, $\widetilde{k}_{1},\widetilde{k}_{2}\in \mathbb{D}^{-}$ are pure imaginary eigenvalues. \\
(1)Unbounded two-soliton solutions

%\noindent\rotatebox{0}{\includegraphics[width=4.5cm,height=4.0cm,angle=0]{1-7.eps}}
%\rotatebox{0}{\includegraphics[width=3.3cm,height=3.2cm,angle=0]{1-7-2.eps}}
%\hspace{-3.5cm}
%\centerline{\begin{tikzpicture}[scale=0.53]
%\draw[-][thick](-3,0)--(-2,0);
%\draw[-][thick](-2,0)--(-1,0);
%\draw[-][thick](-1,0)--(0,0);
%\draw[-][thick](0,0)--(1,0);
%\draw[-][thick](1,0)--(2,0);
%\draw[-][thick](2,0)--(3,0);
%\draw[-][thick](3,0)--(3,2);
%\draw[-][thick](3,2)--(3,5.7);
%\draw[-][thick](3,5.7)--(0,5.7);
%\draw[-][thick](0,5.7)--(-3,5.7);
%\draw[-][thick](-3,5.7)--(-3,2);
%\draw[-][thick](-3,2)--(-3,0);
%\draw[-][thick](-3,0)--(0,0);
%\draw[-][thick](0,0)--(3,0);
%\draw[-][dashed](-3,3)--(0,3);
%\draw[-][dashed](0,3)--(3,3);
%\draw[-][dashed](0,5.7)--(0,3);
%\draw[-][dashed](0,3)--(0,0);
%\draw[fill] (0,3.6)node{$\textcolor[rgb]{1.00,0.00,0.00}{\bullet}$};
%\draw[fill] (0,2.3)node{$\textcolor[rgb]{1.00,0.00,0.00}{\bullet}$};
%\draw[fill] (0,1.5)node{$\textcolor[rgb]{0.00,0.00,1.00}{\bullet}$};
%\draw[fill] (0,4.4)node{$\textcolor[rgb]{0.00,0.00,1.00}{\bullet}$};
%\draw[-][dashed](2,3)--(3,3)node[right]{$Re~z$};
%\draw[-][dashed](0,5.5)--(0,5.7)node[above]{$Im~z$};
%\end{tikzpicture}}
%
%$\qquad~~\quad(\textbf{a})\qquad~~\quad\qquad\qquad\qquad\qquad(\textbf{b})
%\quad\qquad~~~~\qquad\qquad\qquad(\textbf{c})$

\noindent{\rotatebox{0}{\includegraphics[width=3.3cm,height=2.8cm,angle=0]{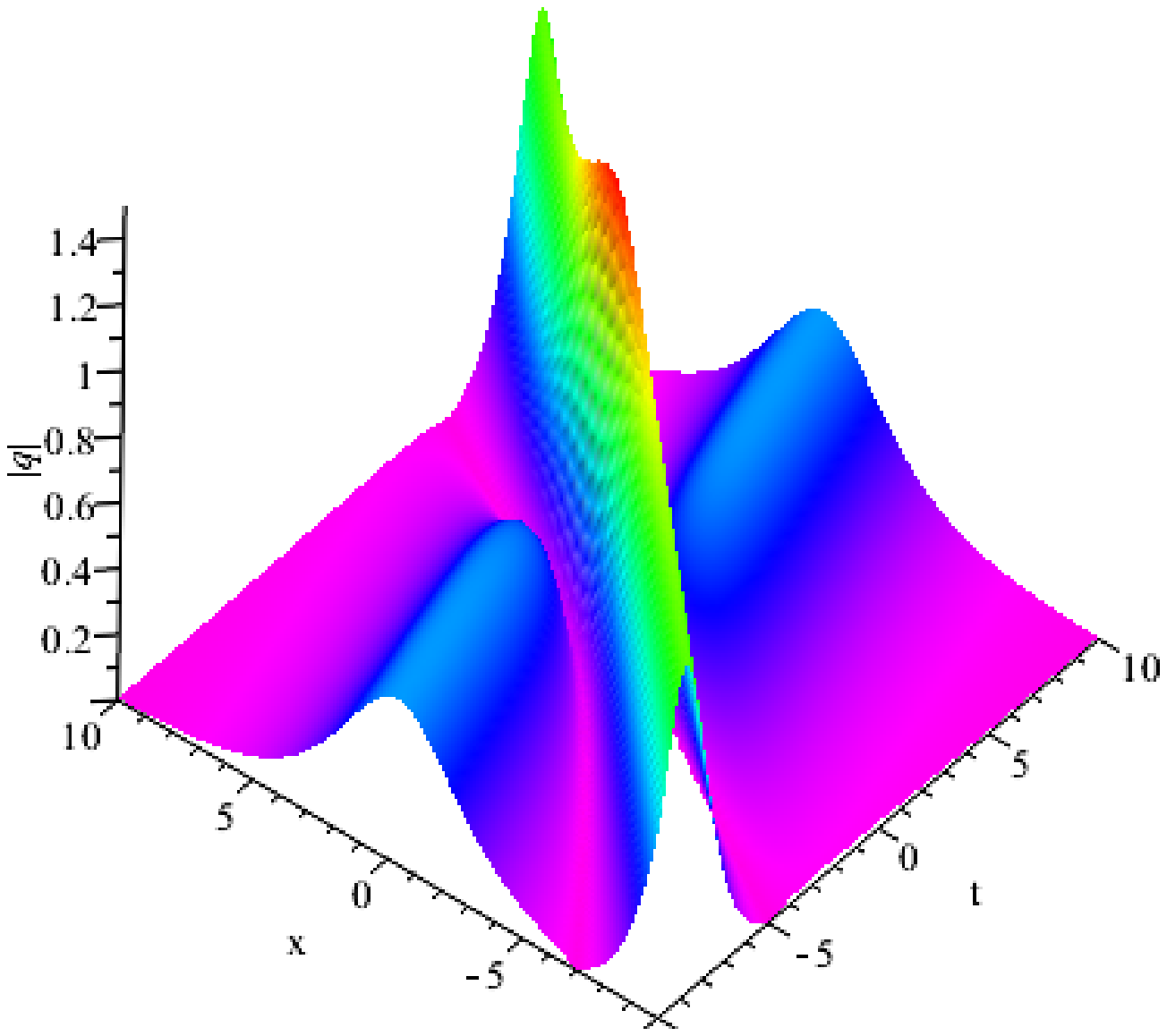}}
~\quad\rotatebox{0}{\includegraphics[width=2.5cm,height=2.5cm,angle=0]{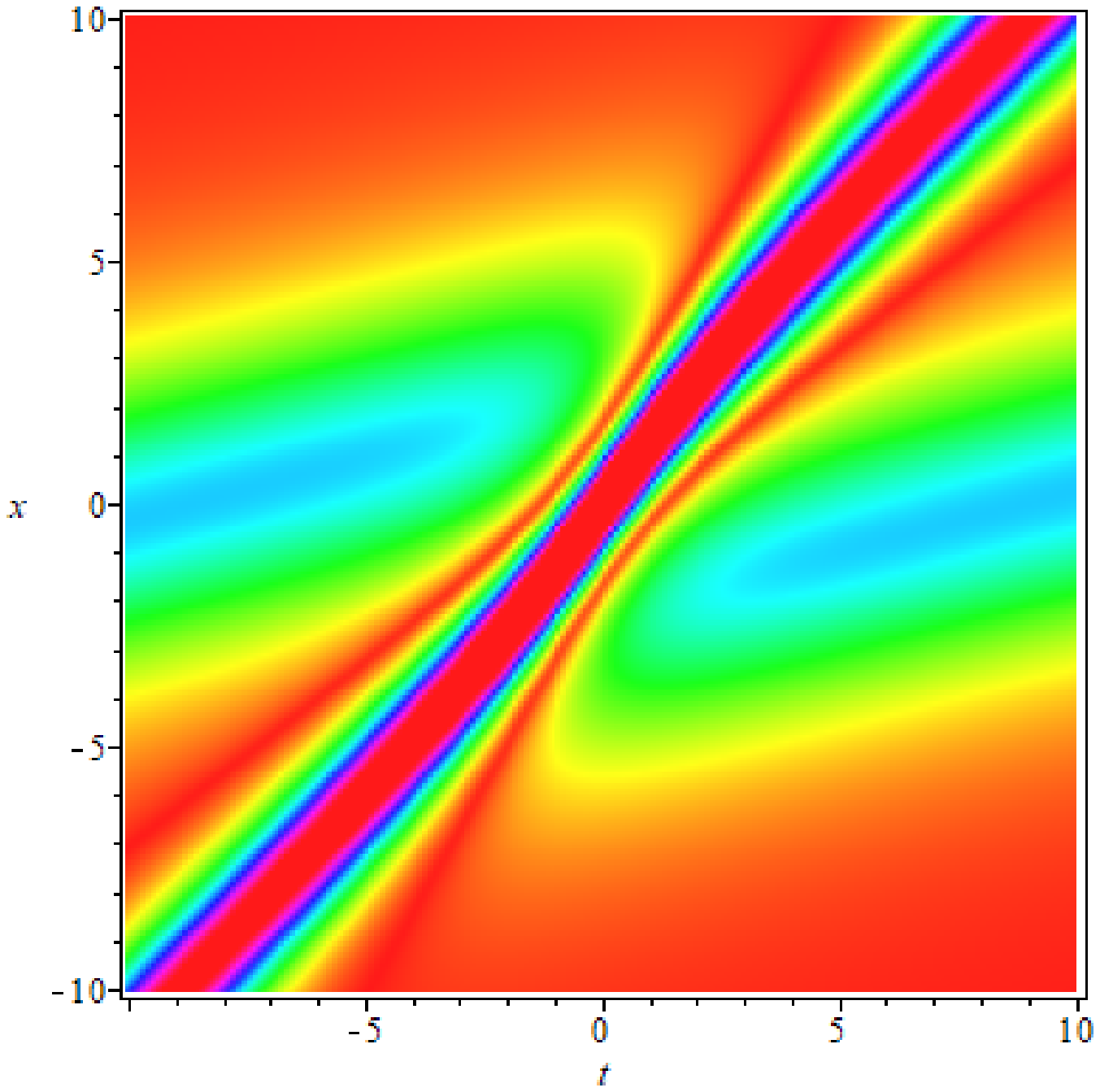}}
~\quad\rotatebox{0}{\includegraphics[width=2.5cm,height=2.5cm,angle=0]{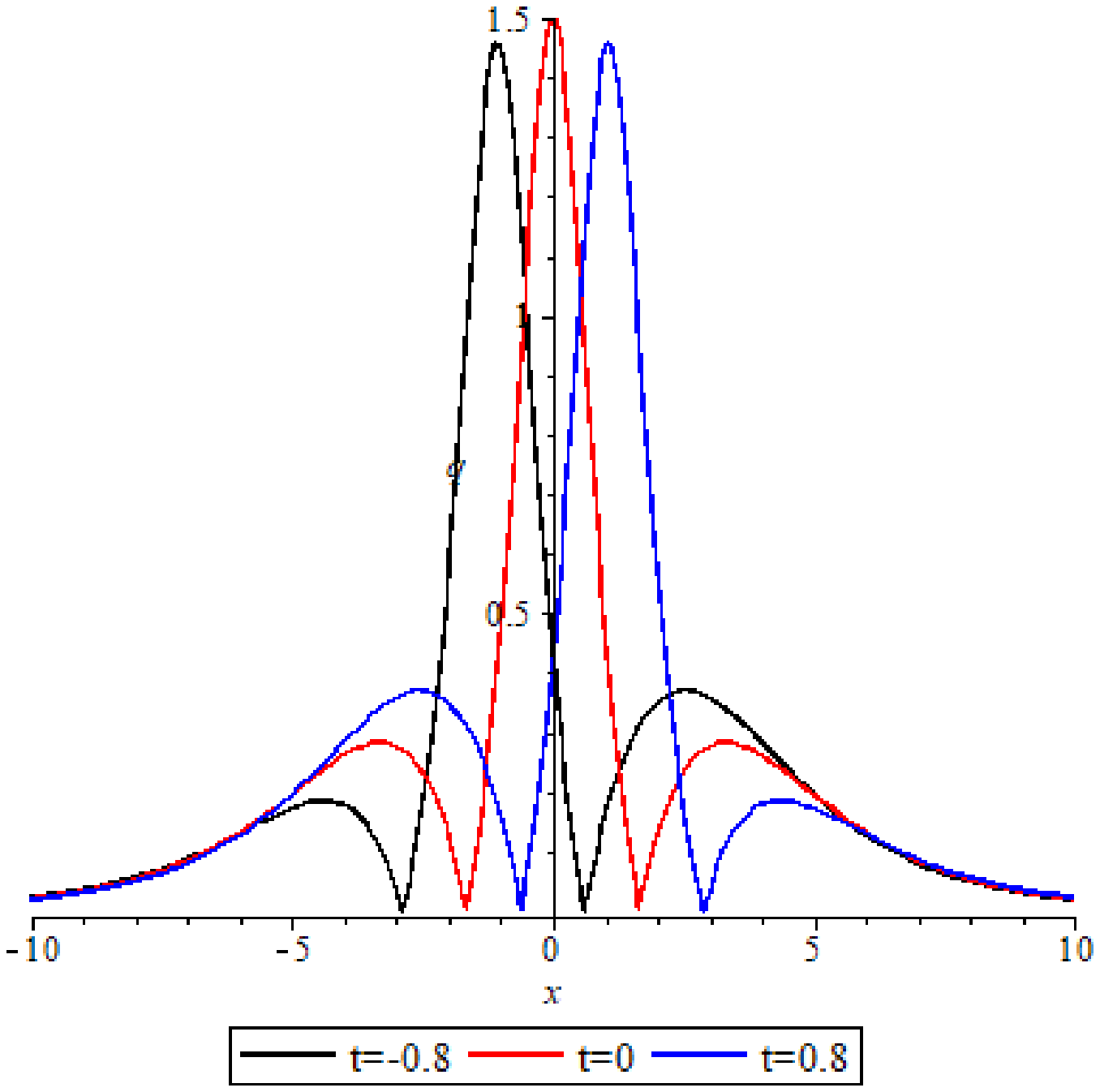}}
~\quad\rotatebox{0}{\includegraphics[width=2.5cm,height=2.5cm,angle=0]{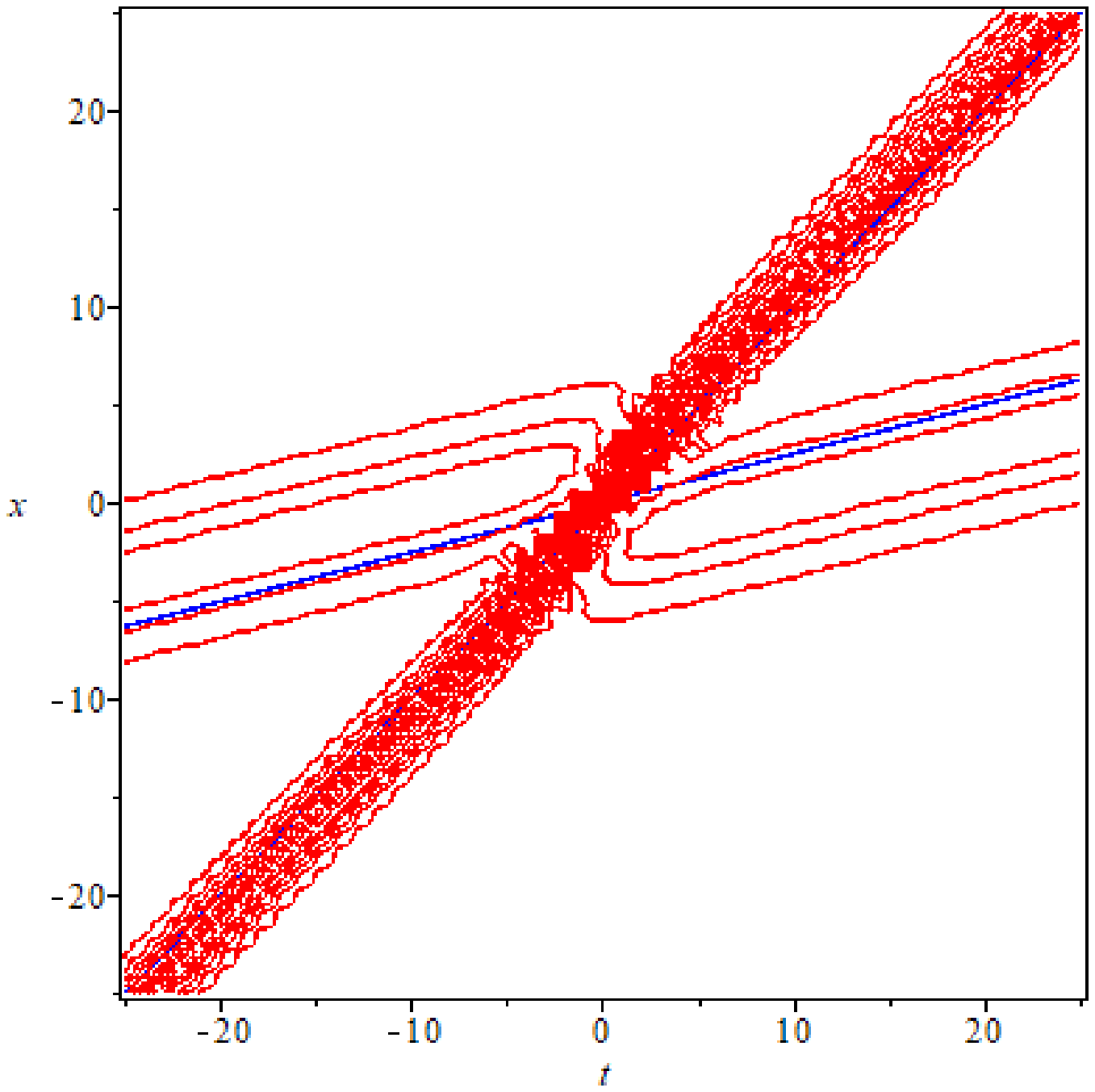}}}

$~~\quad\quad(\textbf{a})\qquad\qquad\qquad\qquad\quad(\textbf{b})
\qquad\qquad\qquad~~~(\textbf{c})\qquad\qquad~~~\qquad\quad(\textbf{d})$\\

\noindent{\rotatebox{0}{\includegraphics[width=3.3cm,height=2.8cm,angle=0]{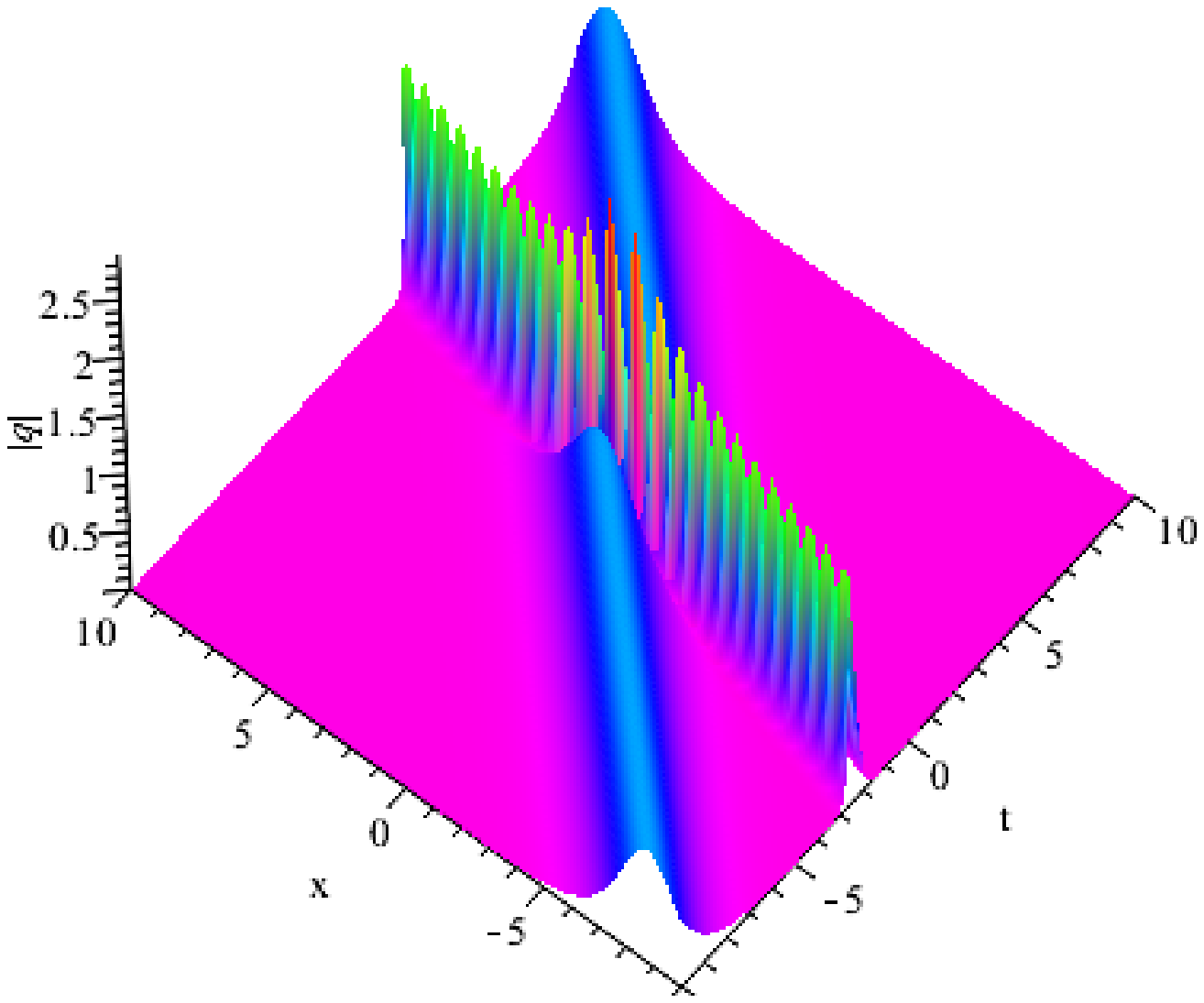}}
~\quad\rotatebox{0}{\includegraphics[width=2.5cm,height=2.5cm,angle=0]{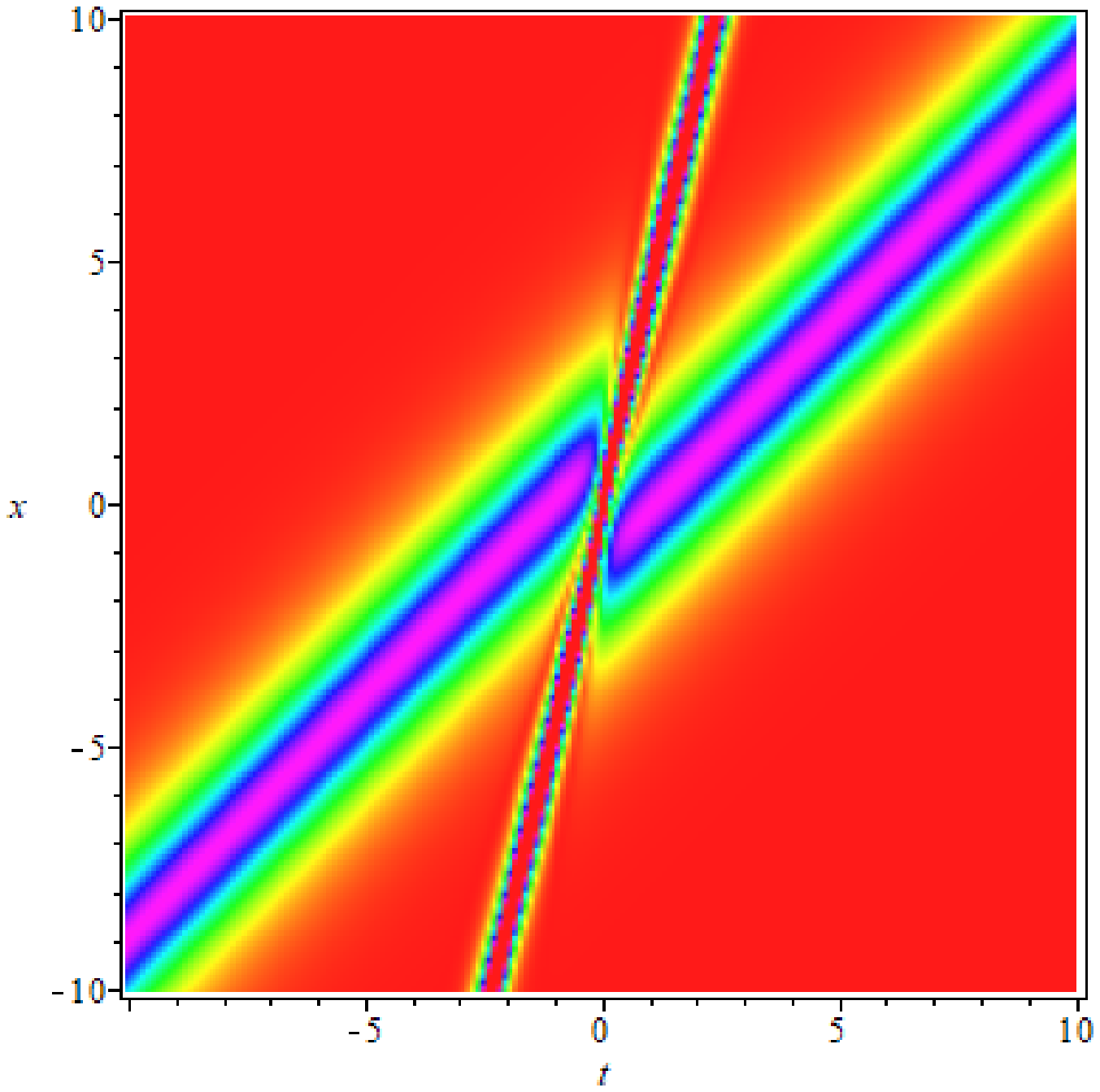}}
~\quad\rotatebox{0}{\includegraphics[width=2.5cm,height=2.5cm,angle=0]{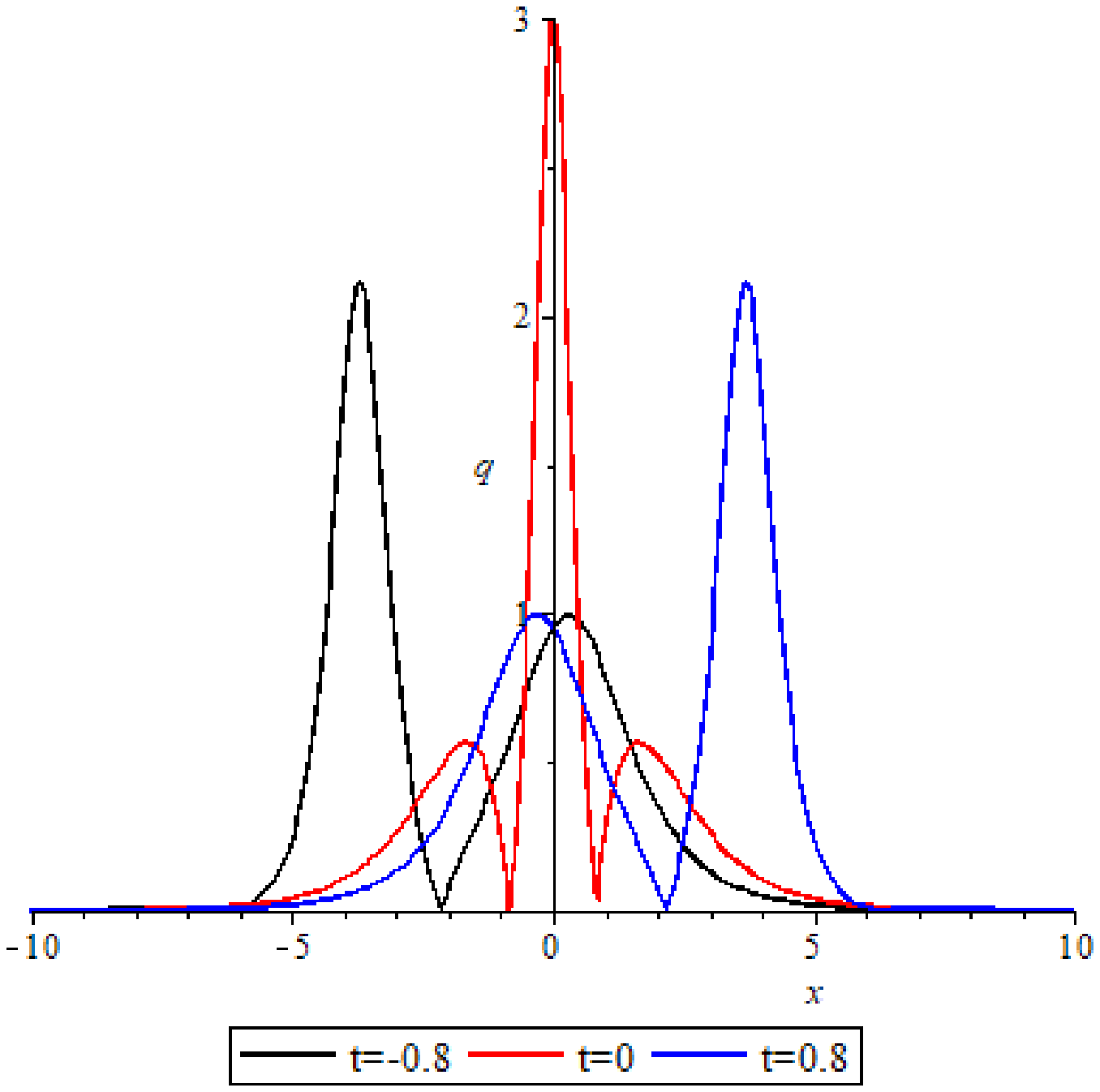}}
~\quad\rotatebox{0}{\includegraphics[width=2.5cm,height=2.5cm,angle=0]{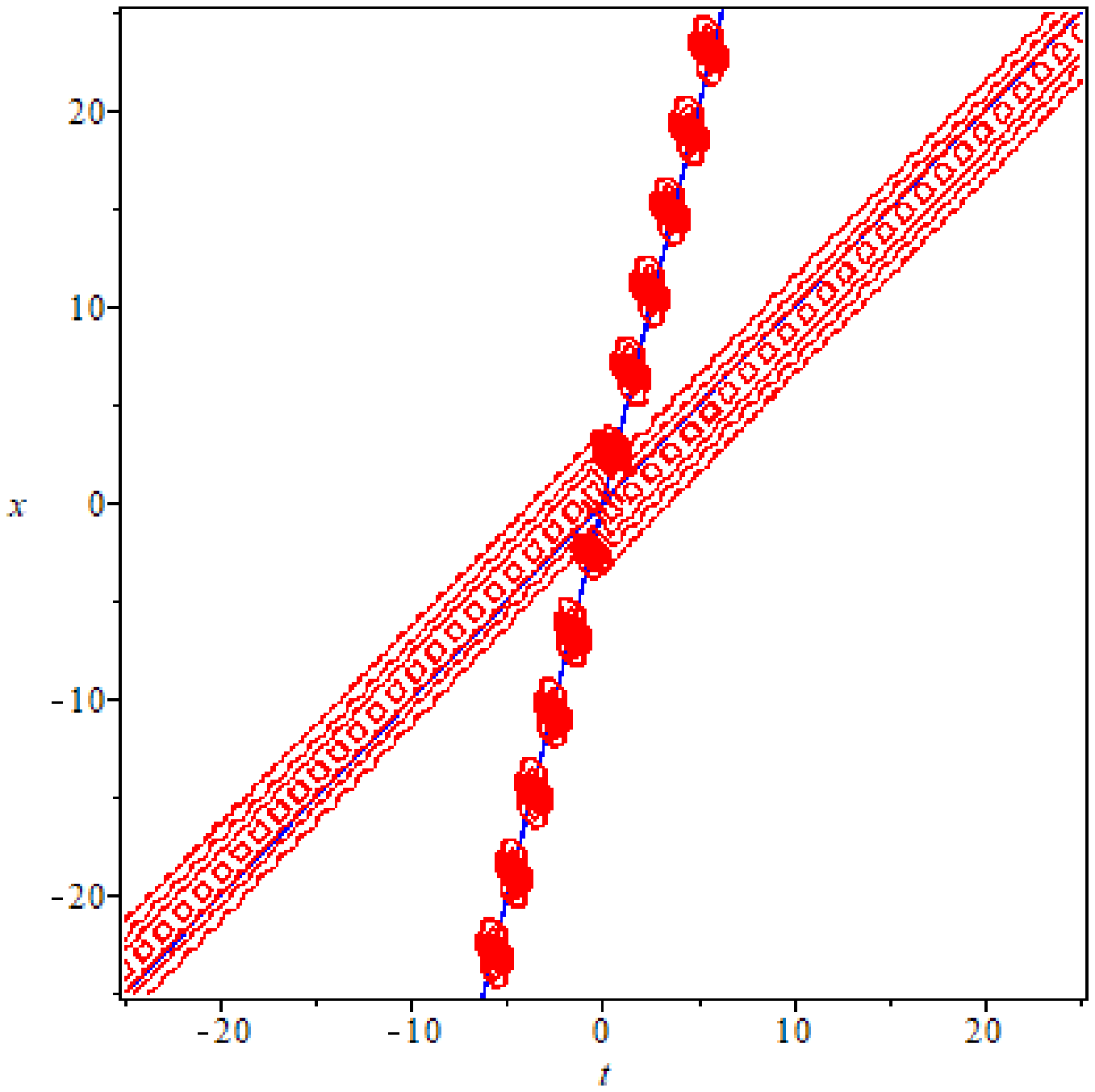}}}

$~~\quad\quad(\textbf{e})\qquad\qquad\qquad\qquad\quad(\textbf{f})
\qquad\qquad\qquad~~~~(\textbf{g})\qquad\qquad~~~\qquad\quad(\textbf{h})$\\

\noindent { \small \textbf{Figure 4.}  The solution \eqref{6} of the equation \eqref{Q1} with the parameters $\textbf{(a)}$ $k_{1}=\frac{i}{2}$, $\widetilde{k}_{1}=-\frac{i}{2}$, $k_{2}=\frac{i}{4}$, $\widetilde{k}_{2}=-\frac{i}{4}$, $\omega_{1}=\widetilde{\omega}_{1}=\omega_{2}=\widetilde{\omega}_{2}=1$; $\textbf{(e)}$  is the image after $\textbf{(a)}$ is rotated; $\textbf{(b)}$ and $\textbf{(f)}$ denote the density of $(\textbf{a})$ and $\textbf{(e)}$, respectively; $\textbf{(c)}$ and $\textbf{(g)}$ represents the dynamic behavior of the two-soliton solutions at different times;  $\textbf{(d)}$ is the characteristic line graph (blue line $L_{1}:x-4t=0$) and contour map of $\textbf{(a)}$; $\textbf{(h)}$  is the characteristic line graph (blue line $L_{2}:x-t=0$) and contour map of $\textbf{(e)}$.}

\noindent{\rotatebox{0}{\includegraphics[width=3.3cm,height=2.8cm,angle=0]{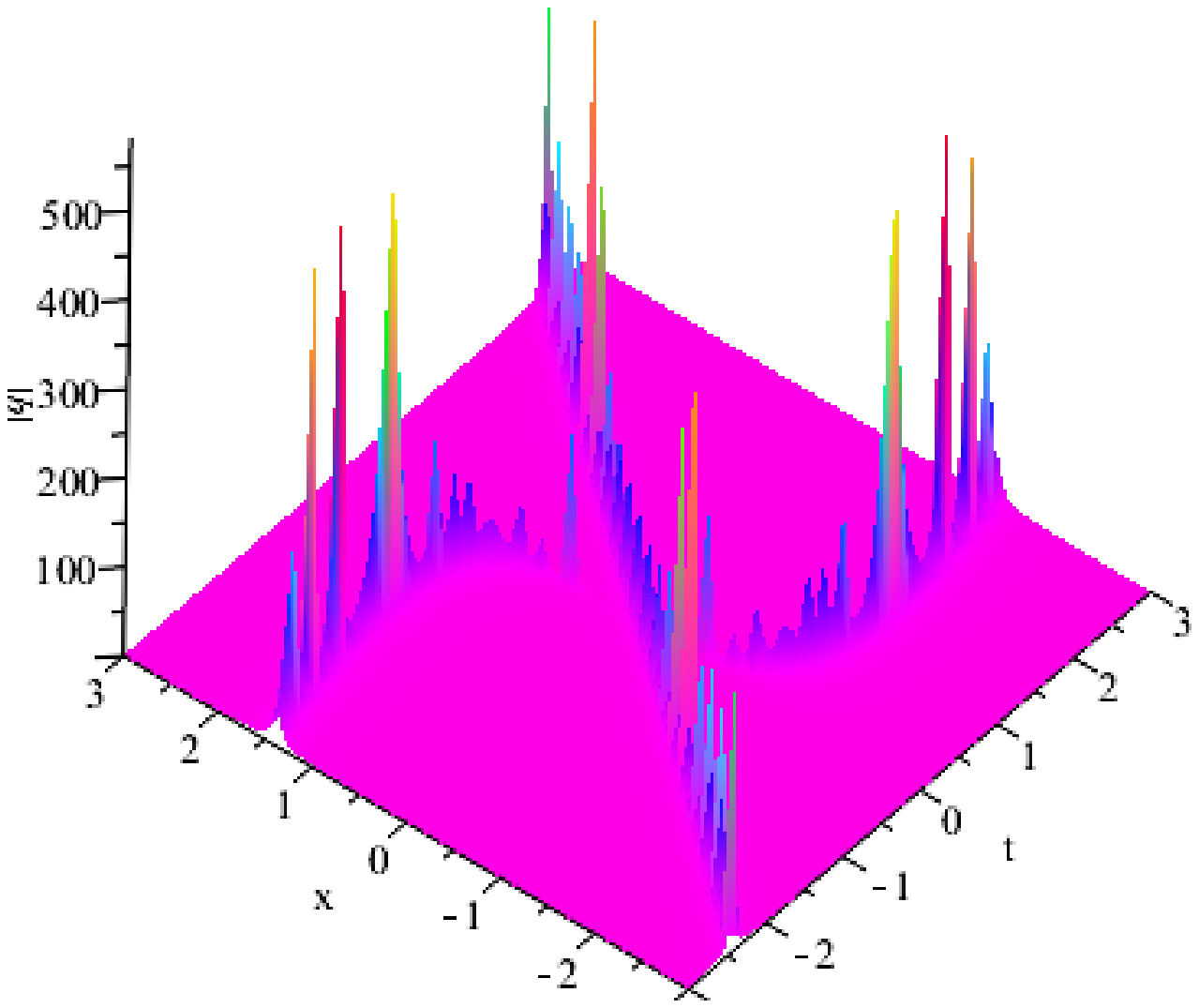}}
~\quad\rotatebox{0}{\includegraphics[width=2.5cm,height=2.5cm,angle=0]{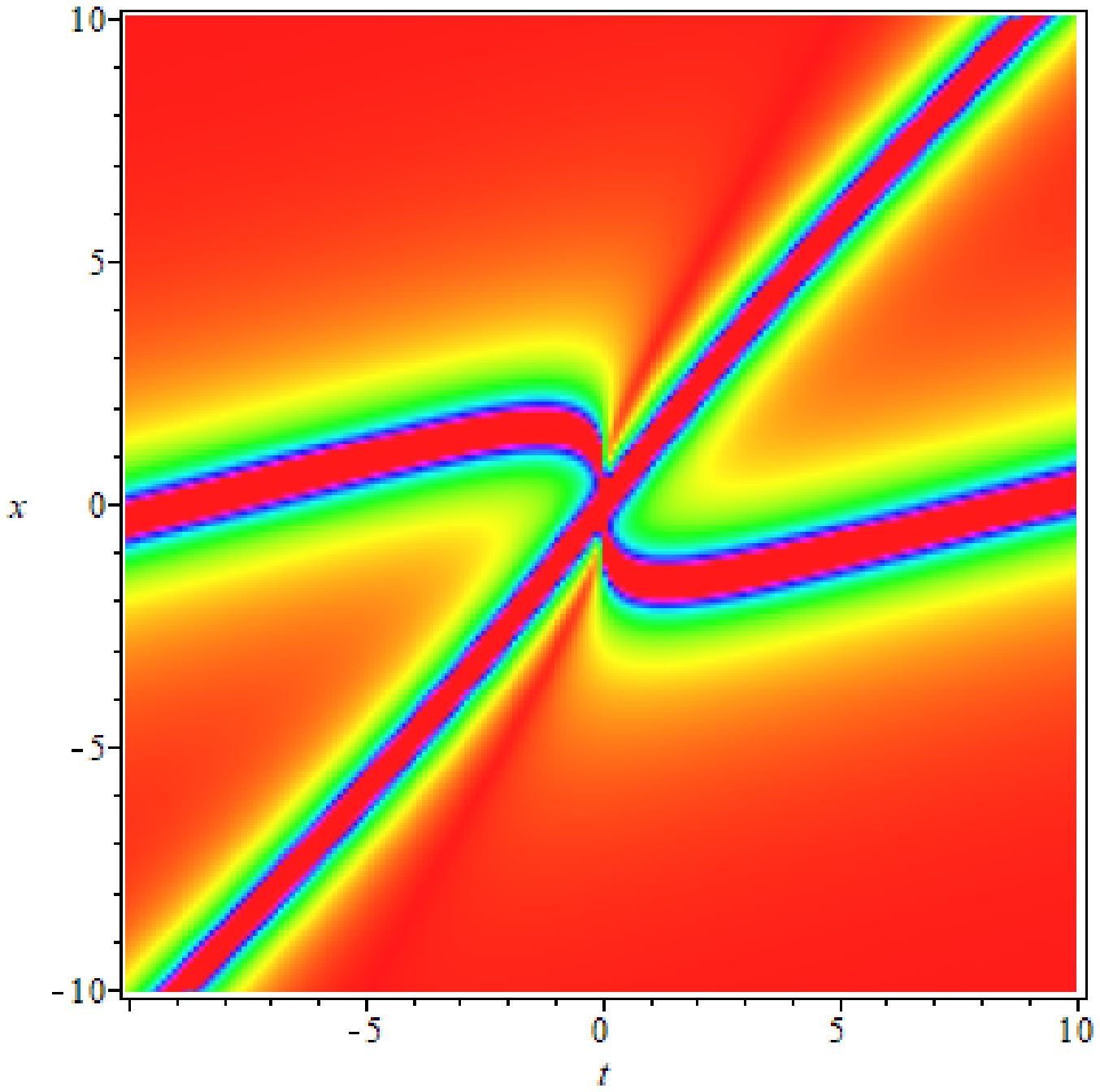}}
~\quad\rotatebox{0}{\includegraphics[width=2.5cm,height=2.5cm,angle=0]{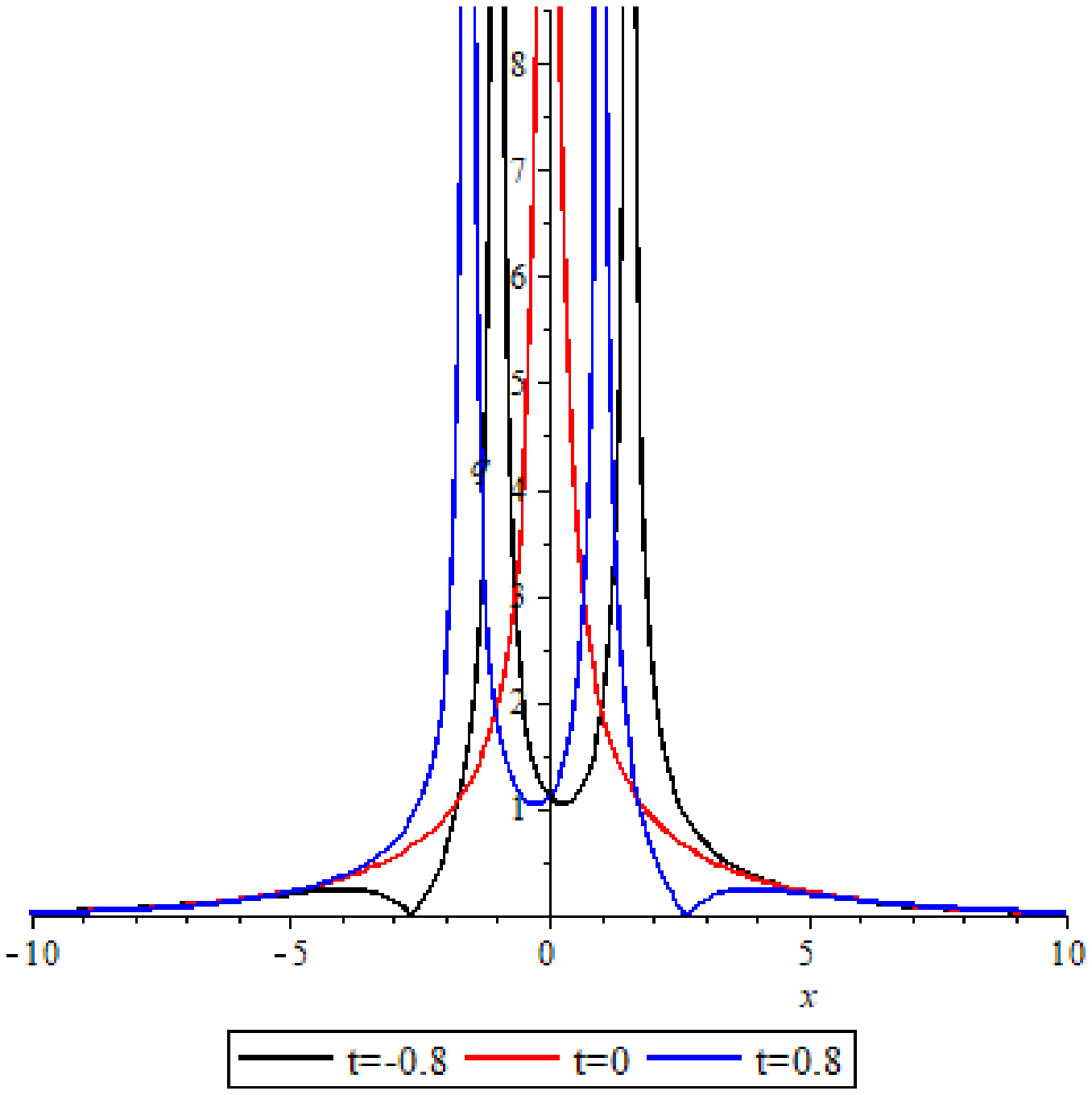}}
~\quad\rotatebox{0}{\includegraphics[width=2.5cm,height=2.5cm,angle=0]{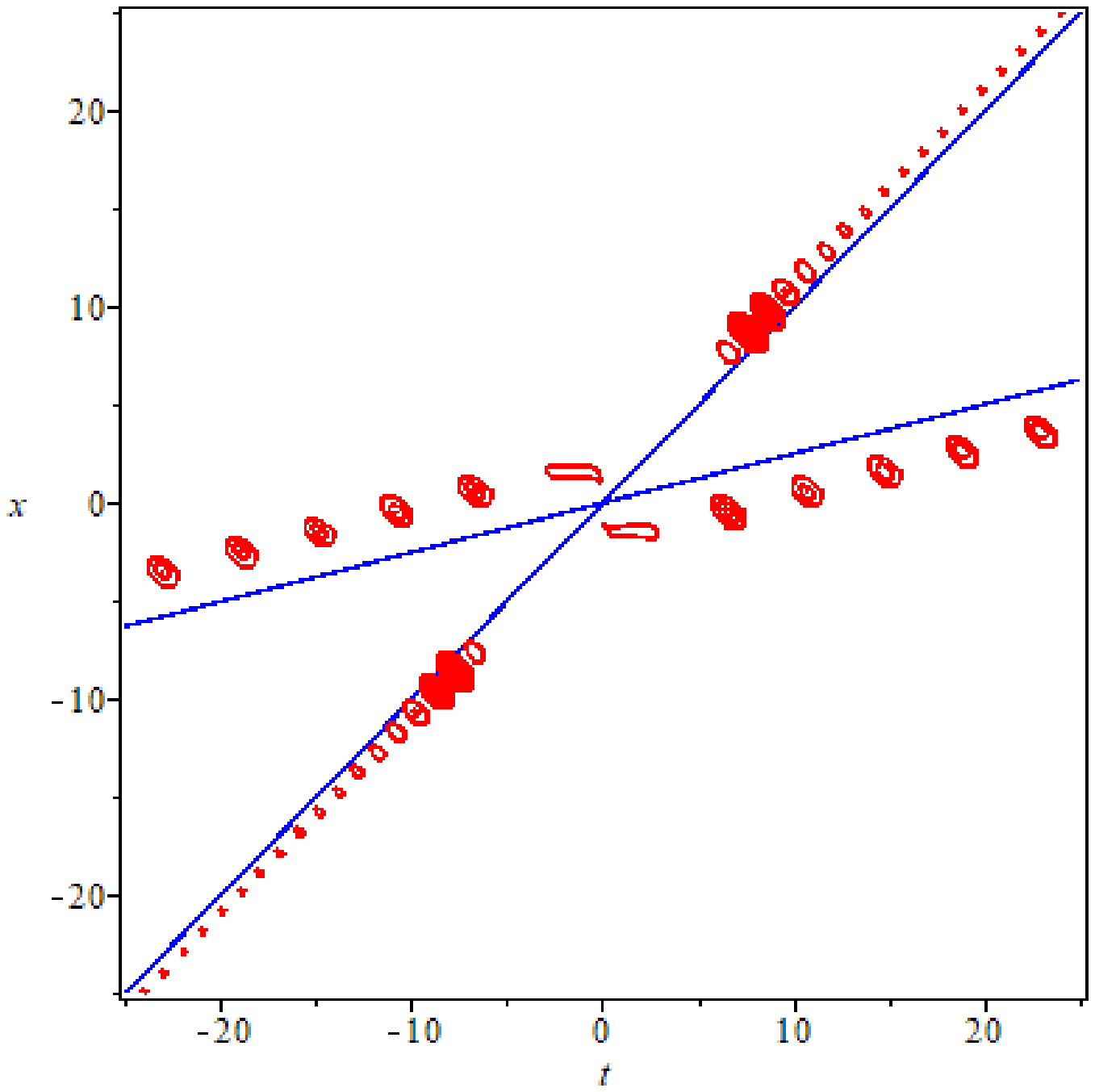}}}

$~~\quad\quad(\textbf{a})\qquad\qquad\qquad\qquad\quad(\textbf{b})
\qquad\qquad\qquad~~~(\textbf{c})\qquad\qquad~~~\qquad\quad(\textbf{d})$\\

\noindent{\rotatebox{0}{\includegraphics[width=3.3cm,height=2.8cm,angle=0]{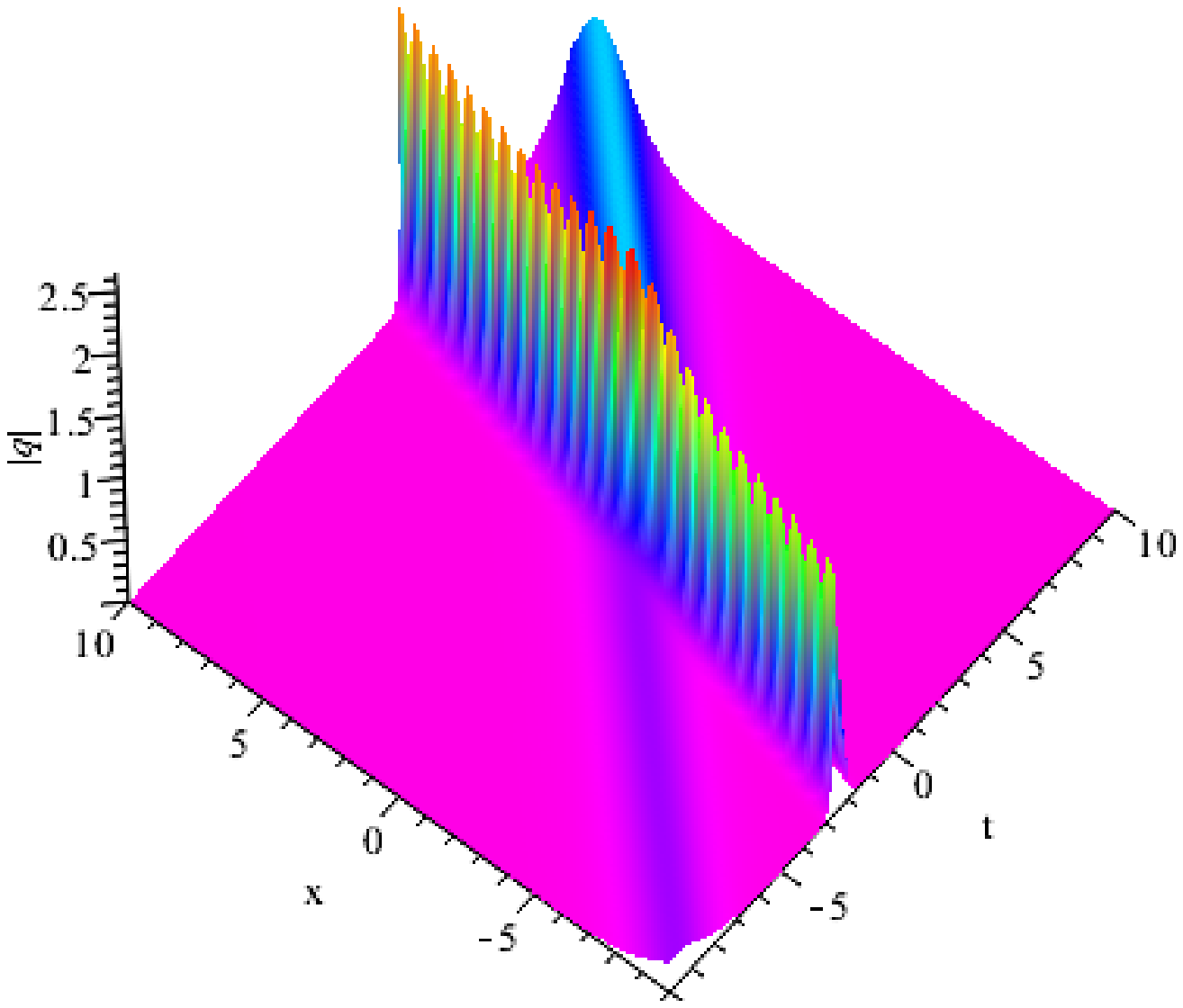}}
~\quad\rotatebox{0}{\includegraphics[width=2.5cm,height=2.5cm,angle=0]{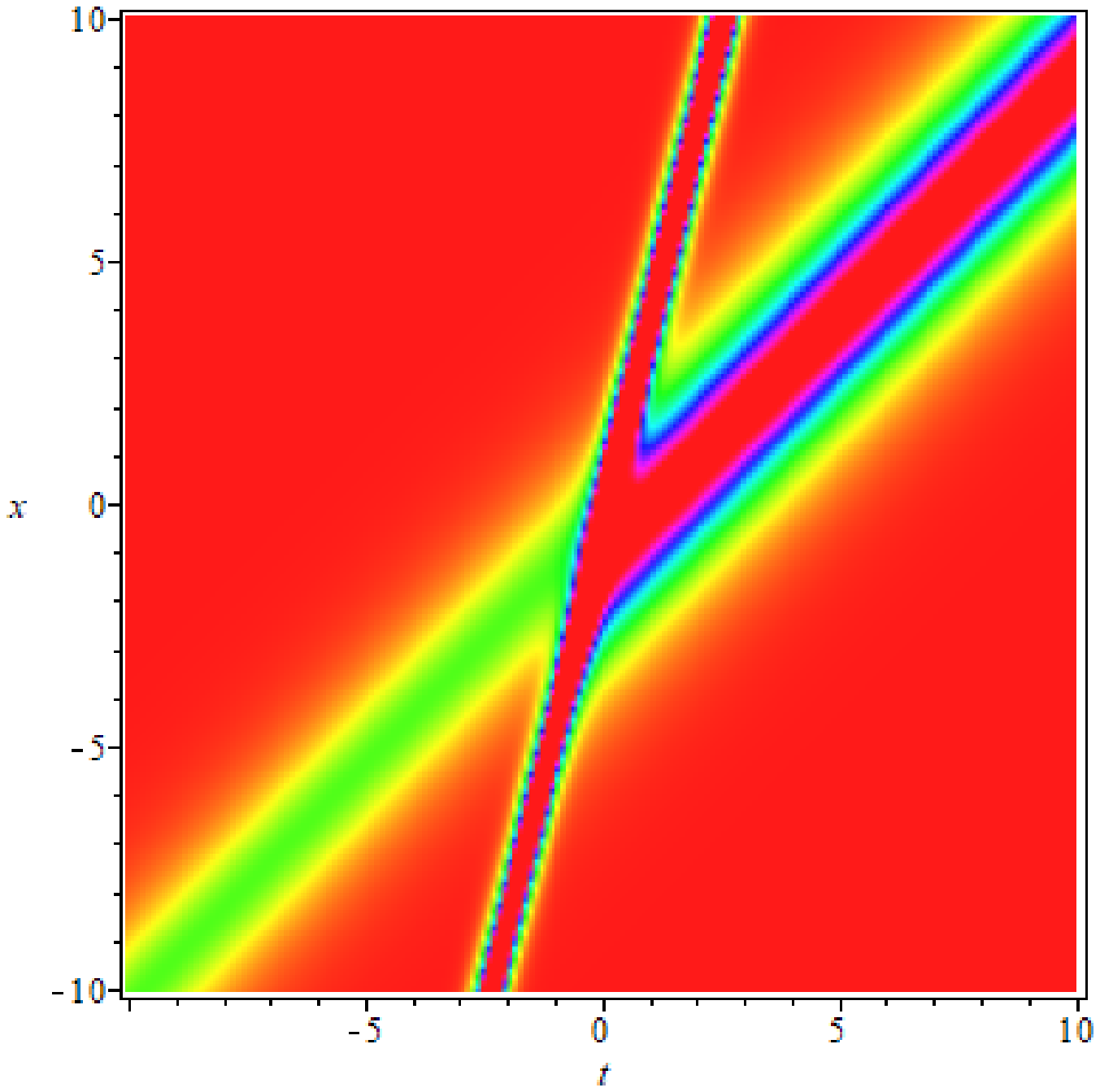}}
~\quad\rotatebox{0}{\includegraphics[width=2.5cm,height=2.5cm,angle=0]{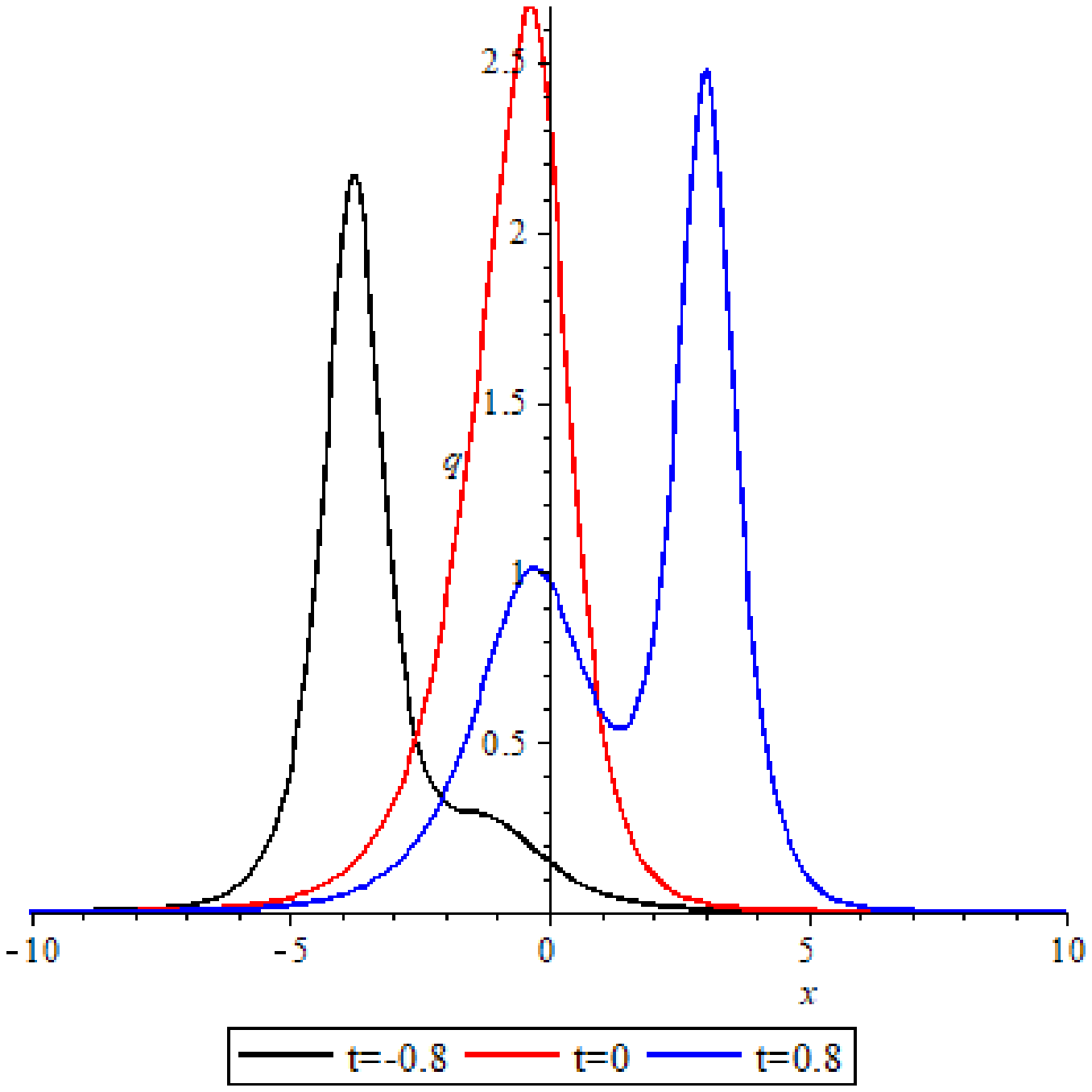}}
~\quad\rotatebox{0}{\includegraphics[width=2.5cm,height=2.5cm,angle=0]{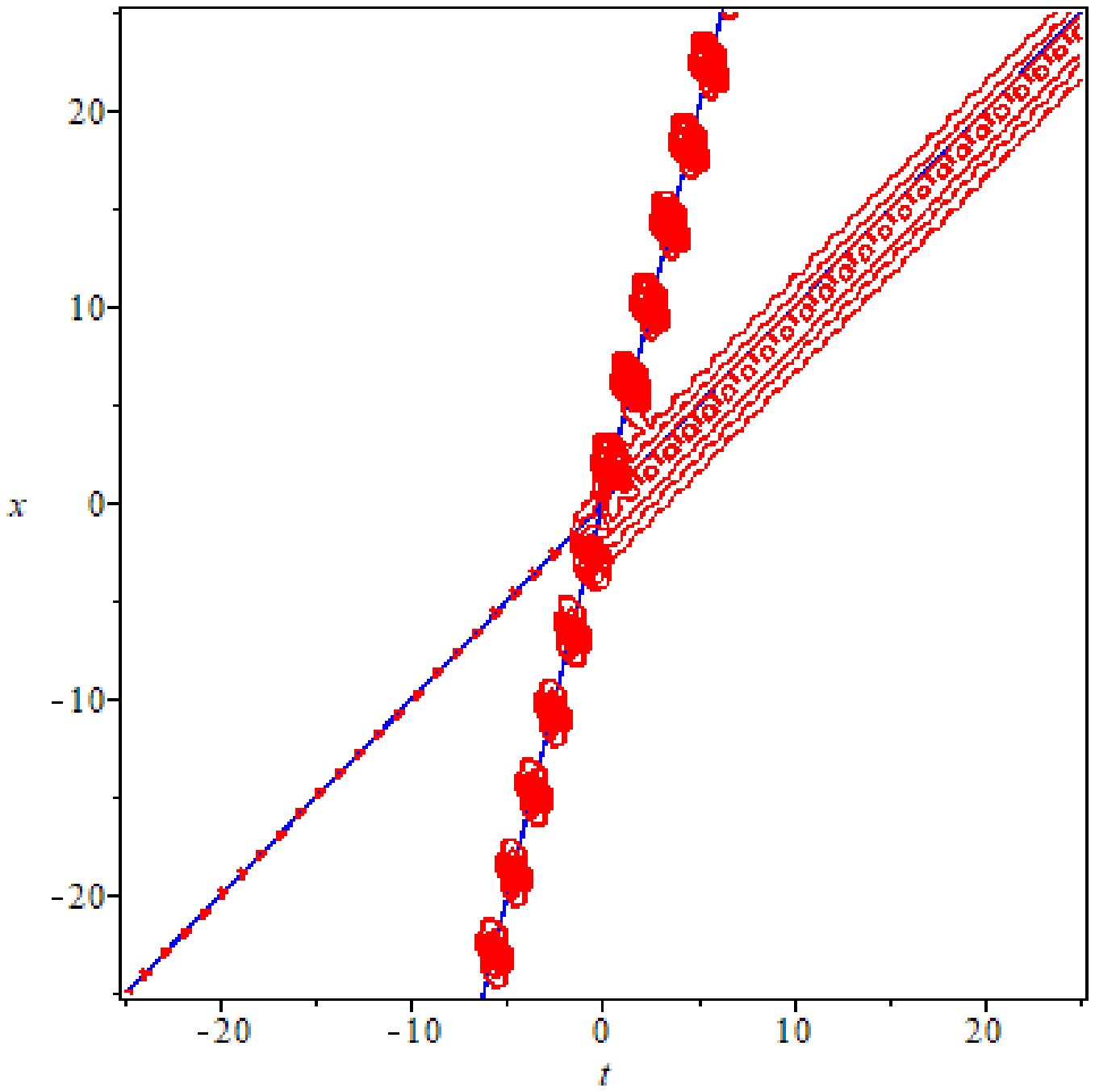}}}

$~~\quad\quad(\textbf{e})\qquad\qquad\qquad\qquad\quad(\textbf{f})
\qquad\qquad\qquad~~~~(\textbf{g})\qquad\qquad~~~\qquad\quad(\textbf{h})$\\

\noindent { \small \textbf{Figure 5.}  The solution \eqref{6} of the equation \eqref{Q1} with the parameters $\textbf{(a)}$ $k_{1}=\frac{i}{2}$, $\widetilde{k}_{1}=-\frac{i}{2}$, $k_{2}=\frac{i}{4}$, $\widetilde{k}_{2}=-\frac{i}{4}$, $\omega_{1}=\omega_{2}=1$, $\widetilde{\omega}_{1}=\widetilde{\omega}_{2}=-1$; $\textbf{(e)}$  is the image after $\textbf{(a)}$ is rotated; $\textbf{(b)}$ and $\textbf{(f)}$ denote the density of $(\textbf{a})$ and $\textbf{(e)}$, respectively; $\textbf{(c)}$ and $\textbf{(g)}$ represents the dynamic behavior of the two-soliton solutions at different times;  $\textbf{(d)}$ is the characteristic line graph (blue line $L_{1}:x-t=0$) and contour map of $\textbf{(a)}$; $\textbf{(h)}$  is the characteristic line graph (blue line $L_{2}:x-\frac{1}{4}t=0$) and contour map of $\textbf{(e)}$.}

As seen in Fig.4, taking $k_{1}=-\widetilde{k}_{1}$, $k_{2}=-\widetilde{k}_{2}$, the solution appears to collapse repeatedly with move in different directions in the process of propagation. In addition, the solutions collapse on the symmetric point pair with $x=0$ as the junction. It is also noted that the four eigenvalues can be divided into two pairs, and each pair can produce a single soliton. From this point of view, the two solitons should be regarded as the nonlinear superposition of two one-soliton, which is the case locally as shown in the Fig. 4.

By comparing Fig. 4 and Fig. 5, it is not difficult to find that  we select the same eigenvalues, the phenomenon are completely different when $\omega$ is taken as the opposite number, which is completely opposite to the phenomenon in the graph of two-solitons.

\rotatebox{0}{\includegraphics[width=4.4cm,height=3.5cm,angle=0]{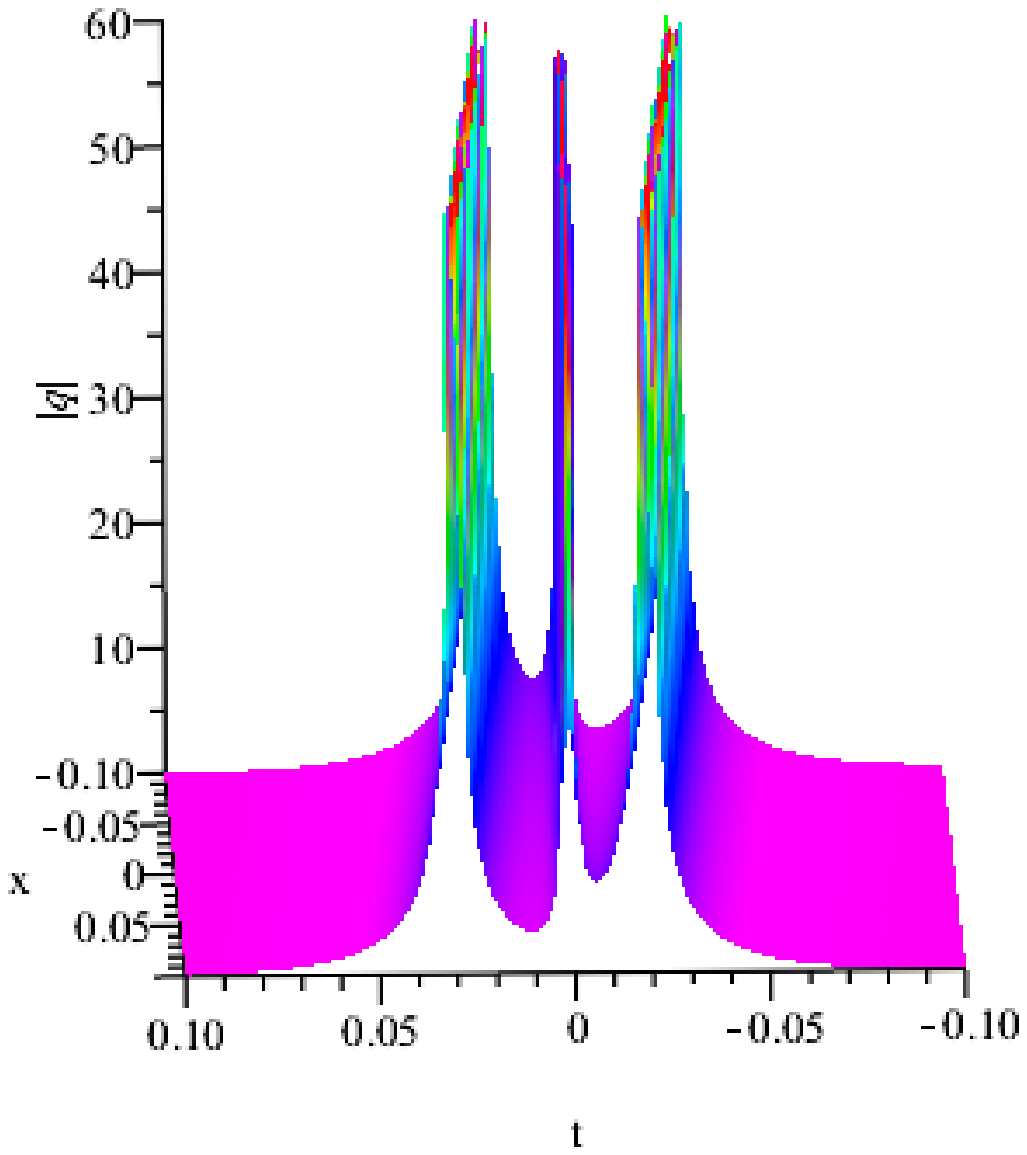}}
\rotatebox{0}{\includegraphics[width=3.3cm,height=3.0cm,angle=0]{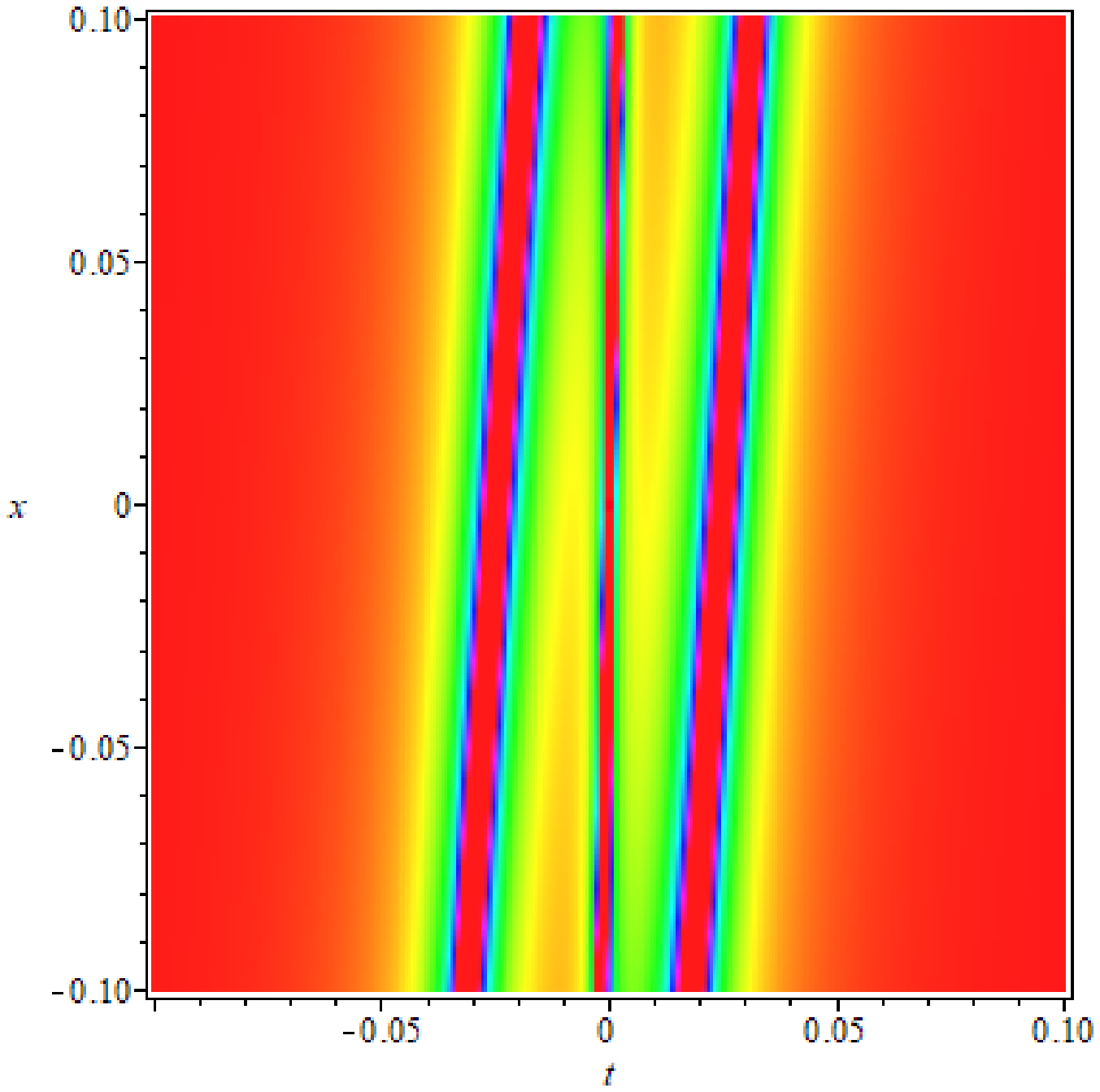}}
\quad\rotatebox{0}{\includegraphics[width=3.3cm,height=3.0cm,angle=0]{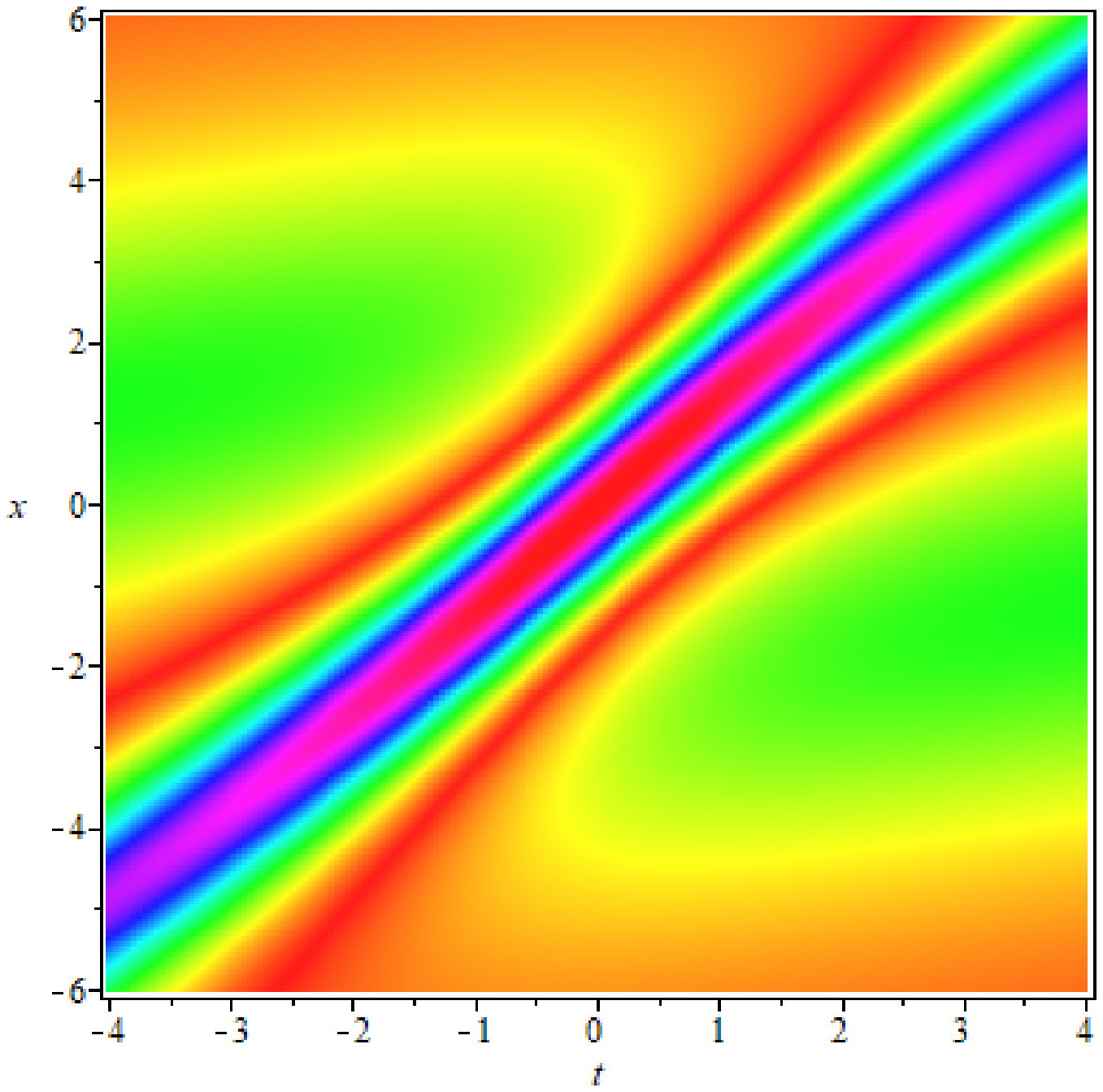}}

$~~~~\quad\qquad\qquad(\textbf{a})~\qquad\qquad\qquad\qquad\qquad(\textbf{b})
~~\qquad\quad\qquad\qquad\qquad(\textbf{c})$

\noindent { \small \textbf{Figure 6.} Bounded solution \eqref{5} of the equation \eqref{Q1} with the parameters $\textbf{(a)}$  $k_{1}=3i$, $\widetilde{k}_{1}=-3i$, $k_{2}=2i$, $\widetilde{k}_{2}=-2i$, $\omega_{1}=\omega_{2}=1$, $\widetilde{\omega}_{1}=\widetilde{\omega}_{2}=-1$; $\textbf{(c)}$  is the characteristic line graph (blue line $L_{1}:x-36t=0$, and $L_{2}:x-16t=0$) and contour map of $\textbf{(a)}$; (\textbf{a})(\textbf{b})(\textbf{c}): the local structure, density, and characteristic line and contour map.}

As seen in Fig. 5, its left-moving wave does not decay in amplitude while its right-moving wave does. From the eigenvalue point of view, the image should be a nonlinear superposition of two-solitons, but the result is not so. Because the same eigenvalue parameters, but $\omega$ value is not the same. Taking $\widetilde{k}_{1}=-k_{1}$, in this case,  it describes the oscillatory $M$-shaped respiratory wave with two overlapping characteristic lines. Unlike $M$-soliton wave, the peak value and amplitude increase with the increase of eigenvalue.

(3)The kink solutions

\noindent{\rotatebox{0}{\includegraphics[width=3.3cm,height=2.8cm,angle=0]{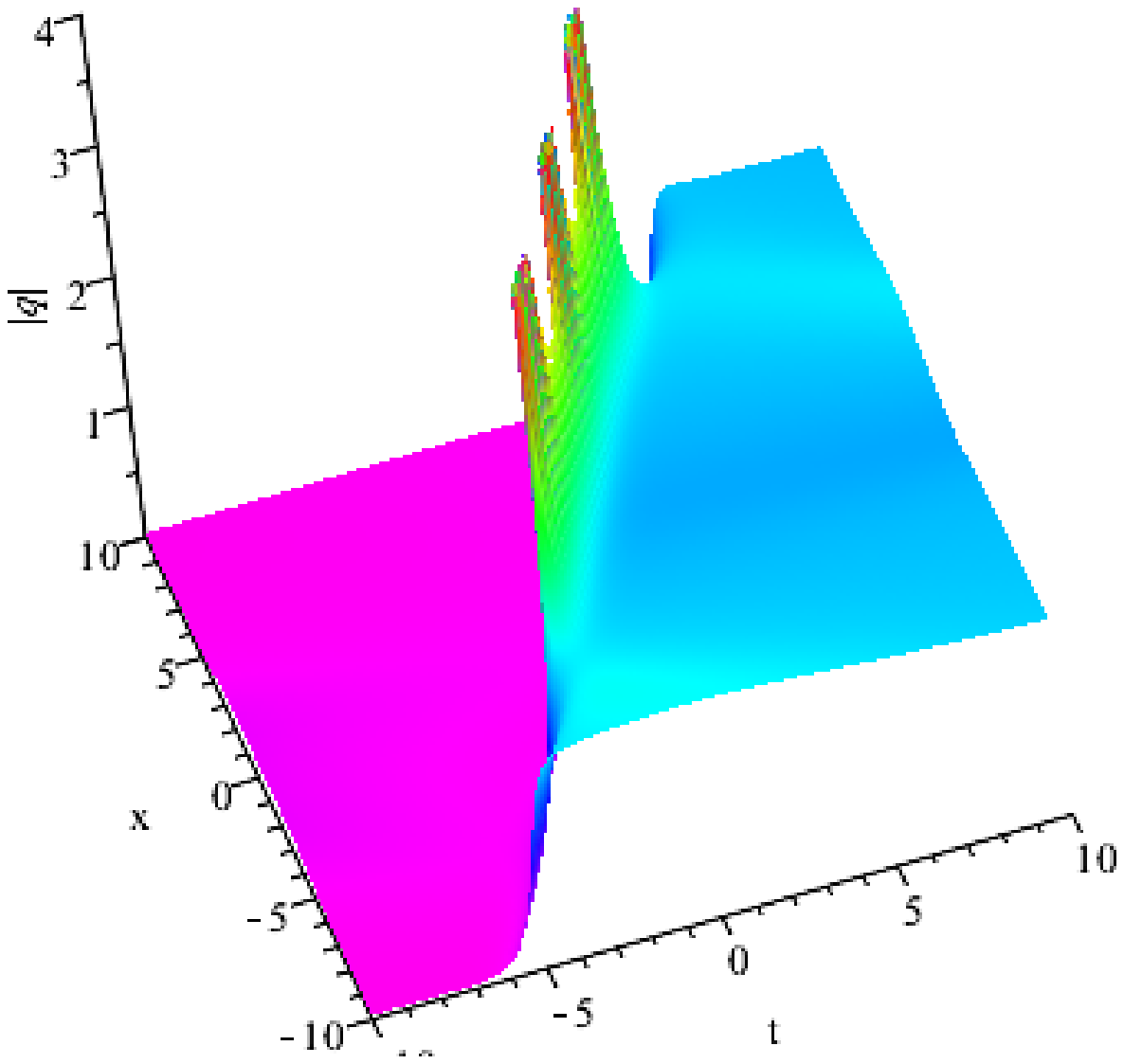}}
\rotatebox{0}{\includegraphics[width=2.5cm,height=2.5cm,angle=0]{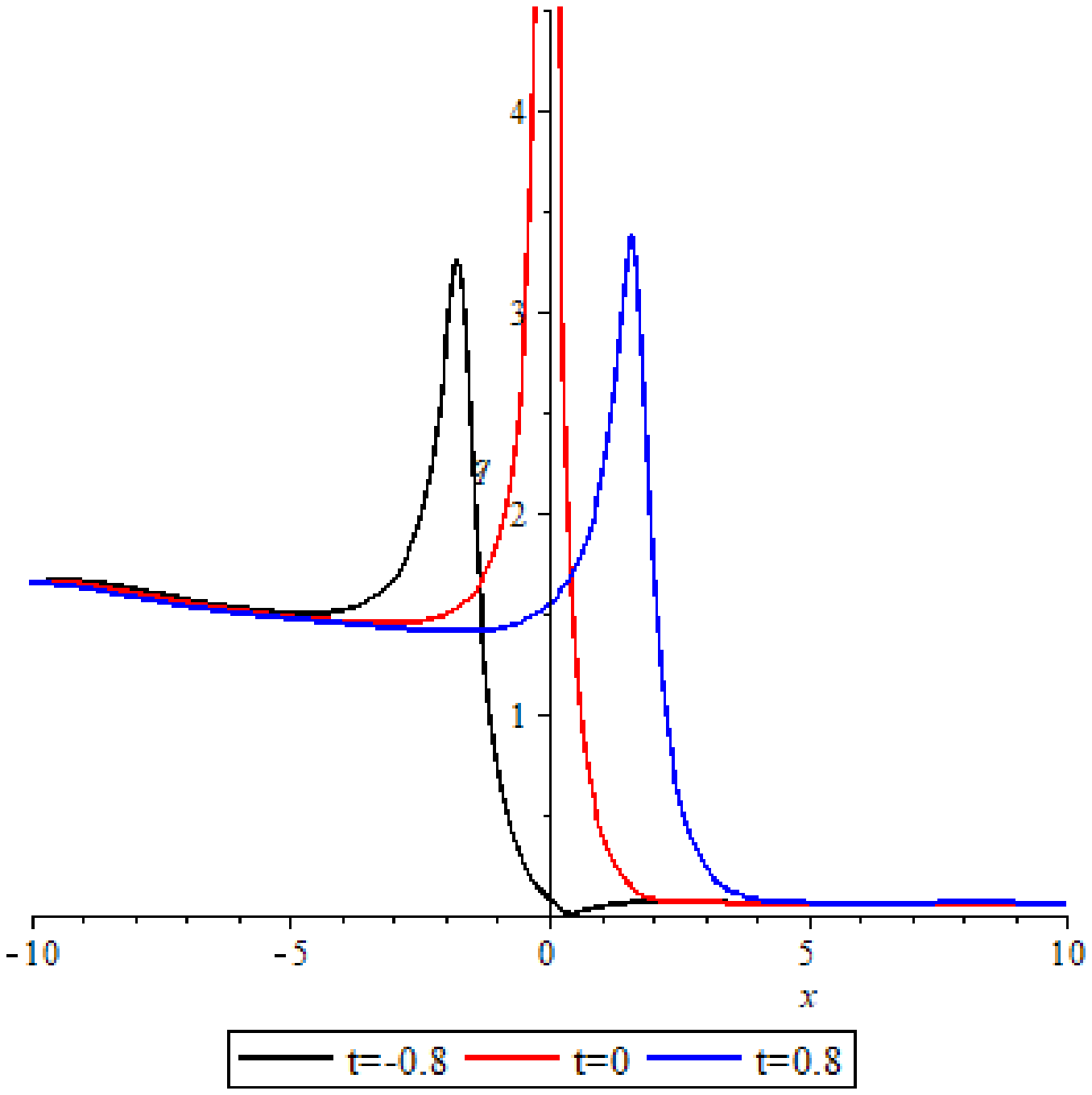}}
\rotatebox{0}{\includegraphics[width=2.5cm,height=2.5cm,angle=0]{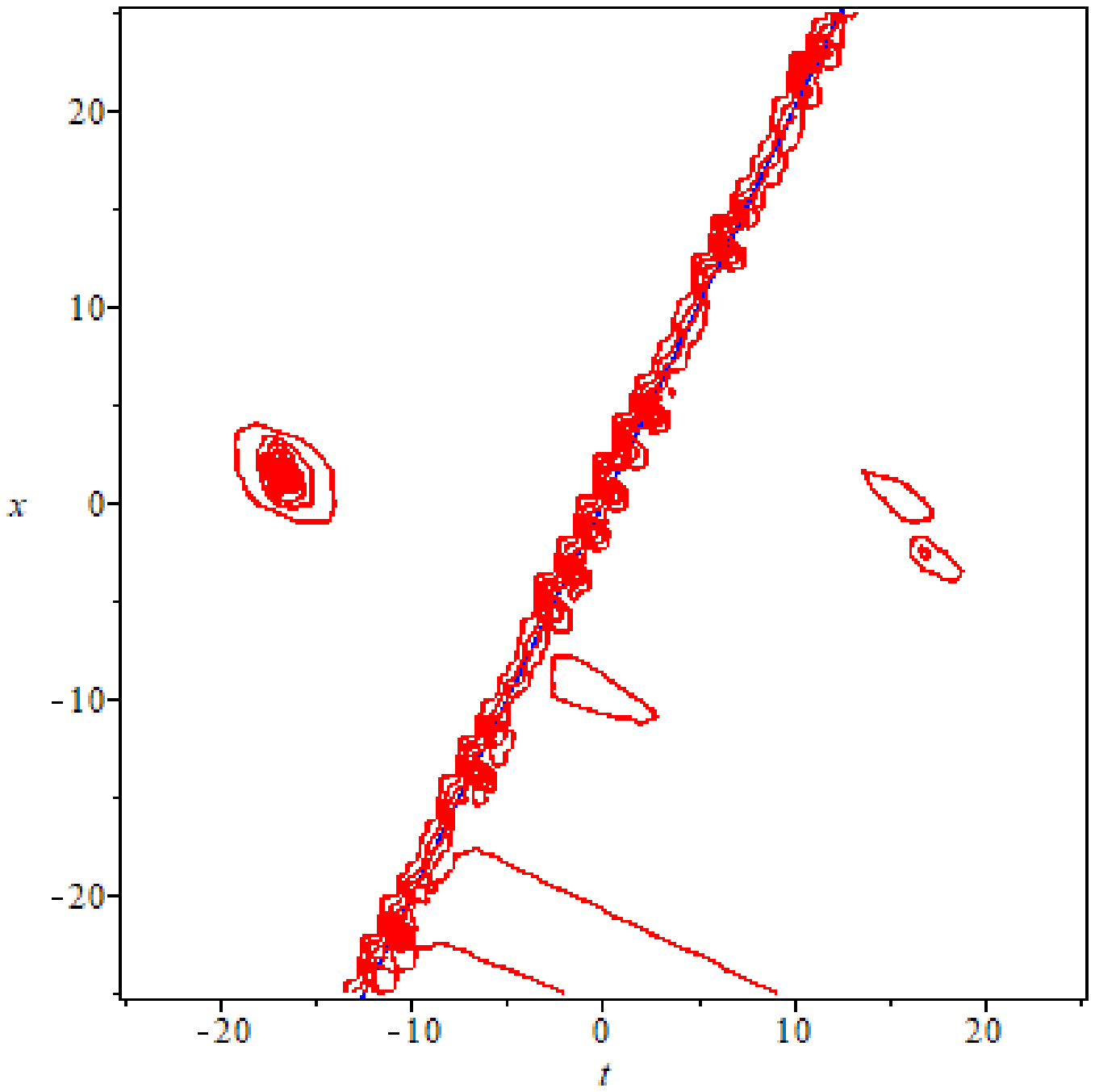}}
\hspace{-3.9cm}
\centerline{\begin{tikzpicture}[scale=0.35]
\draw[-][thick](-3,0)--(-2,0);
\draw[-][thick](-2,0)--(-1,0);
\draw[-][thick](-1,0)--(0,0);
\draw[-][thick](0,0)--(1,0);
\draw[-][thick](1,0)--(2,0);
\draw[-][thick](2,0)--(3,0);
\draw[-][thick](3,0)--(3,2);
\draw[-][thick](3,2)--(3,5.7);
\draw[-][thick](3,5.7)--(0,5.7);
\draw[-][thick](0,5.7)--(-3,5.7);
\draw[-][thick](-3,5.7)--(-3,2);
\draw[-][thick](-3,2)--(-3,0);
\draw[-][thick](-3,0)--(0,0);
\draw[-][thick](0,0)--(3,0);
\draw[-][dashed](-3,3)--(0,3);
\draw[-][dashed](0,3)--(3,3);
\draw[-][dashed](0,5.7)--(0,3);
\draw[-][dashed](0,3)--(0,0);
\draw[fill] (0,3.7)node{$\textcolor[rgb]{0.00,0.00,1.00}{\bullet}$};
\draw[-][dashed](2,3)--(3,3)node[right]{$Re~z$};
\draw[-][dashed](0,5.5)--(0,5.7)node[above]{$Im~z$};
\end{tikzpicture}}}

$~~\quad\qquad(\textbf{a})~~\qquad\qquad\qquad(\textbf{b})
\qquad\qquad\qquad~~(\textbf{c})\qquad\qquad~\qquad\quad(\textbf{d})$\\

\noindent { \small \textbf{Figure 7.} The solution \eqref{5} of the equation \eqref{Q1} with the parameters $\textbf{(a)}$  $k_{2}=\frac{\sqrt2 i}{2}$, $\omega_{1}=\widetilde{\omega}_{1}=\omega_{2}=1$, $\widetilde{\omega}_{2}=-1$;) and contour map of $\textbf{(a)}$; (\textbf{a})(\textbf{b}): the local structure and  distribution pattern of discrete spectral points. }

If $k_{1}, \widetilde{k}_{1}$ and $\widetilde{k}_{2}$ are close to real constants, $k_{2}$ is a pure imaginary eigenvalue and the parameters have the following relations,
\begin{align*}
k_{1}=\widetilde{k}_{1}=\widetilde{k}_{2}\approx0,
\end{align*}
They breathe and collapse at $x=2t$ over time, which amplitude increases slowly near $x=2t$ and reaches infinity at $x=2t$. It is not difficult to find that the amplitude on the right side of the characteristic line $x=2t$ is always higher than that on the left side.
\subsubsection{Non-pure imaginary eigenvalues}
\
\newline
(1) Bounded two-soliton solutions

For the first case, we take the  eigenvalues as follows,
\begin{align*}
\left\{\begin{aligned}k_{1}&=-0.4+0.22i,~\widetilde{k}_{1}=-0.4-0.22i,~\omega_{1}=\widetilde{\omega}_{1}=1\\
k_{2}&=0.6-0.42i,~\widetilde{k}_{2}=0.6+0.42i,~\omega_{2}=\widetilde{\omega}_{2}=1.
\end{aligned}\right.
\end{align*}
then the solution of \eqref{6} can be characterized in Fig. 8. As seen in Fig. 8, the Fig. 8(\textbf{b}) is obtained by the rotation of  Fig. 8\textbf{(a)} with $\tan\theta=-\frac{3}{5}$. The selected eigenvalues are the form of non-pure imaginary numbers, choosing the different coefficients in the symmetry relationship do not affect the dynamic behaviors of the two-soliton solution.

\noindent{\rotatebox{0}{\includegraphics[width=3.3cm,height=2.8cm,angle=0]{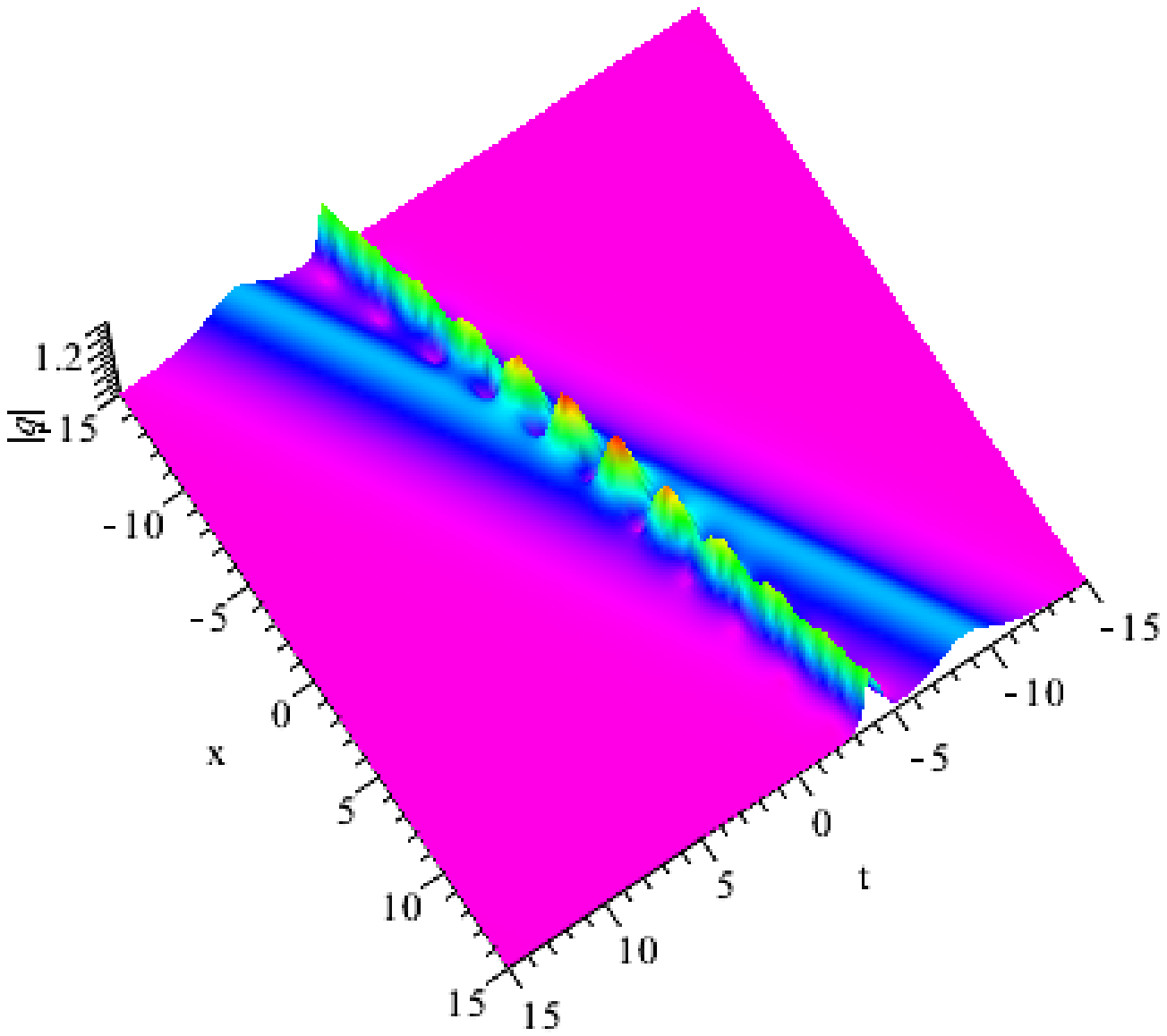}}
~\quad\rotatebox{0}{\includegraphics[width=2.5cm,height=2.5cm,angle=0]{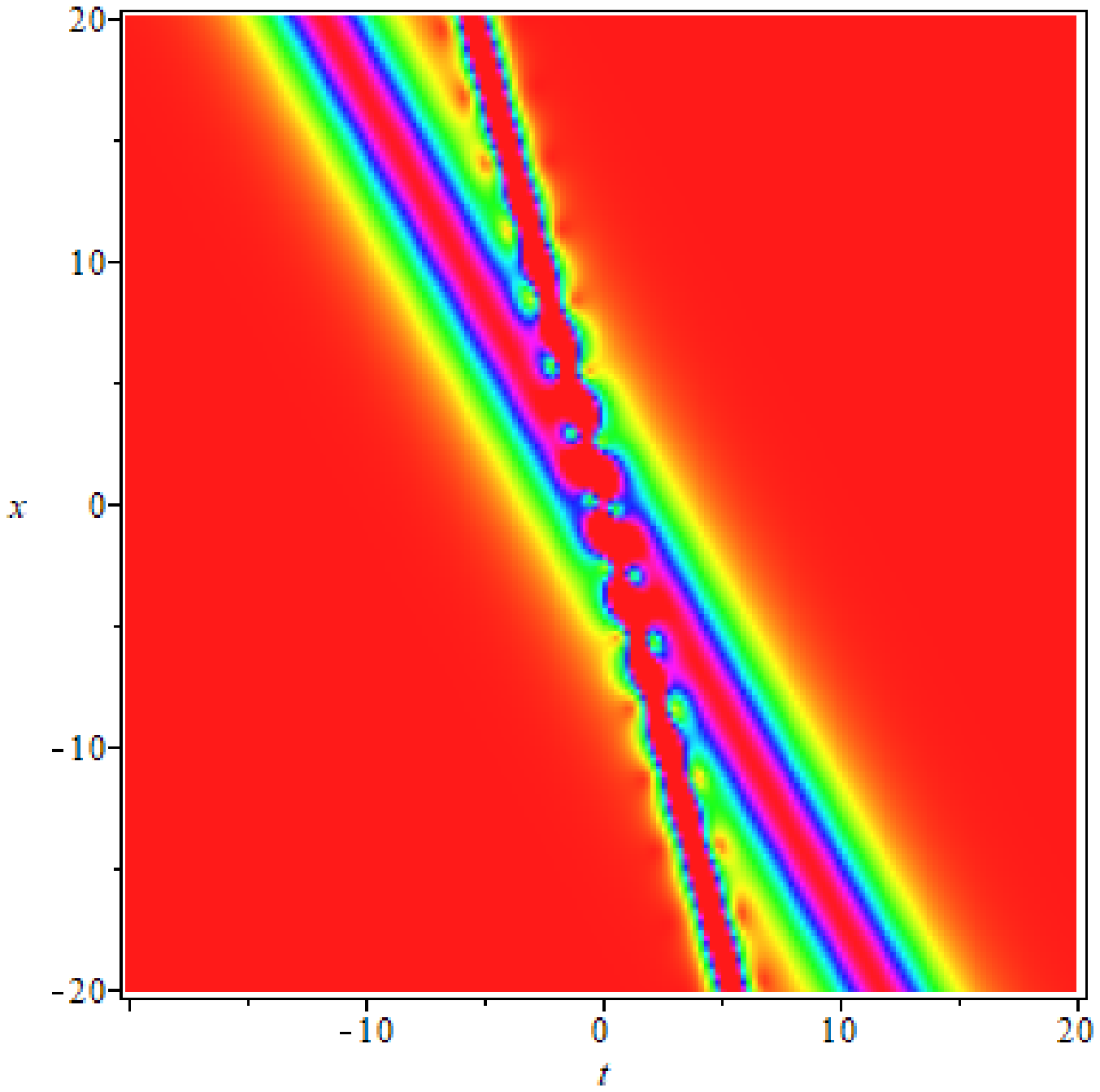}}
~\quad\rotatebox{0}{\includegraphics[width=2.5cm,height=2.5cm,angle=0]{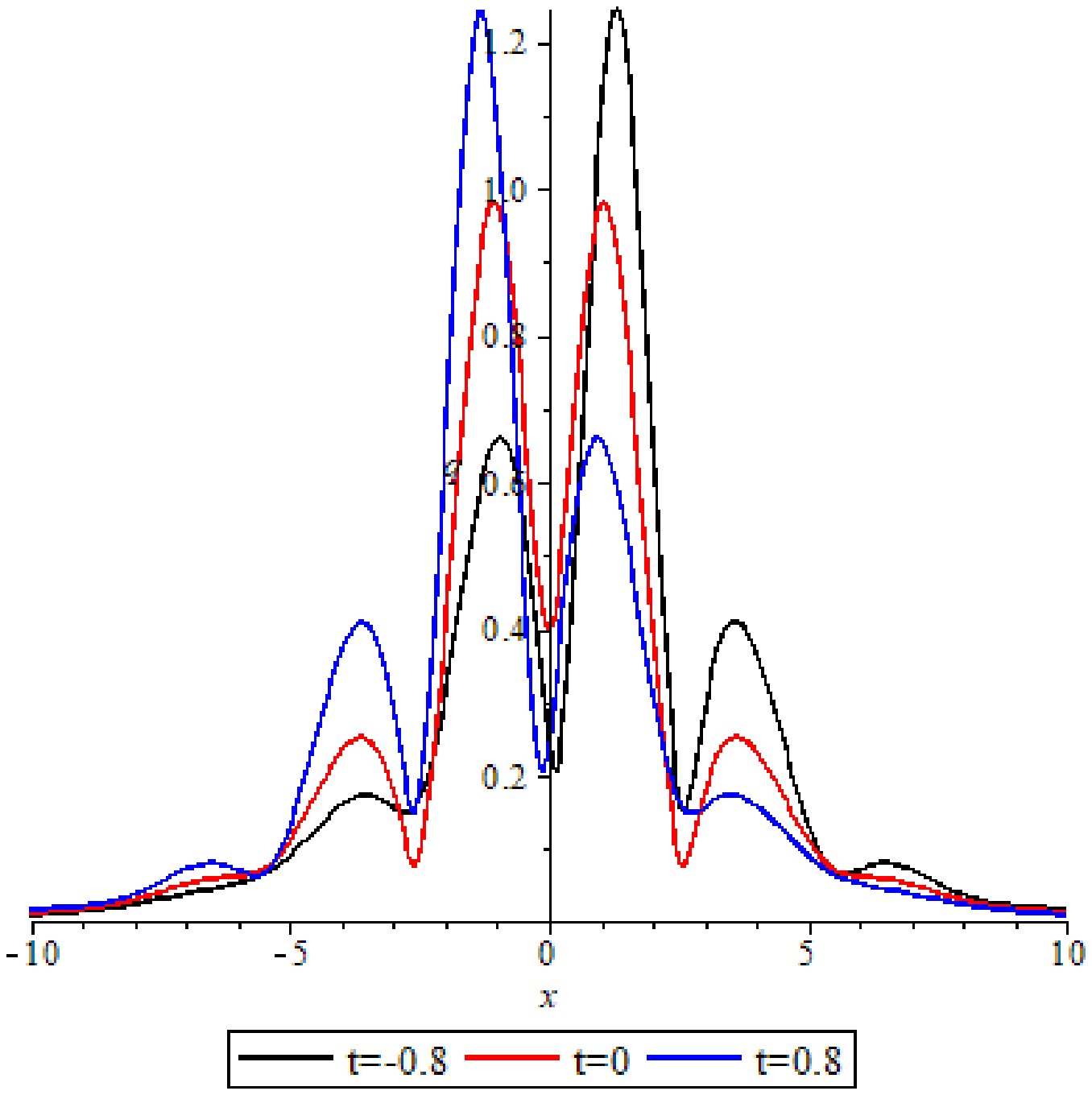}}
~\quad\rotatebox{0}{\includegraphics[width=2.5cm,height=2.5cm,angle=0]{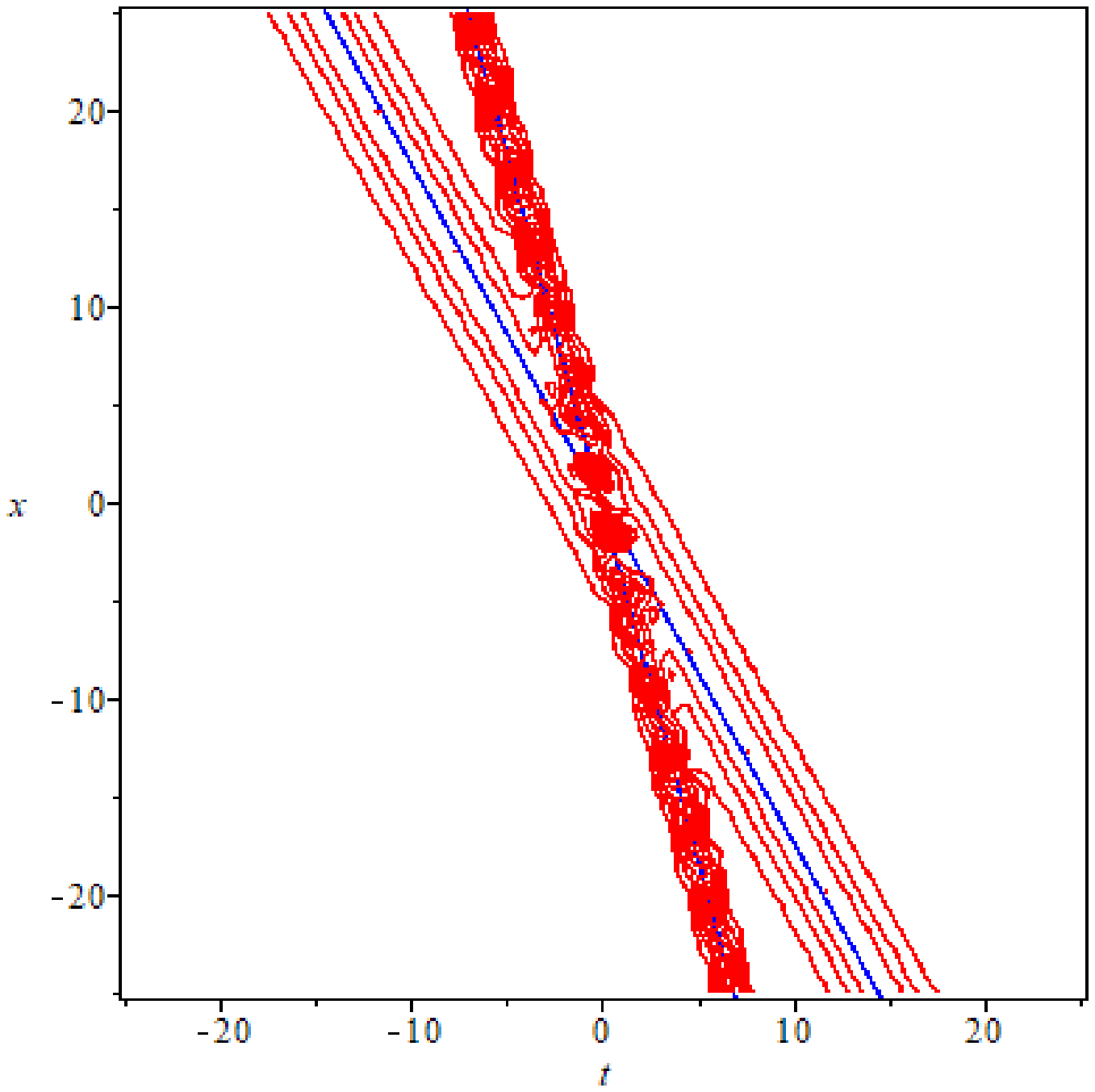}}}

$~~\quad\quad(\textbf{a})\qquad\qquad\qquad\qquad\quad(\textbf{b})
\qquad\qquad\qquad~~~(\textbf{c})\qquad\qquad~~~\qquad\quad(\textbf{d})$\\

\noindent{\rotatebox{0}{\includegraphics[width=3.3cm,height=2.8cm,angle=0]{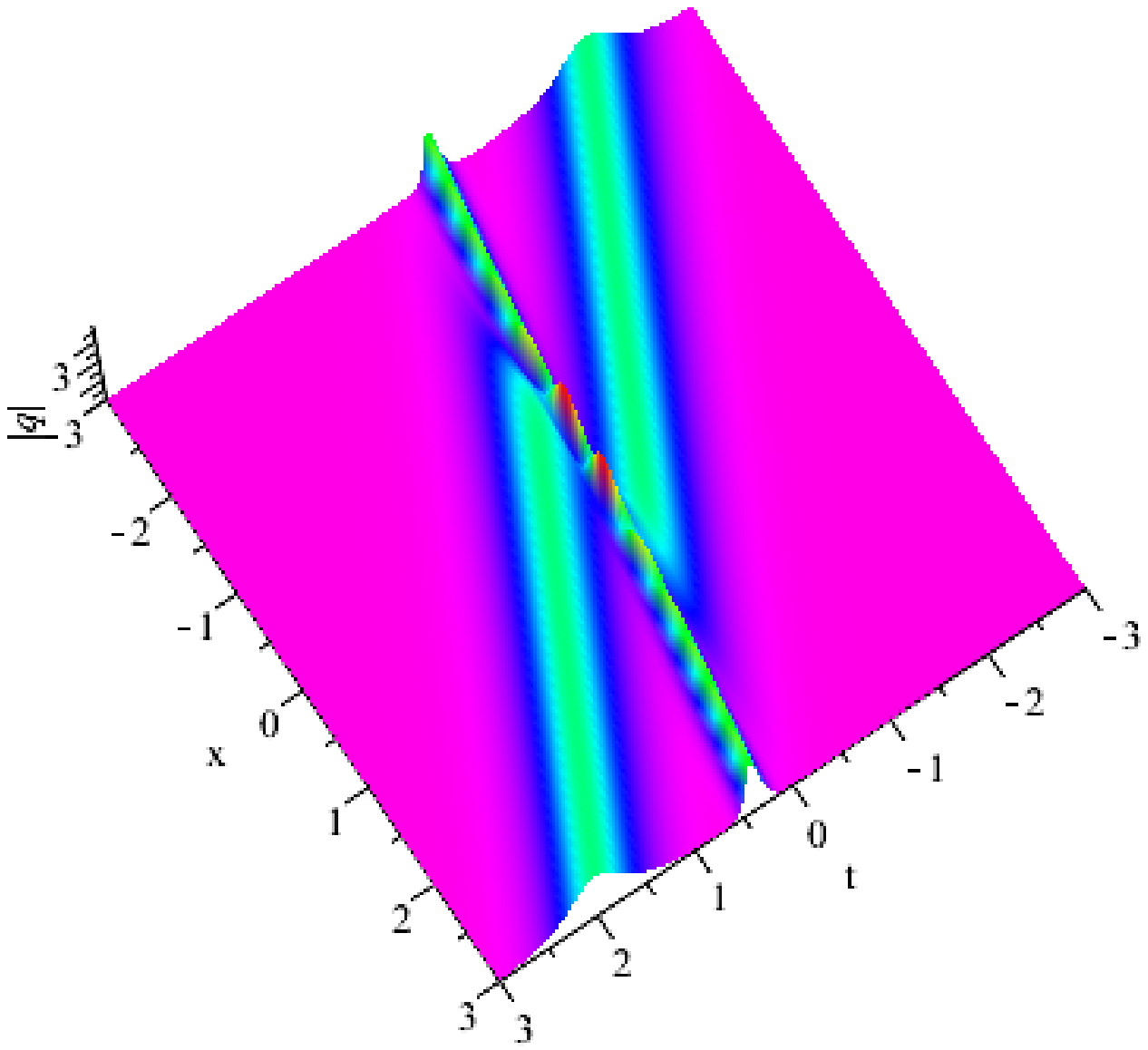}}
~\quad\rotatebox{0}{\includegraphics[width=2.5cm,height=2.5cm,angle=0]{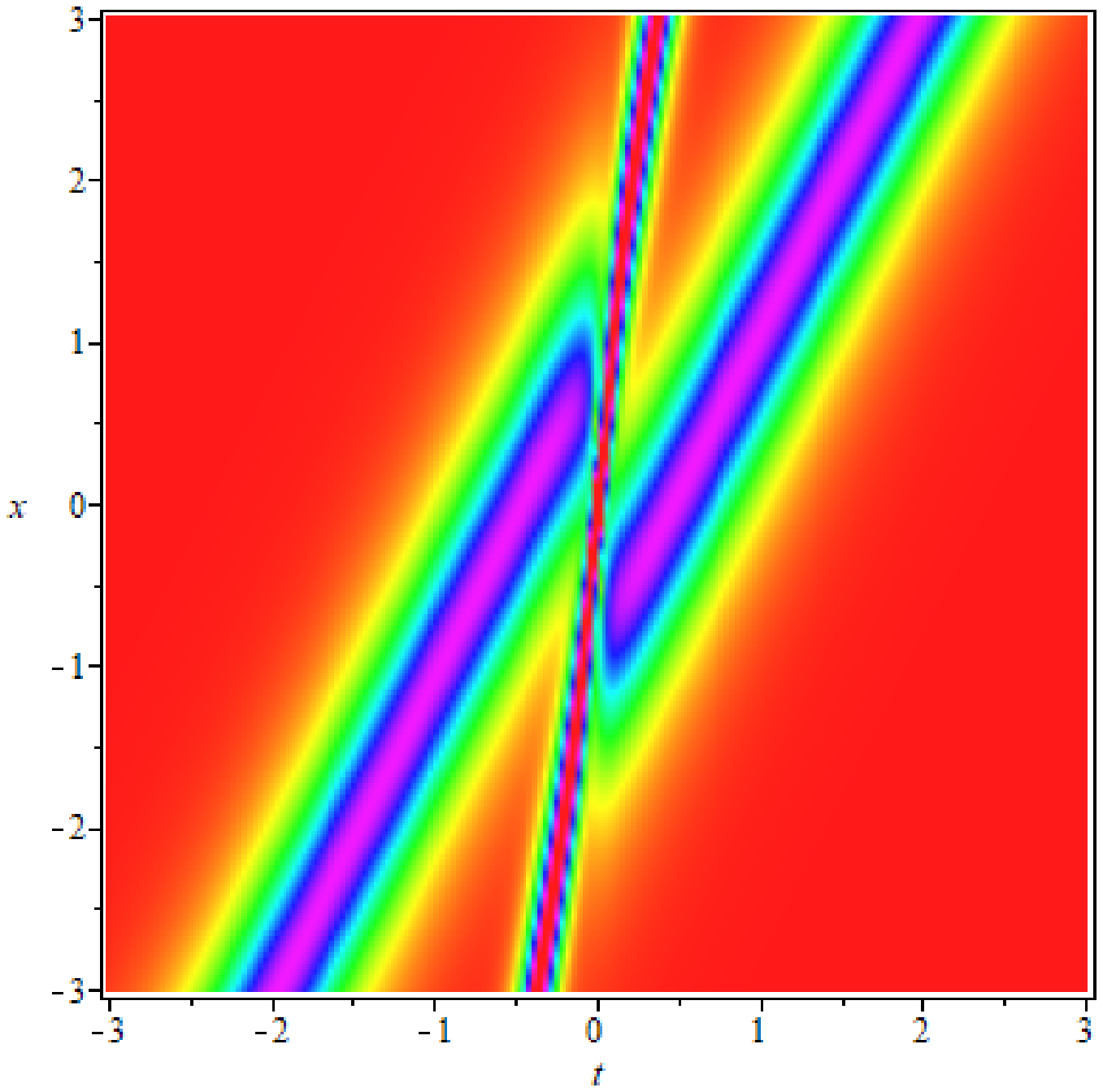}}
~\quad\rotatebox{0}{\includegraphics[width=2.5cm,height=2.5cm,angle=0]{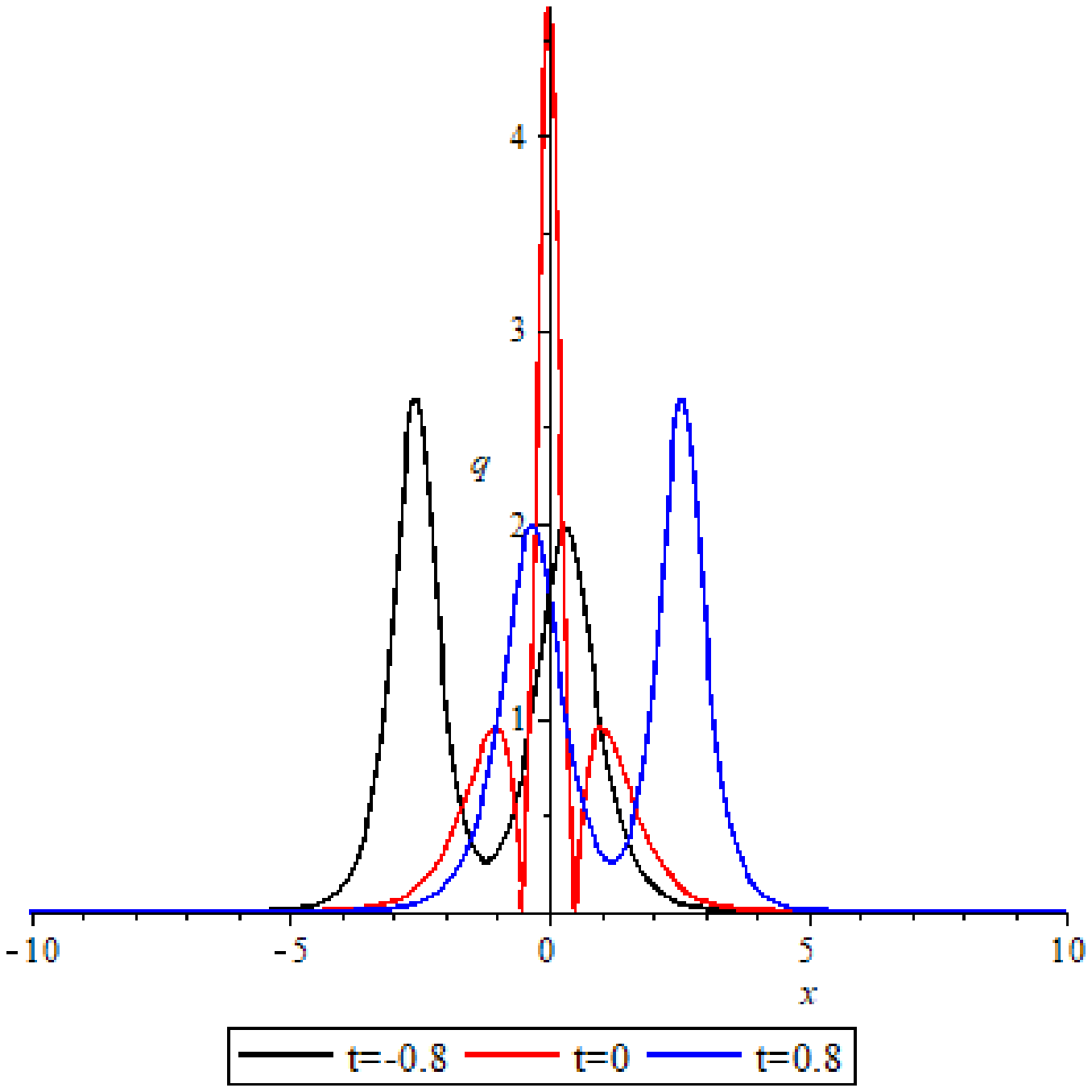}}
~\quad\rotatebox{0}{\includegraphics[width=2.5cm,height=2.5cm,angle=0]{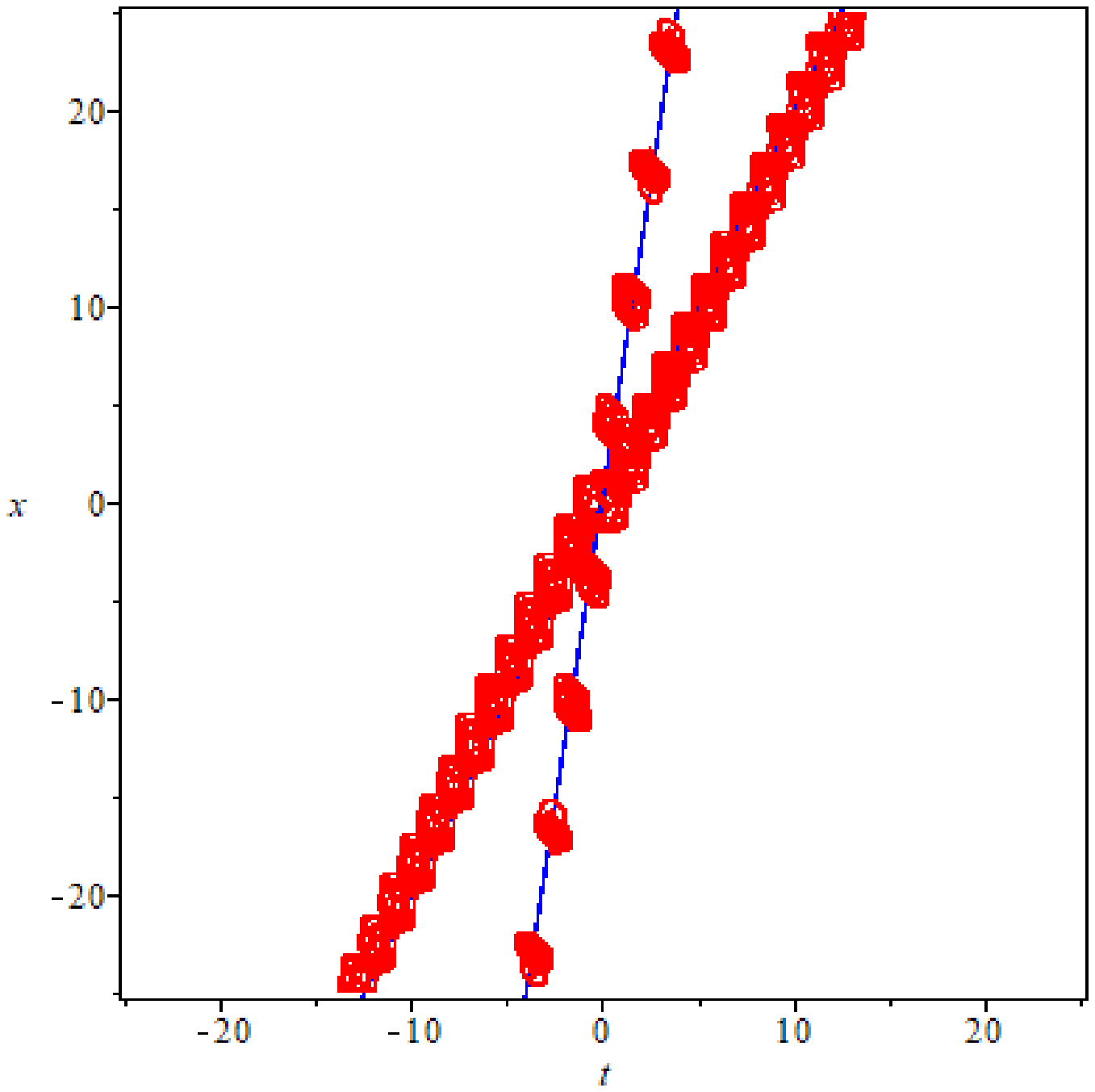}}}

$~~\quad\quad(\textbf{e})\qquad\qquad\qquad\qquad\quad(\textbf{f})
\qquad\qquad\qquad~~~~(\textbf{g})\qquad\qquad~~~\qquad\quad(\textbf{h})$\\

\noindent { \small \textbf{Figure 8.}  The solution \eqref{6} of the equation \eqref{Q1} with the parameters $\textbf{(a)}$ $k_{1}=-0.2+0.9i$, $\widetilde{k}_{1}=-0.2-0.9i$, $k_{2}=0.5+0.6i$, $\widetilde{k}_{2}=0.5-0.6i$, $\omega_{1}=\widetilde{\omega}_{1}=\omega_{2}=\widetilde{\omega}_{2}=1$; $\textbf{(e)}$ $k_{1}=\frac{\sqrt{6}}{6}+i$, $\widetilde{k}_{1}=\frac{\sqrt{6}}{6}-i$, $k_{2}=\frac{\sqrt{434}}{\sqrt{251}}+\sqrt{5}i$, $\widetilde{k}_{2}=\frac{\sqrt{434}}{\sqrt{251}}-\sqrt{5}i$, $\omega_{1}=\omega_{2}=1$, $\widetilde{\omega}_{1}=\widetilde{\omega}_{2}=-1$; $\textbf{(b)}$ and $\textbf{(f)}$ denote the density of $(\textbf{a})$ and $\textbf{(e)}$, respectively; $\textbf{(c)}$ and $\textbf{(g)}$ represents the dynamic behavior of the two-soliton solutions at different times;  $\textbf{(d)}$ is the characteristic line graph (blue line $L_{1}:x-2.76t=0$ and $L_{2}:x-t=0$) and contour map of $\textbf{(a)}$; $\textbf{(h)}$  is the characteristic line graph (blue line $L_{3}=x-2t$ and $L_{4}=x+\frac{4204}{251}t$) and contour map of $\textbf{(e)}$.}

It is not hard to see that after the  interaction between two solitons, the amplitude of change is not lower or higher than the original amplitude of the two waves. These phenomena indicates that the collision is elastic, which means that there is no energy loss in the collision process.

For the subtle difference in eigenvalues, namely $\Re(k_{1})=\Re(k_{2})$, but $\Re(\widetilde{k}_{1})\neq\Re(\widetilde{k}_{2})$, we find that the moving wave on one side does not change after the collision, while the moving wave on the other side decays slowly. This kind of graph is not given here and can be verified by itself. When we find out the relation between $x$ and $t$, we also can get the characteristic line graph of the two-soliton.

For the second case, if $k_{1}=-\widetilde{k}_{1}$, $k_{2}=-\widetilde{k}_{2}$, tend to real number, and the solitons reduce to
\begin{align*}
q(x,t)=-\frac{2}{sin(8t+2x)}.
\end{align*}

When the eigenvalue is close to a constant in this case, it presents a periodic wave, which is a breathing wave with a period. It is observed  that this solution present a stable and bounded periodic behavior wave. From Wang's work \cite{WL-2020-Chaos}, we know that the solution of this soliton is called pure solution, which is represented by hyperbola or trigonometric function. In addition, there is only one characteristic line in the pure solution, which ensures the wave to keep its original state under the influence of time evolution and collision.

(2) Unbounded two-soliton solutions

\noindent{\rotatebox{0}{\includegraphics[width=3.3cm,height=2.8cm,angle=0]{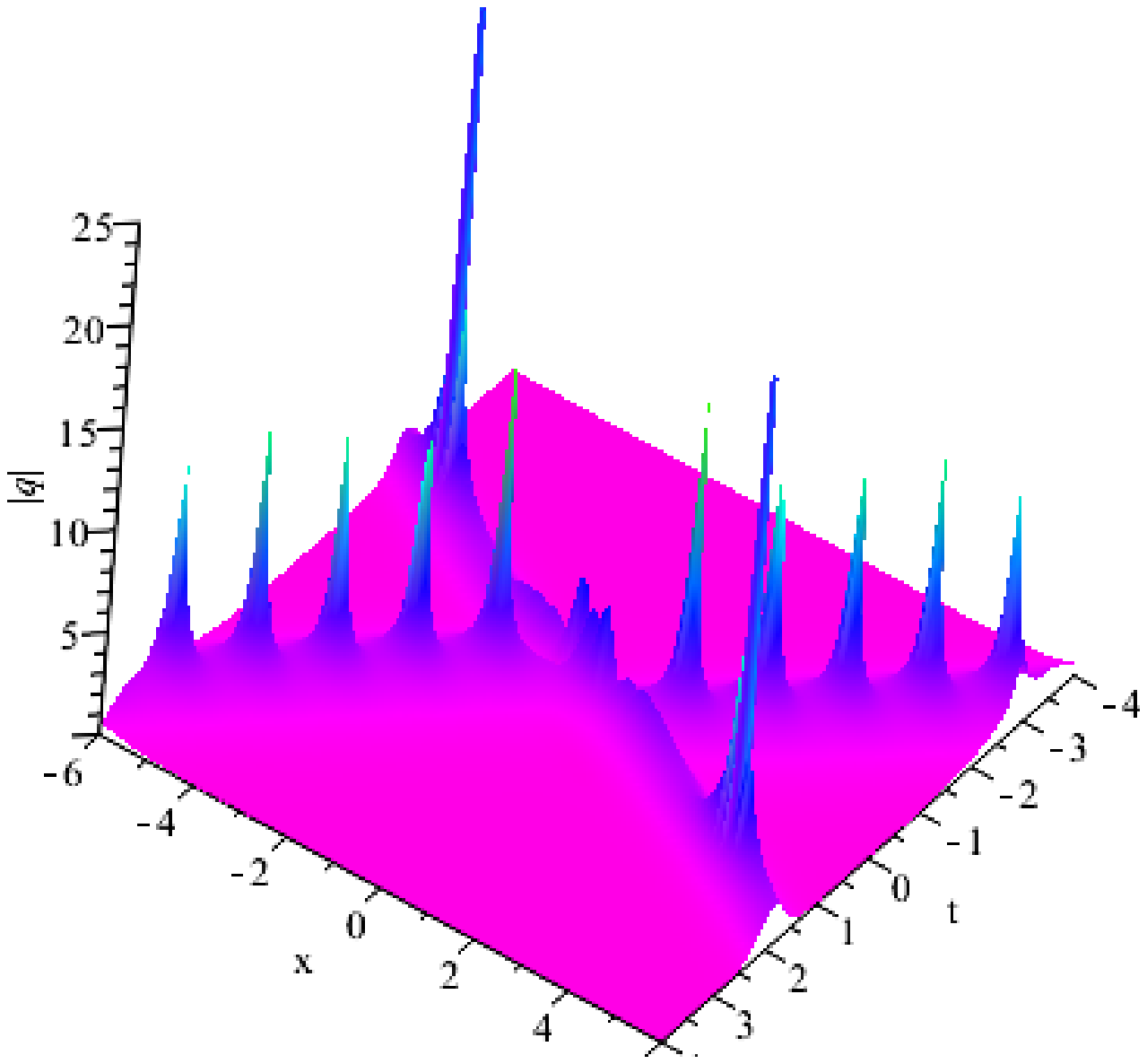}}
\rotatebox{0}{\includegraphics[width=2.5cm,height=2.5cm,angle=0]{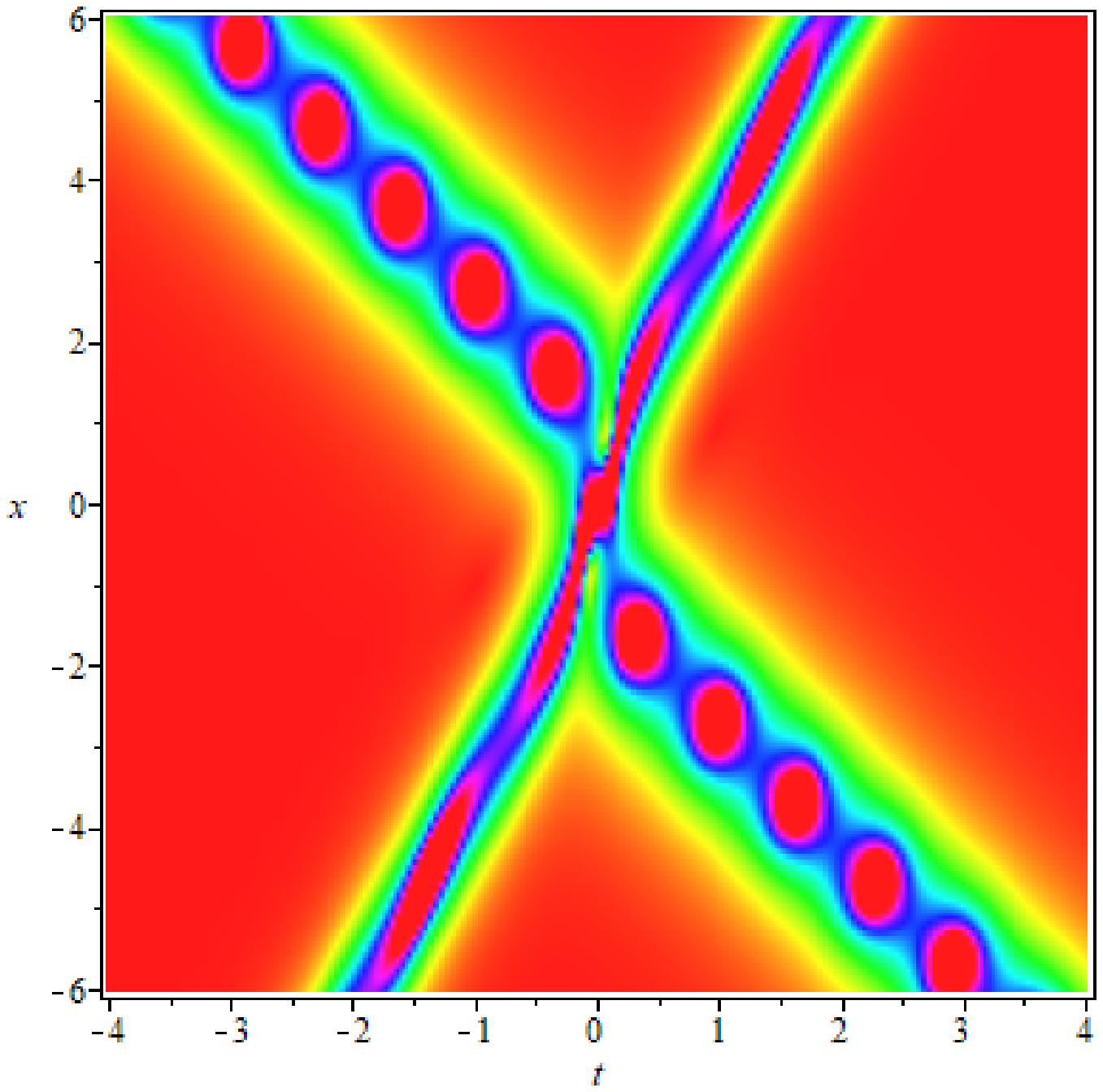}}
~\rotatebox{0}{\includegraphics[width=2.5cm,height=2.5cm,angle=0]{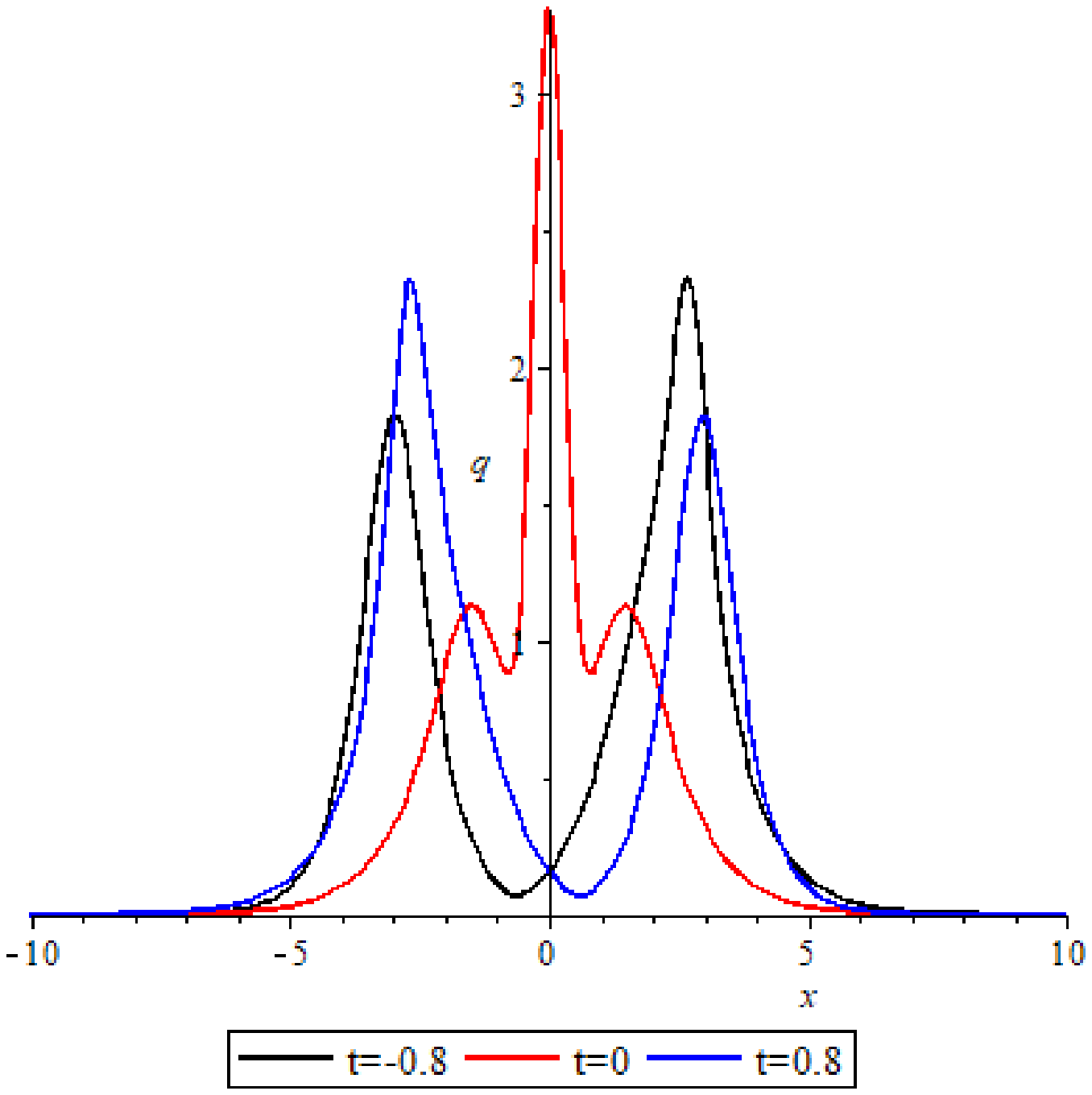}}
\hspace{-4.2cm}
\centerline{\begin{tikzpicture}[scale=0.43]
\draw[-][thick](-3,0)--(-2,0);
\draw[-][thick](-2,0)--(-1,0);
\draw[-][thick](-1,0)--(0,0);
\draw[-][thick](0,0)--(1,0);
\draw[-][thick](1,0)--(2,0);
\draw[-][thick](2,0)--(3,0);
\draw[-][thick](3,0)--(3,2);
\draw[-][thick](3,2)--(3,5.7);
\draw[-][thick](3,5.7)--(0,5.7);
\draw[-][thick](0,5.7)--(-3,5.7);
\draw[-][thick](-3,5.7)--(-3,2);
\draw[-][thick](-3,2)--(-3,0);
\draw[-][thick](-3,0)--(0,0);
\draw[-][thick](0,0)--(3,0);
\draw[-][dashed](-3,3)--(0,3);
\draw[-][dashed](0,3)--(3,3);
\draw[-][dashed](0,5.7)--(0,3);
\draw[-][dashed](0,3)--(0,0);
\draw[fill] (-0.3,4)node{$\textcolor[rgb]{1.00,0.00,0.00}{\bullet}$};
\draw[fill] (-0.3,2)node{$\textcolor[rgb]{1.00,0.00,0.00}{\bullet}$};
\draw[fill] (0.7,3.5)node{$\textcolor[rgb]{0.00,0.00,1.00}{\bullet}$};
\draw[fill] (-0.7,2.5)node{$\textcolor[rgb]{0.00,0.00,1.00}{\bullet}$};
\draw[-][dashed](2,3)--(3,3)node[right]{$Re~z$};
\draw[-][dashed](0,5.5)--(0,5.7)node[above]{$Im~z$};
\end{tikzpicture}}}

$~~\quad(\textbf{a})\qquad\qquad\qquad\qquad~~~(\textbf{b})
\qquad\qquad\quad~~~~(\textbf{c})\qquad\quad~~~\qquad\quad(\textbf{d})$\\

\noindent { \small \textbf{Figure 9.} The solution \eqref{5} of the equation \eqref{Q1} with the parameters $\textbf{(a)}$  $k_{1}=-0.2+0.9i$, $\widetilde{k}_{1}=-0.2-0.9i$, $k_{2}=0.5+0.6i$, $\widetilde{k}_{2}=-0.5-0.6i$. (\textbf{a})(\textbf{b})(\textbf{c}): the local structure, density, the dynamic behavior of the two-soliton solutions at different times and distribution pattern of discrete spectral points.}

Soliton solutions exhibit nonsingular morphology under certain parameters, which  always collapse repeatedly and are asymmetric structures with phase difference. As we mentioned earlier, the two-soliton is indeed a nonlinear superposition of two fundamental solitons.

\subsubsection{The degenerate solutions}
\
\newline
(1) The position solutions

\noindent\rotatebox{0}{\includegraphics[width=4.4cm,height=3.5cm,angle=0]{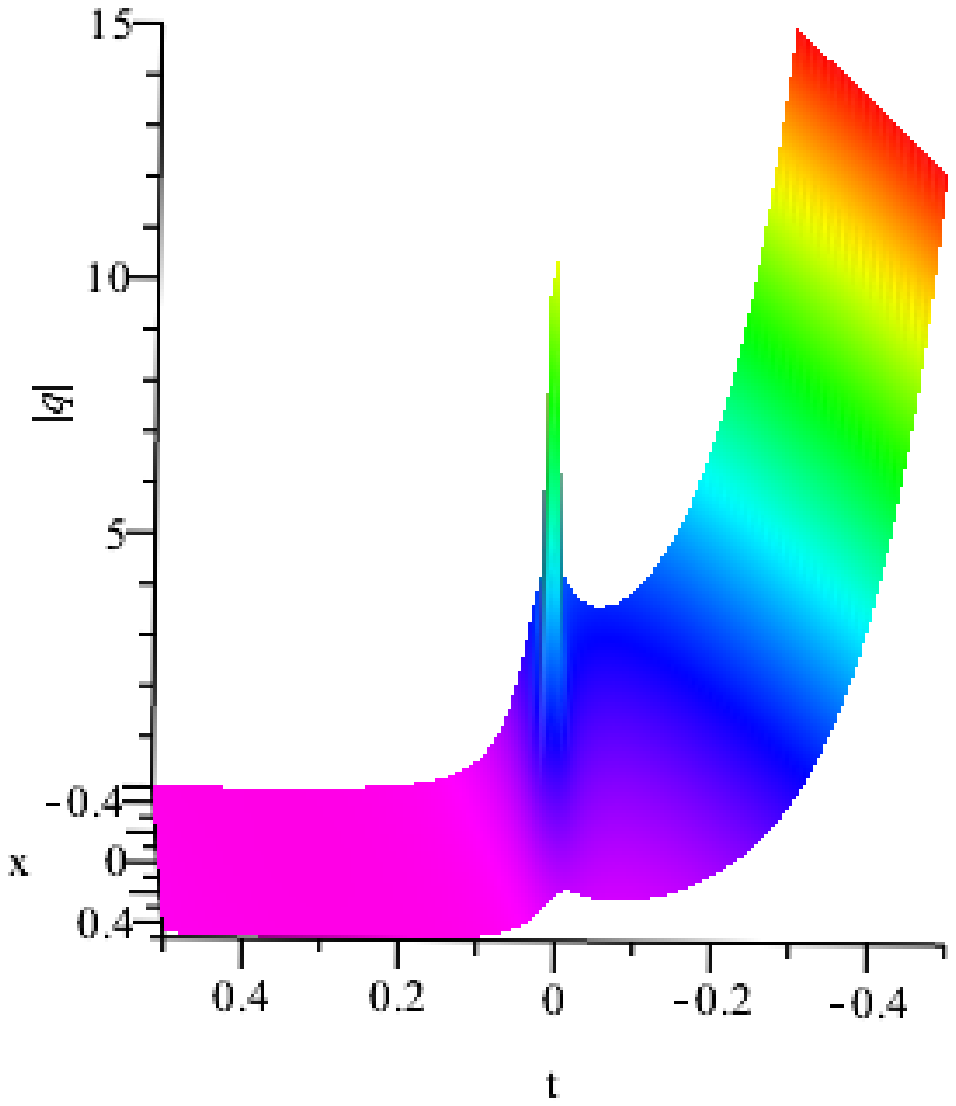}}
\rotatebox{0}{\includegraphics[width=3.3cm,height=2.8cm,angle=0]{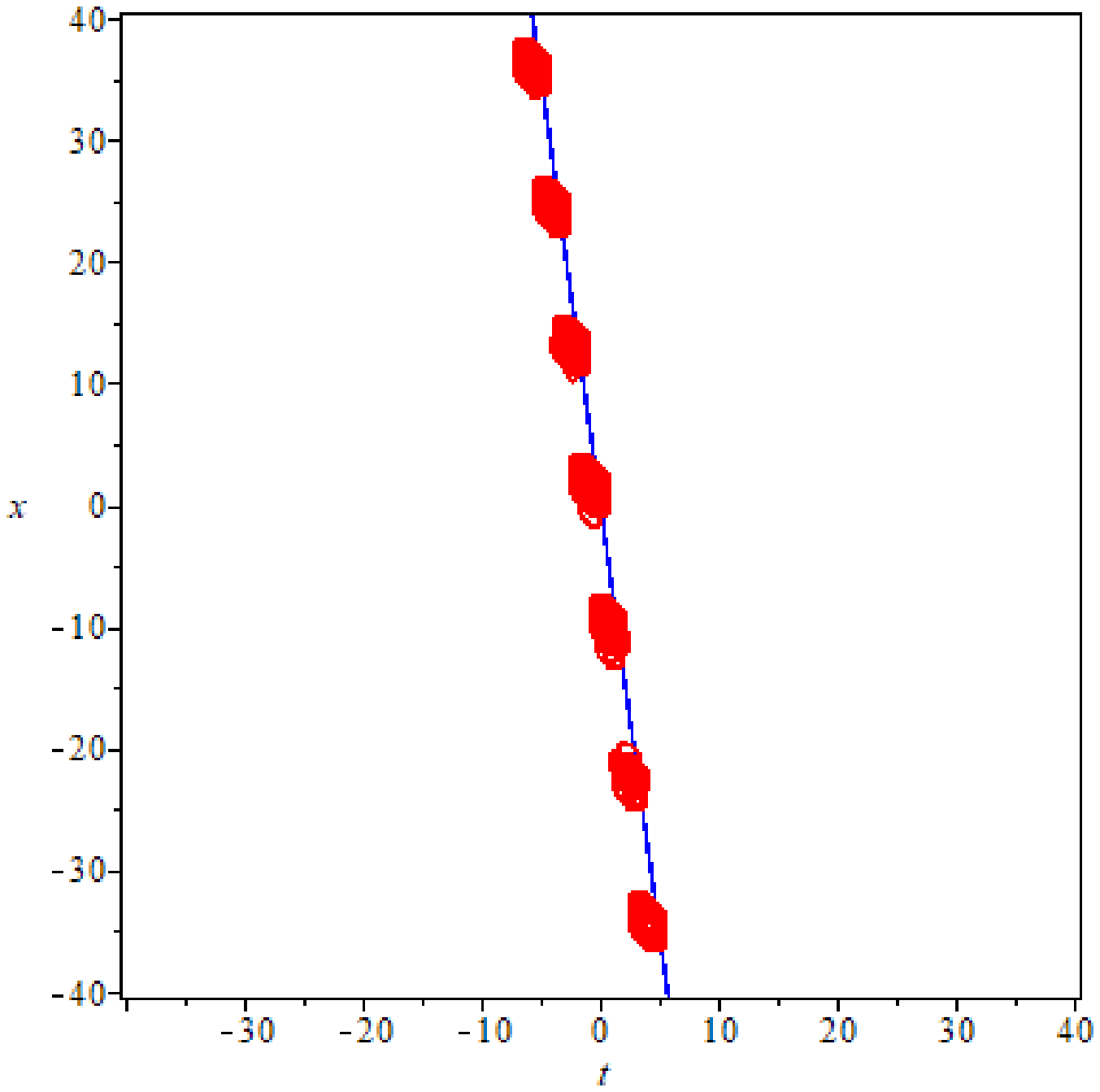}}
\hspace{-3.4cm}
\centerline{\begin{tikzpicture}[scale=0.47]
\draw[-][thick](-3,0)--(-2,0);
\draw[-][thick](-2,0)--(-1,0);
\draw[-][thick](-1,0)--(0,0);
\draw[-][thick](0,0)--(1,0);
\draw[-][thick](1,0)--(2,0);
\draw[-][thick](2,0)--(3,0);
\draw[-][thick](3,0)--(3,2);
\draw[-][thick](3,2)--(3,5.7);
\draw[-][thick](3,5.7)--(0,5.7);
\draw[-][thick](0,5.7)--(-3,5.7);
\draw[-][thick](-3,5.7)--(-3,2);
\draw[-][thick](-3,2)--(-3,0);
\draw[-][thick](-3,0)--(0,0);
\draw[-][thick](0,0)--(3,0);
\draw[-][dashed](-3,3)--(0,3);
\draw[-][dashed](0,3)--(3,3);
\draw[-][dashed](0,5.7)--(0,3);
\draw[-][dashed](0,3)--(0,0);
\draw[fill] (0.9,3.8)node{$\textcolor[rgb]{1.00,0.00,0.00}{\bullet}$};
\draw[fill] (-0.9,2.2)node{$\textcolor[rgb]{1.00,0.00,0.00}{\bullet}$};
\draw[fill] (1.0,4.0)node{$\textcolor[rgb]{0.00,0.00,1.00}{\bullet}$};
\draw[fill] (-1.0,2.0)node{$\textcolor[rgb]{0.00,0.00,1.00}{\bullet}$};
\draw[-][dashed](2,3)--(3,3)node[right]{$Re~z$};
\draw[-][dashed](0,5.5)--(0,5.7)node[above]{$Im~z$};
\end{tikzpicture}}

$~\qquad\quad\quad(\textbf{a})\qquad\qquad~\qquad\qquad\qquad(\textbf{b})
\qquad~~~~\quad\qquad\qquad\qquad(\textbf{c})$

\noindent { \small \textbf{Figure 10.} The solution \eqref{5} of the equation \eqref{Q1} with the parameters $\textbf{(a)}$  $k_{1}=\frac{9}{10}+\frac{4i}{5}$, $\widetilde{k}_{1}=-\frac{9}{10}-\frac{4i}{5}$. (\textbf{a})(\textbf{b})(\textbf{c}): the local structure, characteristic line and  distribution pattern of discrete spectral points. }

The four characteristic parameters we choose are all non pure imaginary numbers and present the relations $\widetilde{k}_{1}=-\widetilde{k}_{1}$. In addition, taking ${k}_{2}\rightarrow{k}_{1}$, the degenerate solutions can be derived. It can be observed that these solutions collapse at $(x,t)=(0,0)$. For a given $x$, they are close to infinity regardless of any parameter as $x\rightarrow -\infty$.

\subsection{Three-soliton solutions}
We take $N$=3 in the formulae \eqref{5} in order to get three-soliton solutions.  We need to note that the characteristic lines in this case are different from the characteristic lines of one-soliton and two-soliton. Similarly, as a comparison, we discuss the following two situations.
\subsubsection{Pure imaginary eigenvalues}
\
\newline
Bounded three-soliton solutions

For the first case, we take the eigenvalues as follows,
\begin{align*}
\left\{\begin{aligned}k_{1}&=-i,~\widetilde{k}_{1}=i, ~\omega_{1}=\widetilde{\omega}_{1}=1,\\
k_{2}&=-\frac{i}{5},~ \widetilde{k}_{2}=\frac{i}{5},~ \omega_{2}=\widetilde{\omega}_{2}=1,\\
k_{3}&=-\frac{i}{2},~ \widetilde{k}_{3}=\frac{i}{2}, ~\omega_{3}=\widetilde{\omega}_{3}=1,
\end{aligned}\right.
\end{align*}
then the solution of \eqref{5} can be characterized in Fig. 11.

\noindent{\rotatebox{0}{\includegraphics[width=3.3cm,height=2.7cm,angle=0]{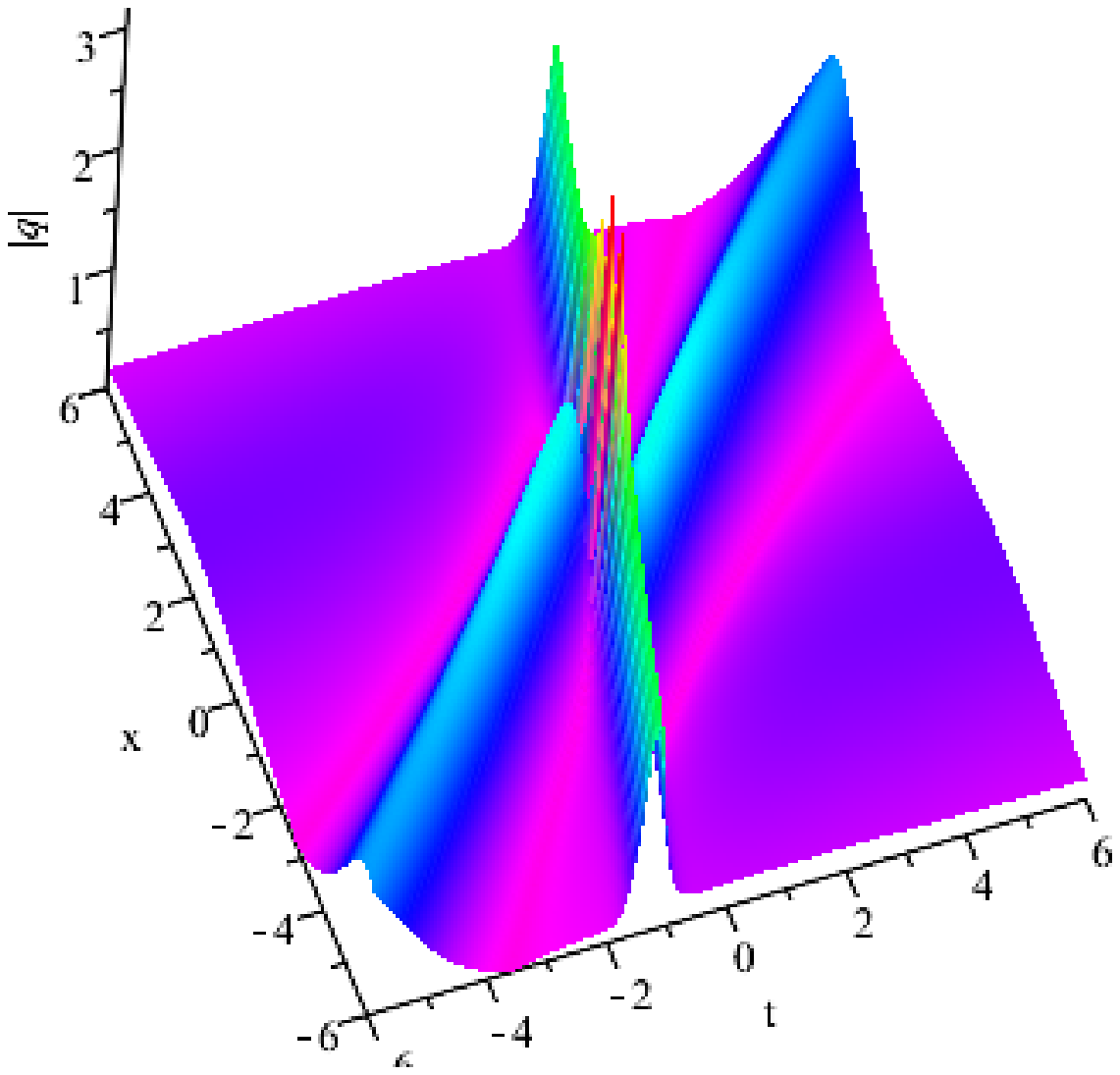}}
~\quad\rotatebox{0}{\includegraphics[width=2.5cm,height=2.4cm,angle=0]{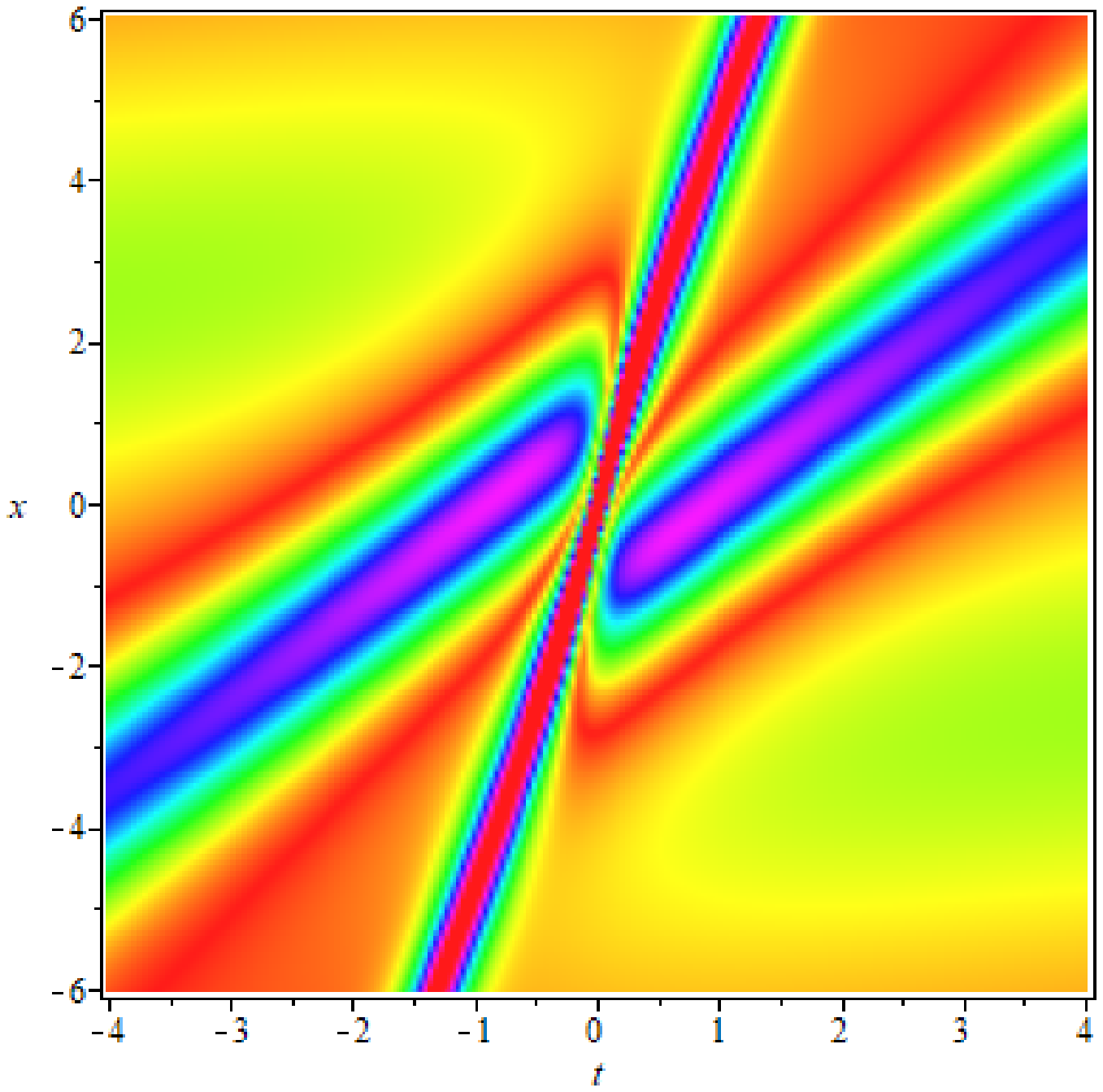}}
~\quad\rotatebox{0}{\includegraphics[width=2.5cm,height=2.4cm,angle=0]{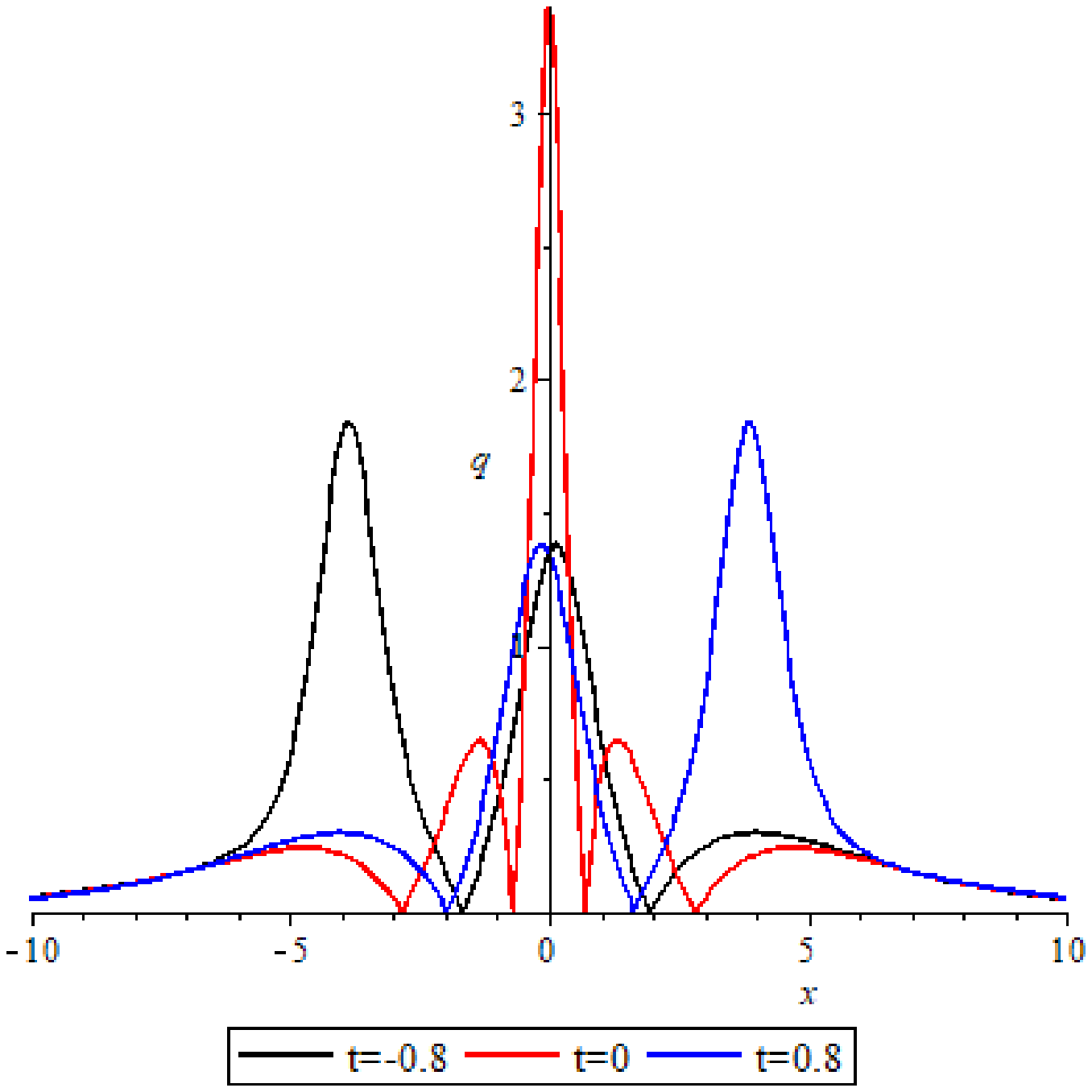}}
~\quad\rotatebox{0}{\includegraphics[width=2.5cm,height=2.4cm,angle=0]{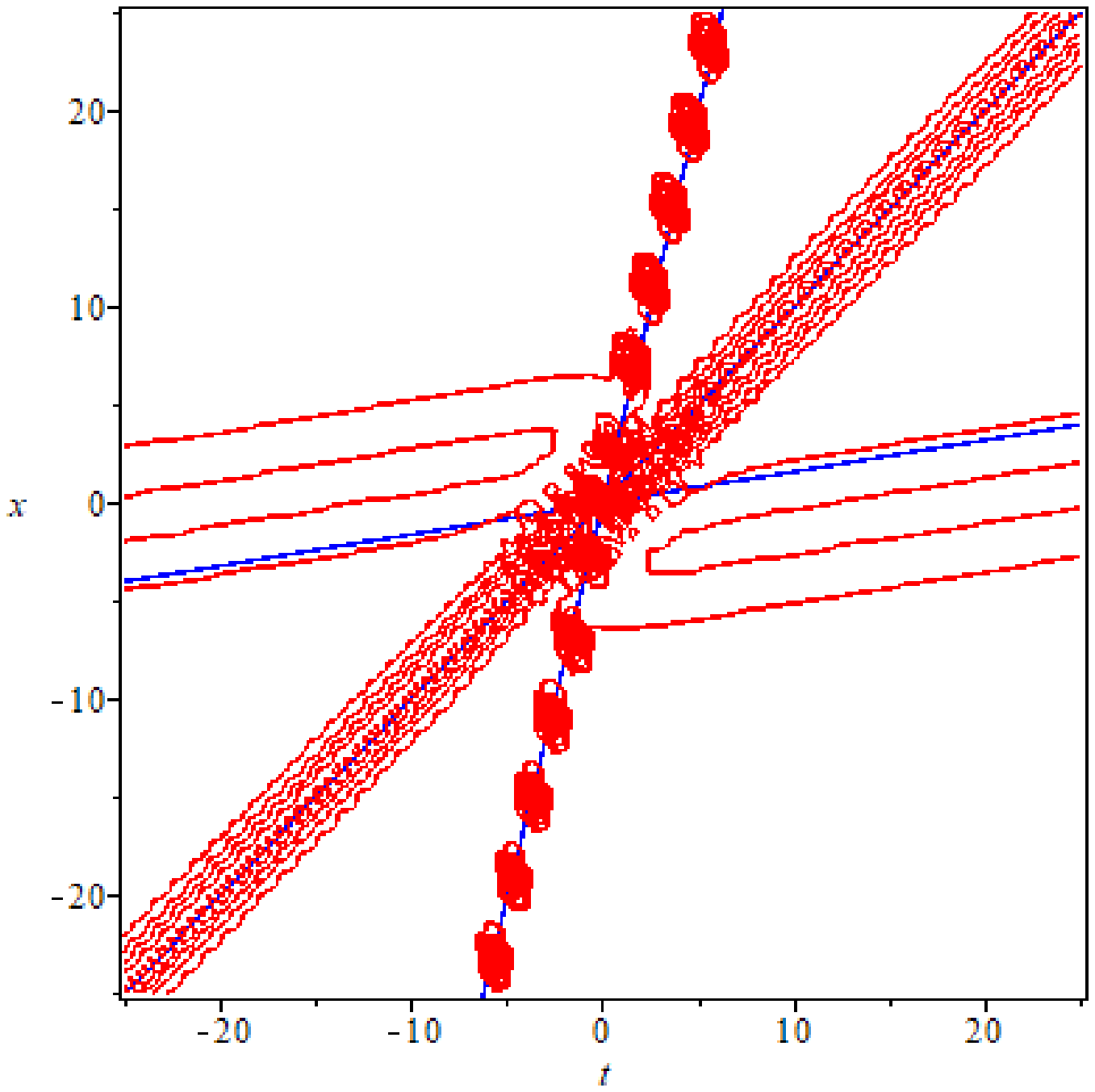}}}

$~~\quad\quad(\textbf{a})\qquad\qquad\qquad\qquad\quad(\textbf{b})
\qquad\qquad\qquad~~~(\textbf{c})\qquad\qquad~~~\qquad\quad(\textbf{d})$\\

\noindent{\rotatebox{0}{\includegraphics[width=3.3cm,height=2.7cm,angle=0]{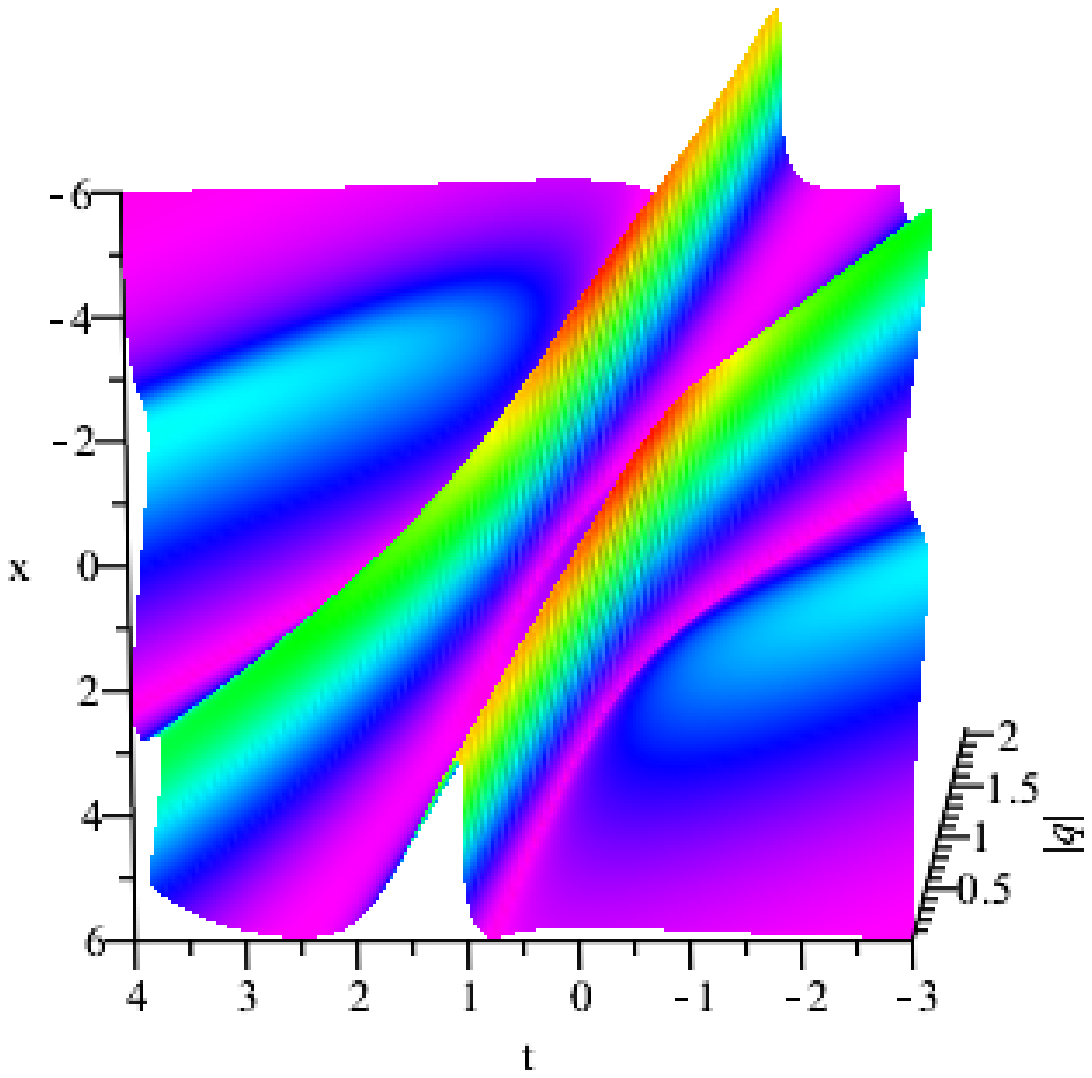}}
~\quad\rotatebox{0}{\includegraphics[width=2.5cm,height=2.4cm,angle=0]{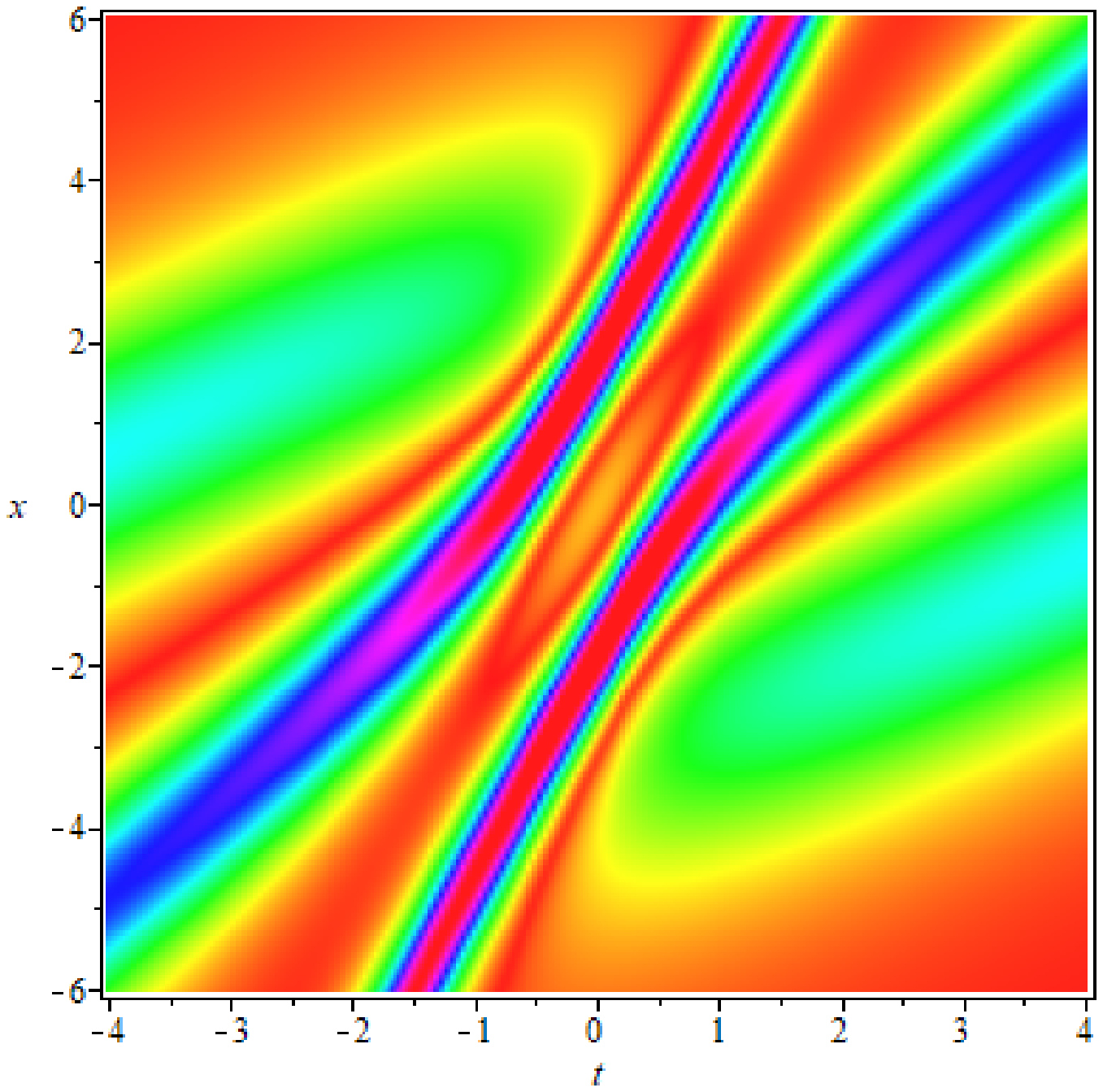}}
~\quad\rotatebox{0}{\includegraphics[width=2.5cm,height=2.4cm,angle=0]{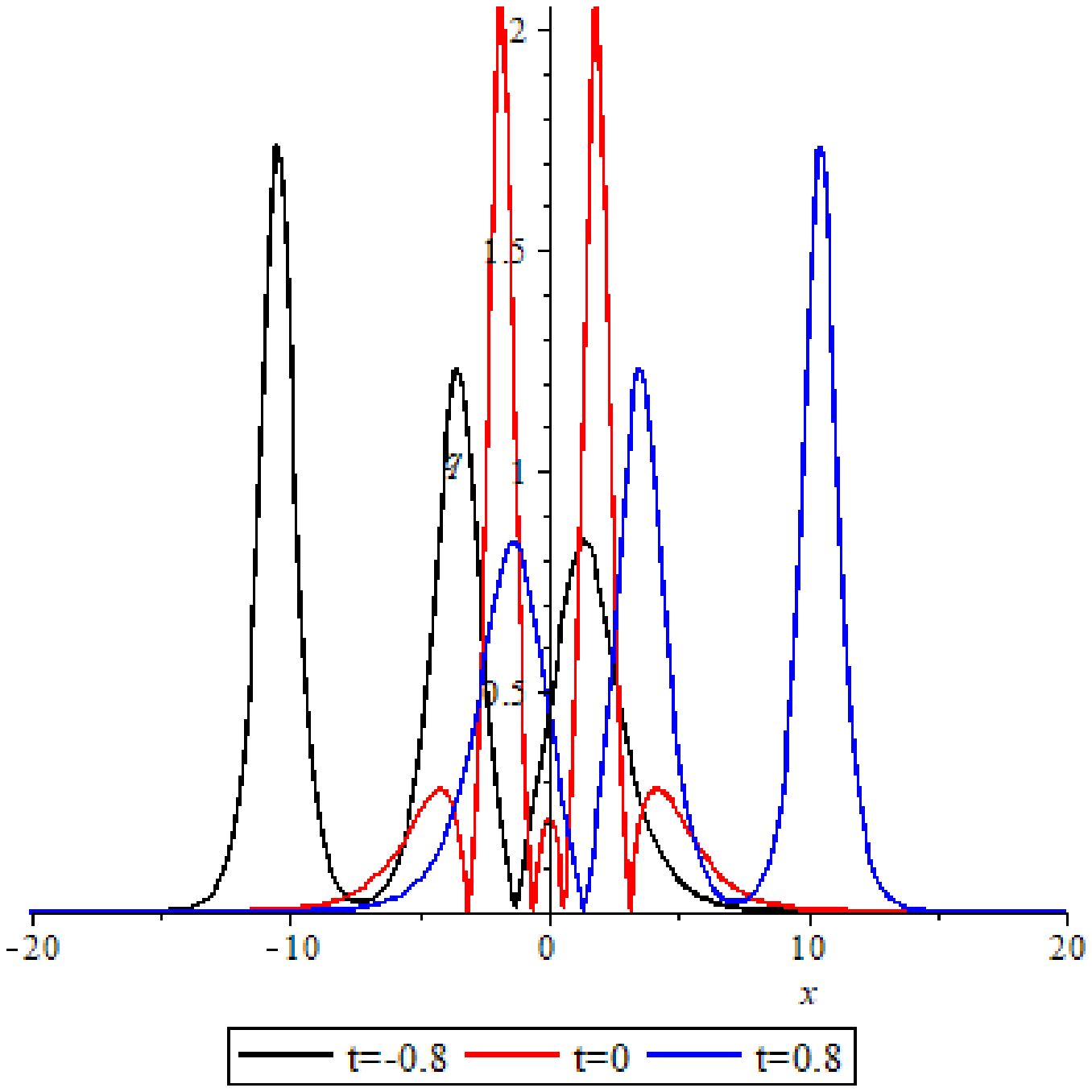}}
~\quad\rotatebox{0}{\includegraphics[width=2.5cm,height=2.4cm,angle=0]{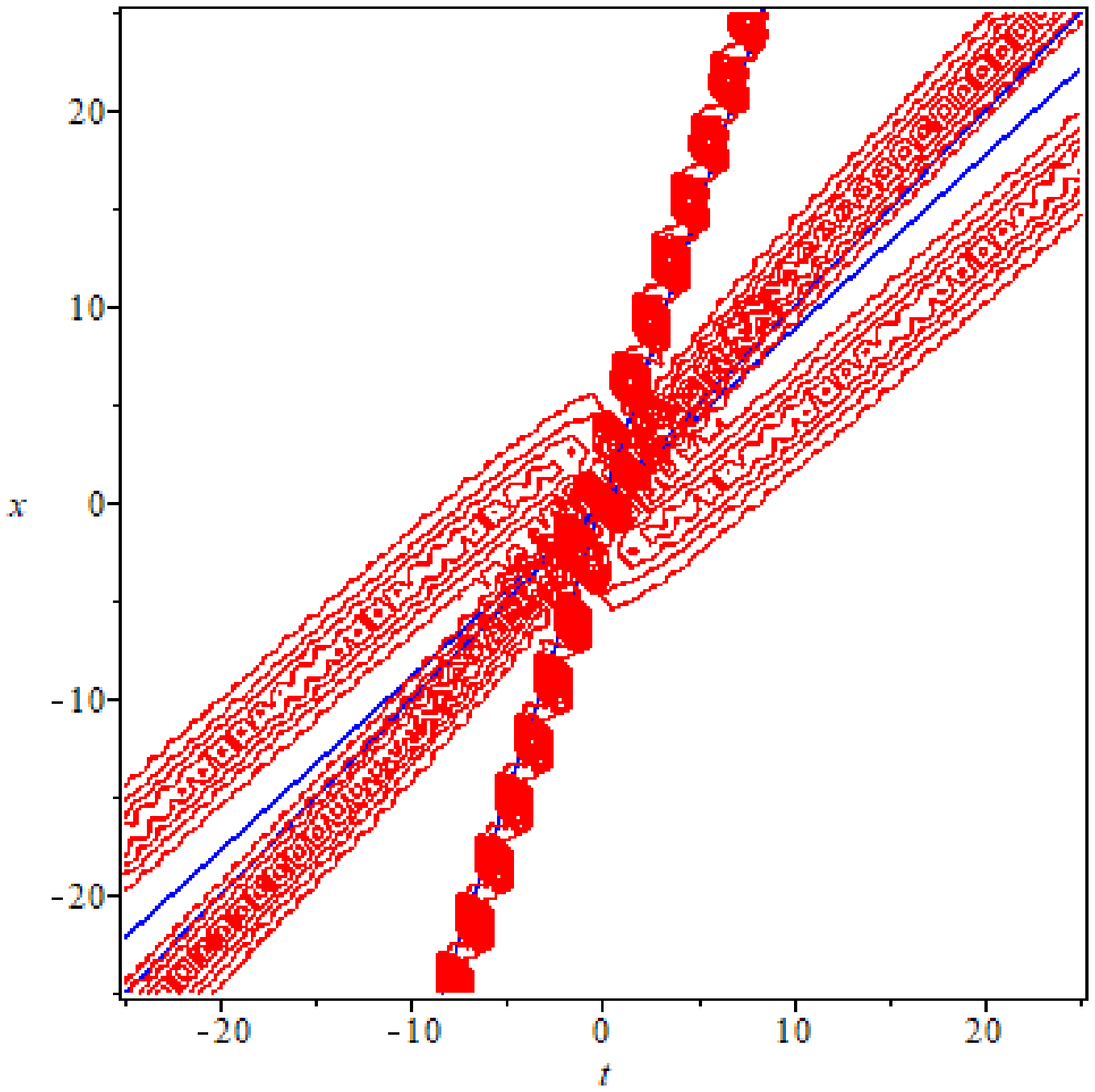}}}

$~~\quad\quad(\textbf{e})\qquad\qquad\qquad\qquad\quad(\textbf{f})
\qquad\qquad\qquad~~~~(\textbf{g})\qquad\qquad~~~\qquad\quad(\textbf{h})$\\

\noindent{\rotatebox{0}{\includegraphics[width=3.3cm,height=2.7cm,angle=0]{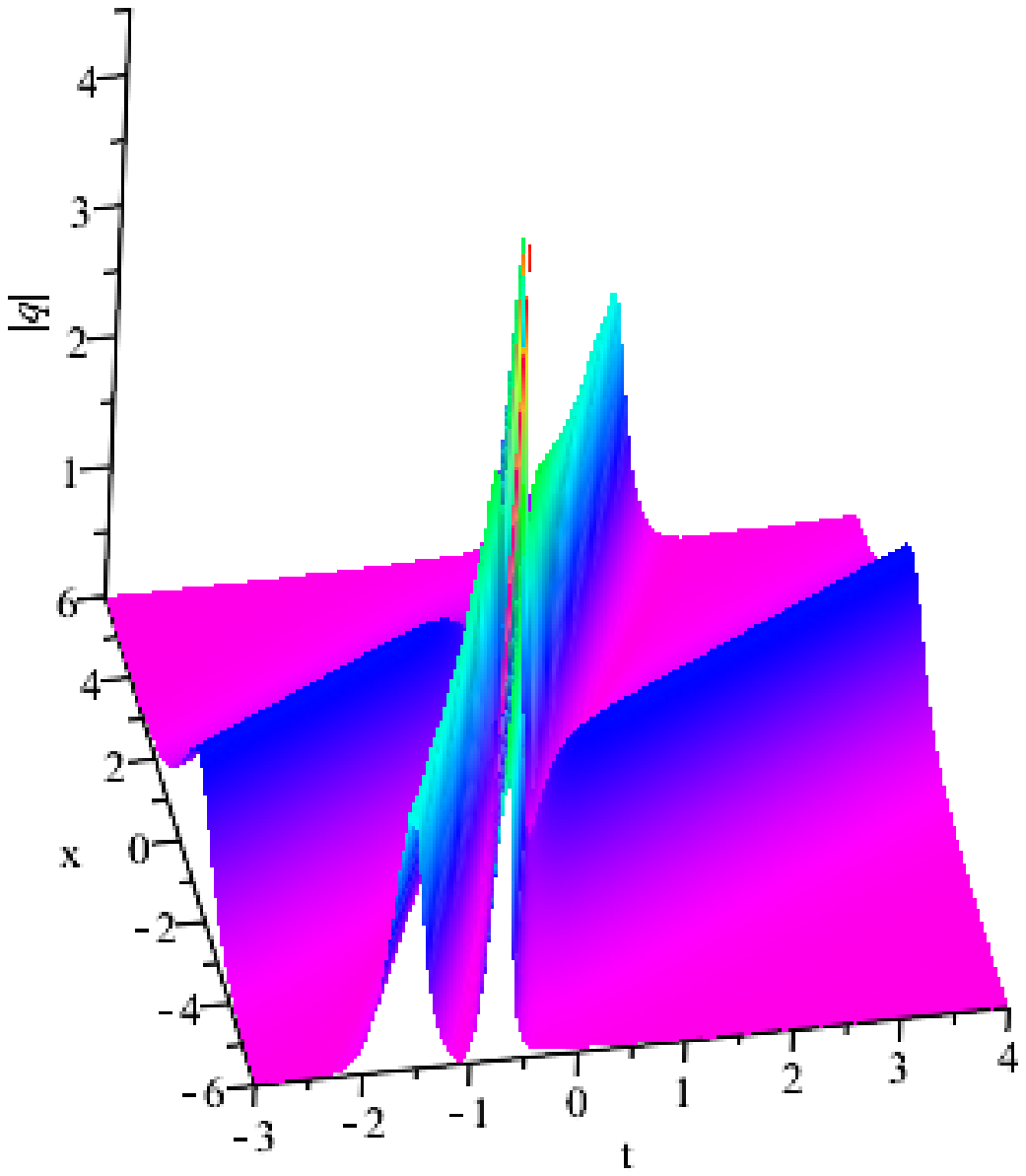}}
~\quad\rotatebox{0}{\includegraphics[width=2.5cm,height=2.4cm,angle=0]{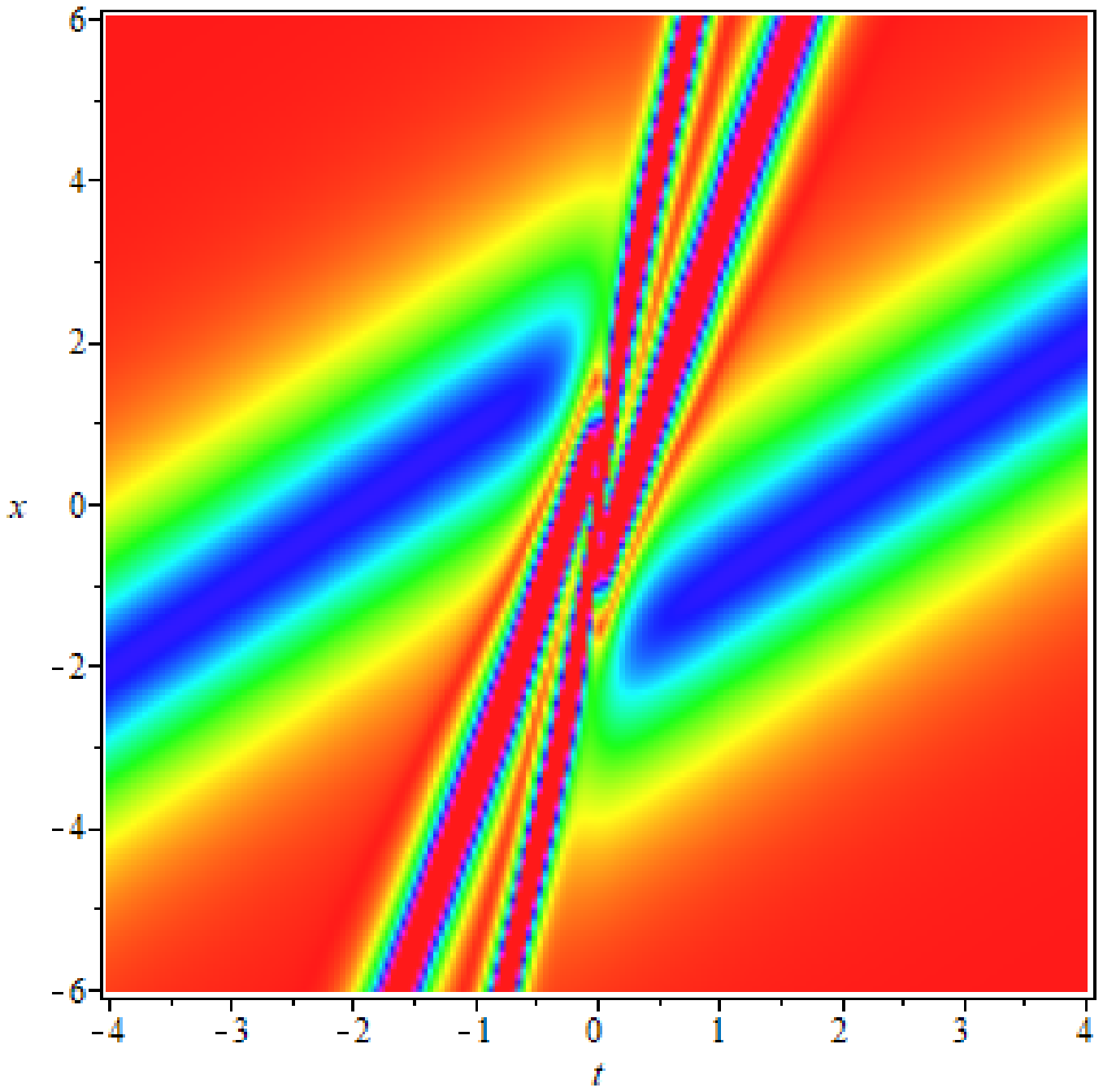}}
~\quad\rotatebox{0}{\includegraphics[width=2.5cm,height=2.4cm,angle=0]{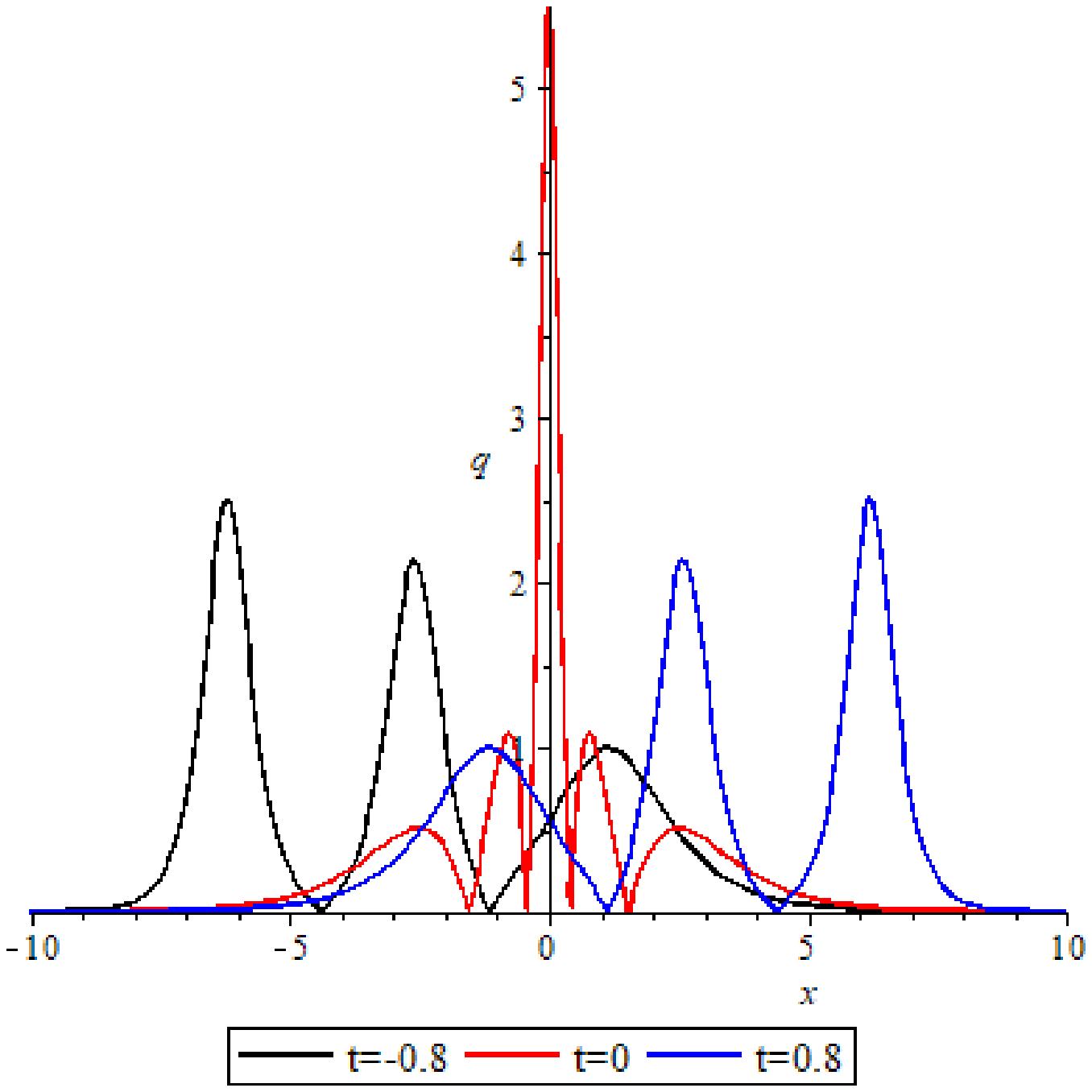}}
~\quad\rotatebox{0}{\includegraphics[width=2.5cm,height=2.4cm,angle=0]{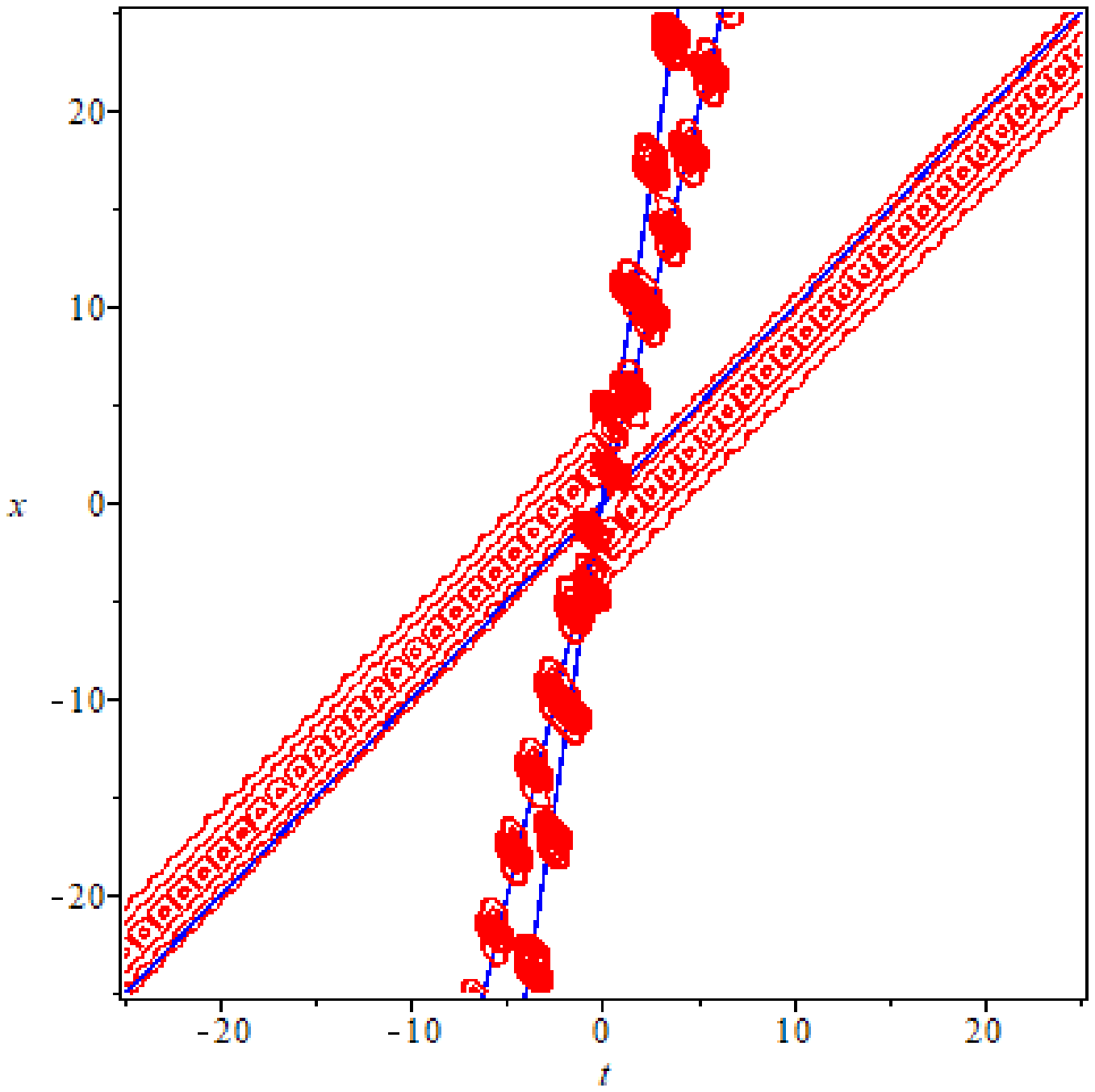}}}

$~~\quad\quad(\textbf{i})\qquad\qquad\qquad\qquad\quad(\textbf{j})
\qquad\qquad\qquad~~~~(\textbf{k})\qquad\qquad~~~\qquad\quad(\textbf{l})$\\

\noindent { \small \textbf{Figure 11.}  The solution \eqref{6} of the equation \eqref{Q1} with the parameters $\textbf{(a)}$ $k_{1}=-i$, $\widetilde{k}_{1}=i$, $k_{2}=-\frac{i}{5}$, $\widetilde{k}_{2}=\frac{i}{5}$, $k_{3}=-\frac{i}{2}$, $\widetilde{k}_{3}=\frac{i}{2}$, $\omega_{1}=\omega_{2}=\omega_{3}=1$, $\widetilde{\omega}_{1}=\widetilde{\omega}_{2}=\widetilde{\omega}_{3}=1$; $\textbf{(e)}$ $k_{1}=-\frac{\sqrt{219}i}{\sqrt{988}}$, $\widetilde{k}_{1}=\frac{\sqrt{219}i}{\sqrt{988}}$, $k_{2}=\frac{\sqrt{3}i}{2}$, $\widetilde{k}_{2}=-\frac{\sqrt{3}i}{2}$, $k_{3}=-\frac{i}{2}$, $\widetilde{k}_{3}=-\frac{i}{2}$, $\omega_{1}=\omega_{2}=\omega_{3}=1$, $\widetilde{\omega}_{1}=\widetilde{\omega}_{2}=\widetilde{\omega}_{3}=1$;
$\textbf{(i)}$  $k_{1}=-i$, $\widetilde{k}_{1}=i$, $k_{2}=-\frac{i}{2}$, $\widetilde{k}_{2}=\frac{i}{2}$, $k_{3}=-\frac{5i}{4}$, $\widetilde{k}_{3}=\frac{5i}{4}$, $\omega_{1}=\omega_{2}=\omega_{3}=1$, $\widetilde{\omega}_{1}=\widetilde{\omega}_{2}=\widetilde{\omega}_{3}=1$;
$\textbf{(b)}$, $\textbf{(f)}$ and $\textbf{(j)}$ denote the density of $(\textbf{a})$, $\textbf{(e)}$  and $(\textbf{i})$, respectively; $\textbf{(c)}$, $\textbf{(g)}$ and  $\textbf{(k)}$ represents the dynamic behavior of the three-soliton solutions at different times;  $\textbf{(d)}$ is the characteristic line graph (blue line $L_{1}:x-4t=0$, $L_{2}:x-\frac{4}{25}t=0$ and $L_{3}:x-t=0$) and contour map of $\textbf{(a)}$; $\textbf{(h)}$  is the characteristic line graph (blue line $L_{4}:x-\frac{219}{247}t=0$, $L_{5}:x-3t=0$, $L_{6}:x-t=0$) and contour map of $\textbf{(e)}$.$\textbf{(l)}$  is the characteristic line graph (blue line $L_{7}:x-4t=0$, $L_{8}:x-t=0$, $L_{9}:x-\frac{25t}{4}=0$) and contour map of $\textbf{(i)}$.}

As seen in  Fig. 11,  Figs. 11(\textbf{e,i}) are obtained by the rotation of graph Fig. 11(\textbf{a}) with $\tan\theta_{1}=-\frac{96}{41}$ and $\tan\theta_{2}=\frac{21}{29}$.  For this group of pictures, we can see that the graph of Fig. 11(\textbf{e}) and Fig. 11(\textbf{i}) do not change obviously after the rotation angle, and their waveforms and velocities do not change after  colliding with each other. For the remaining rotation, we do not show the change of its dynamic behavior here. Here we explain the situation that  one side of the moving wave remains the original state after the collision, while the other side of the moving wave has repeatedly collapsed. This phenomenon is called semi-elastic collision. The reason for this phenomenon may be that $\tan\theta$ are taken as the approximate value, which leads to the error of solving slope, and finally leads to the error of the eigenvalue. Comparing with the two-soliton, the change of the phenomenon is caused by small errors. In addition, although the error is very small, it effects the propagation of solutions.

\noindent{\rotatebox{0}{\includegraphics[width=3.3cm,height=2.7cm,angle=0]{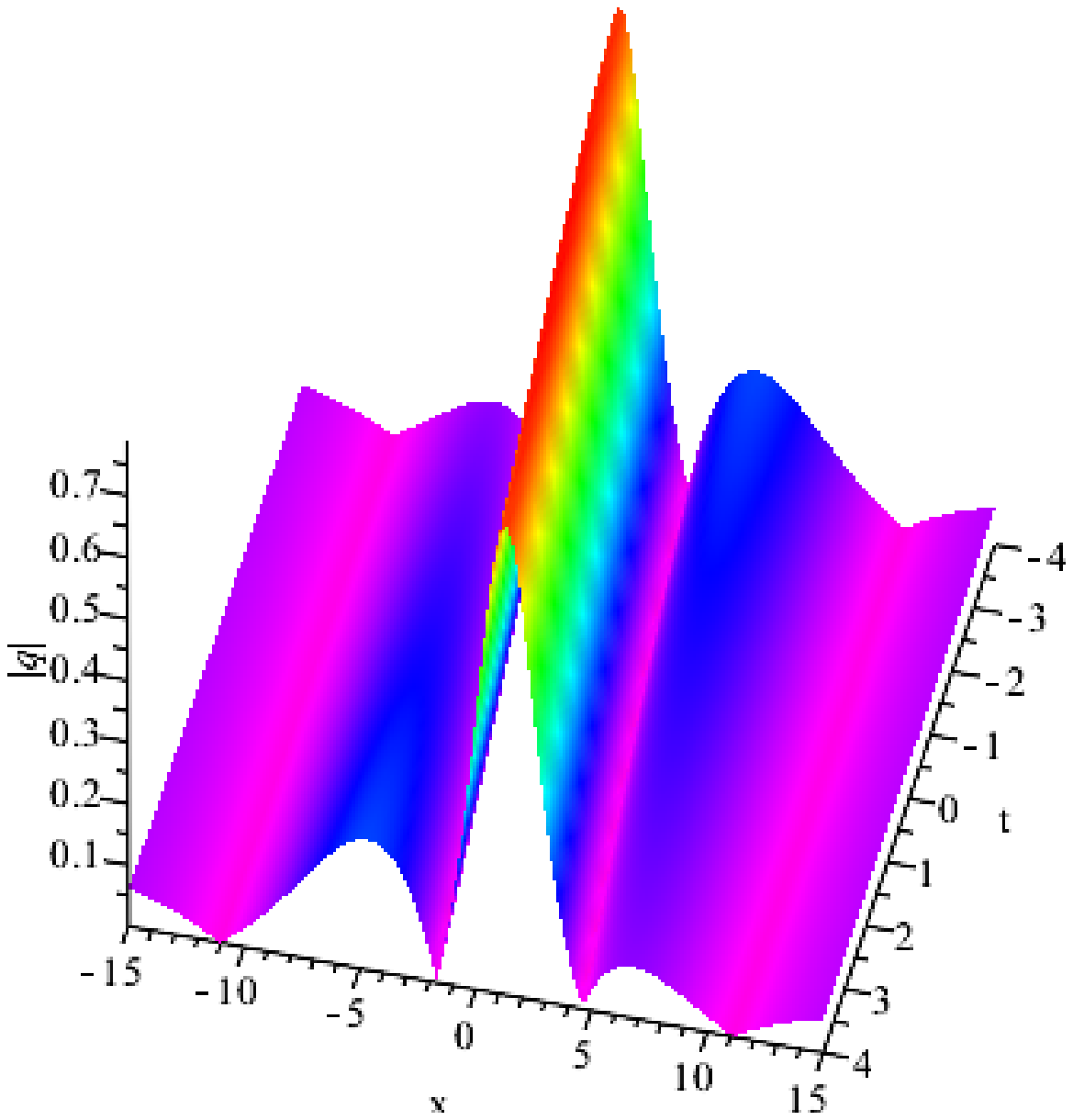}}
~\quad\rotatebox{0}{\includegraphics[width=2.5cm,height=2.4cm,angle=0]{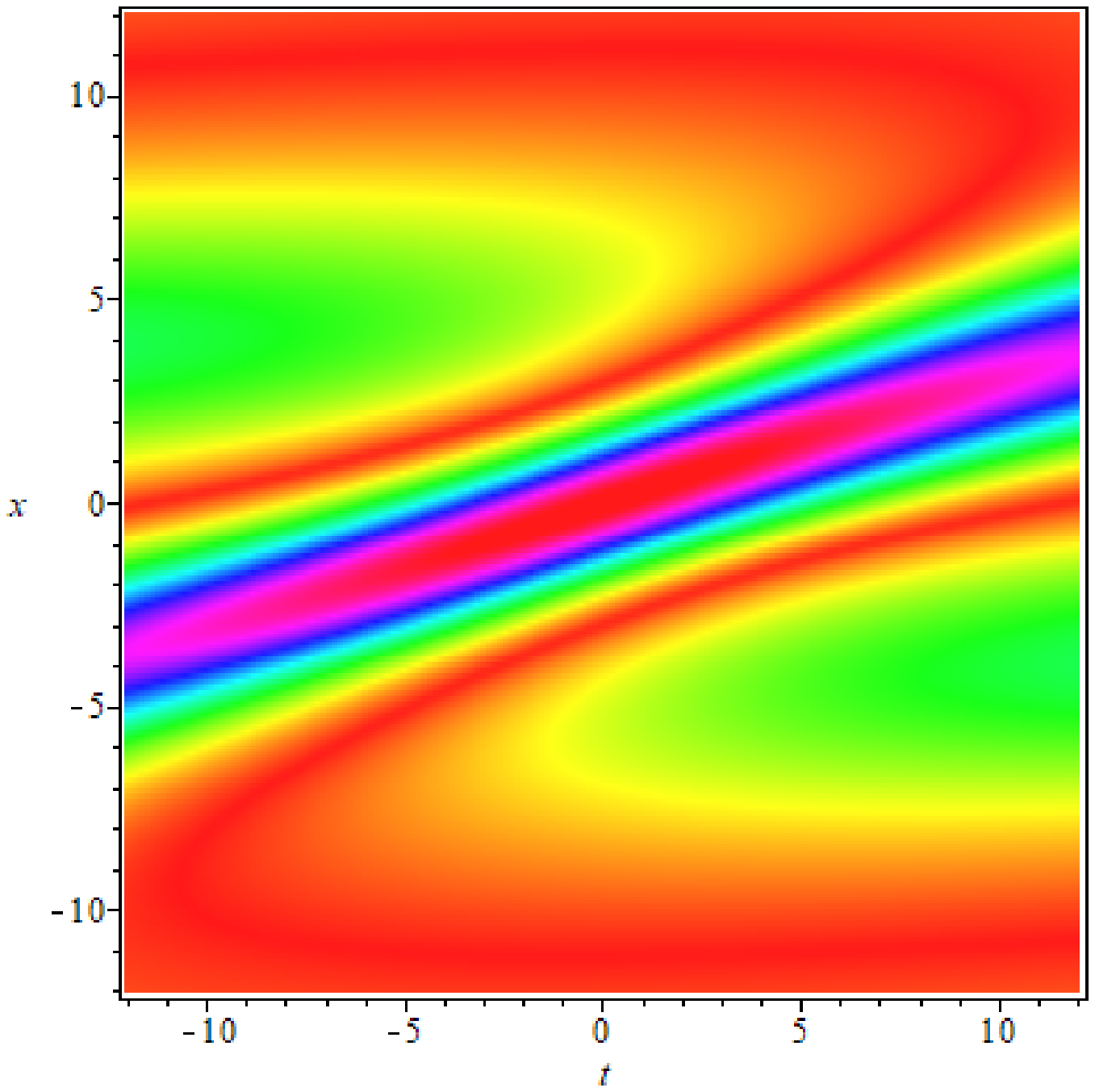}}
~\quad\rotatebox{0}{\includegraphics[width=2.5cm,height=2.4cm,angle=0]{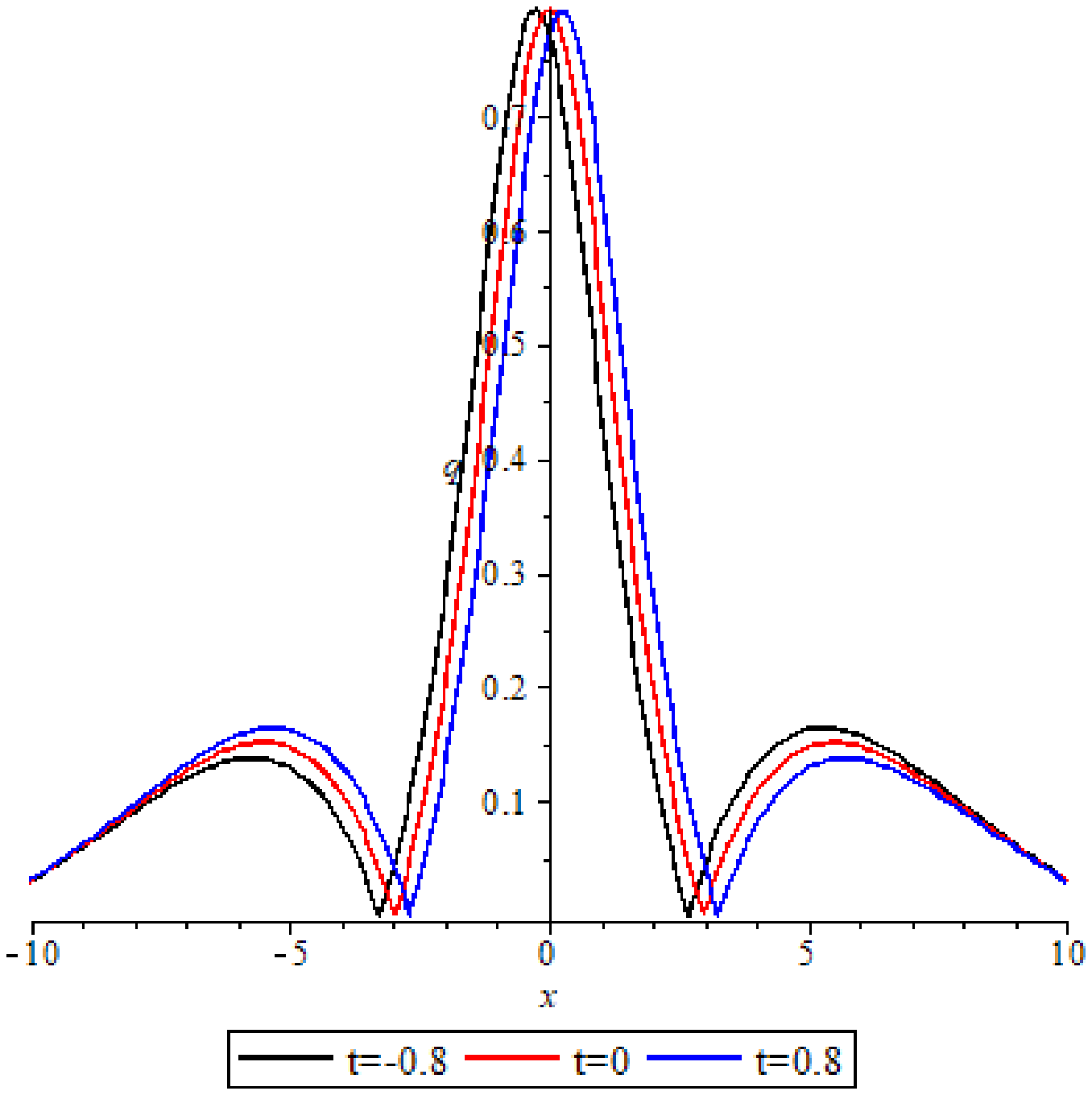}}
~\quad\rotatebox{0}{\includegraphics[width=2.5cm,height=2.4cm,angle=0]{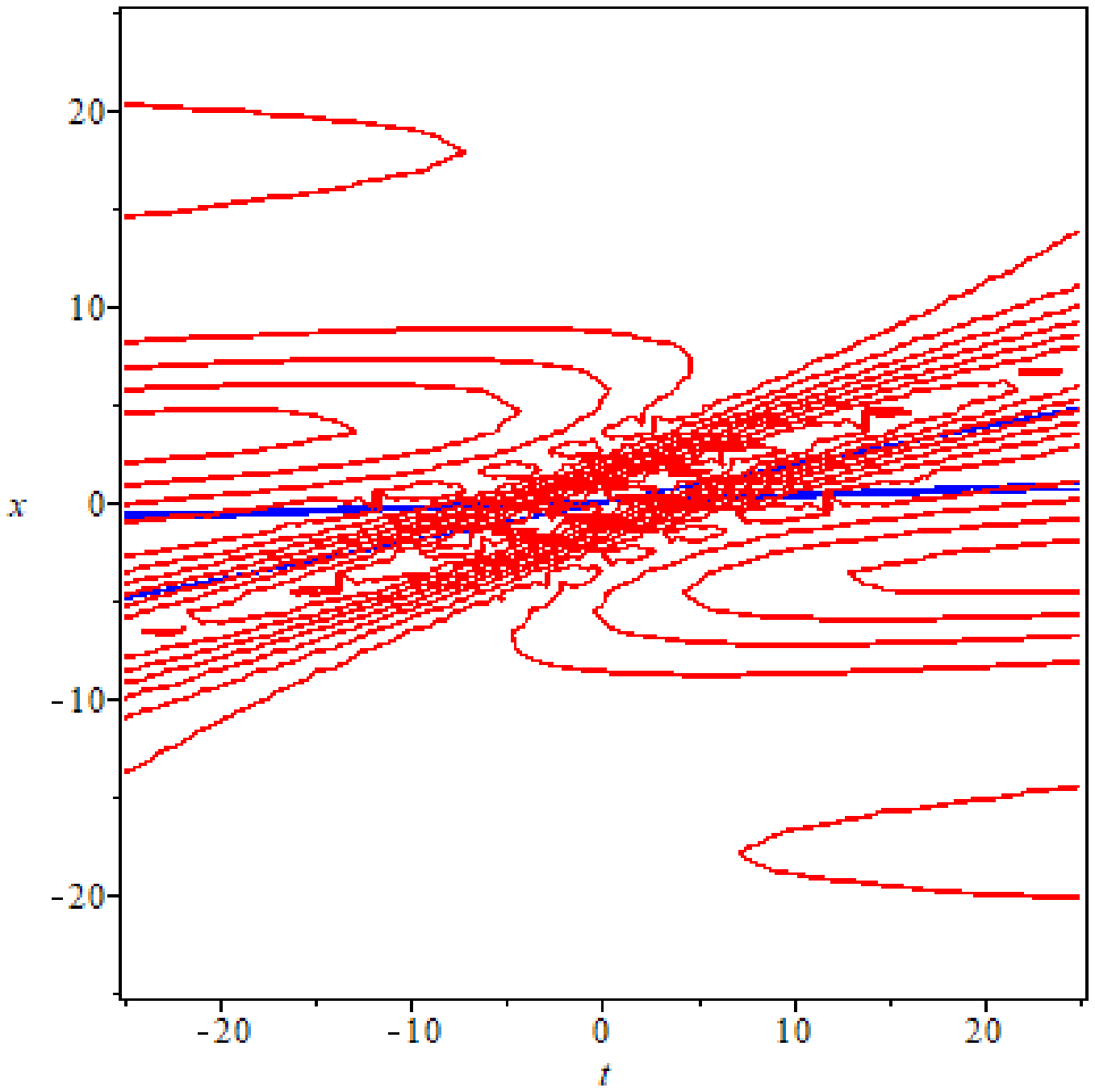}}}

$~~\quad\quad(\textbf{a})\qquad\qquad\qquad\qquad\quad(\textbf{b})
\qquad\qquad\qquad~~~(\textbf{c})\qquad\qquad~~~\qquad\quad(\textbf{d})$\\

\noindent { \small \textbf{Figure 12.} The solution \eqref{5} of the equation \eqref{Q1} with the parameters $\textbf{(a)}$ $k_{1}=\frac{2i}{25}$, $k_{2}=\frac{11i}{50}$, $k_{3}=\frac{3i}{31}$ $\widetilde{k}_{1}=-\frac{2i}{25}$, $\widetilde{k}_{2}=-\frac{11i}{50}$, $\widetilde{k}_{3}=-\frac{3i}{31}$, $\omega_{1}=\widetilde{\omega}_{1}=\omega_{2}=1$, $\widetilde{\omega}_{2}=-1$; $\textbf{(d)}$  is the characteristic line graph (blue line $L_{1}:x-\frac{16}{625}t=0$, $L_{2}:x-\frac{121}{625}t=0$ and $L_{3}:x-\frac{36}{961}t=0$) and contour map of $\textbf{(a)}$;  (\textbf{a})(\textbf{b})(\textbf{c}) and \textbf{(d)}: the local structure, density, the dynamic behavior of the three-soliton solutions at different times and characteristic line. }

For the second case, as shown in the figure, the bounded triple soliton graph has several peaks, and reaches the highest peak near $x=0$. In addition, we can see that with $x=0$ as the dividing line, the other small peaks are not completely symmetrical, but can overlap after rotation with $\theta=\pi$.

\subsubsection{Non-pure imaginary eigenvalues}
\
\newline
(1) Bounded three-soliton solutions

For the first case, we take the  eigenvalues as follows,
\begin{align*}
\left\{\begin{aligned}k_{1}&=-\frac{1}{2}-\frac{i}{4},~ \widetilde{k}_{1}=-\frac{1}{2}+\frac{i}{4},~ \omega_{1}=\widetilde{\omega}_{1}=1\\
k_{2}&=-1+\frac{i}{2},~ \widetilde{k}_{2}=-1-\frac{i}{2},~ \omega_{2}=\widetilde{\omega}_{2}=1,\\
k_{3}&=\frac{1}{2}+\frac{i}{5},~ \widetilde{k}_{3}=\frac{1}{2}-\frac{i}{5},~ \omega_{3}=\widetilde{\omega}_{3}=1,
\end{aligned}\right.
\end{align*}
then the solution of \eqref{6} can be characterized in Fig. 13. As seen in Fig. 13, the Figs. 13(\textbf{e,i}) are obtained by the rotation of graph Fig. 13(\textbf{a}) with $\tan\theta_{1}=-\frac{96}{41}$ and $\tan\theta_{2}=-\frac{3}{227}$.
% and $\tan\theta_{3}=\frac{21}{29}$.

\noindent{\rotatebox{0}{\includegraphics[width=3.3cm,height=2.7cm,angle=0]{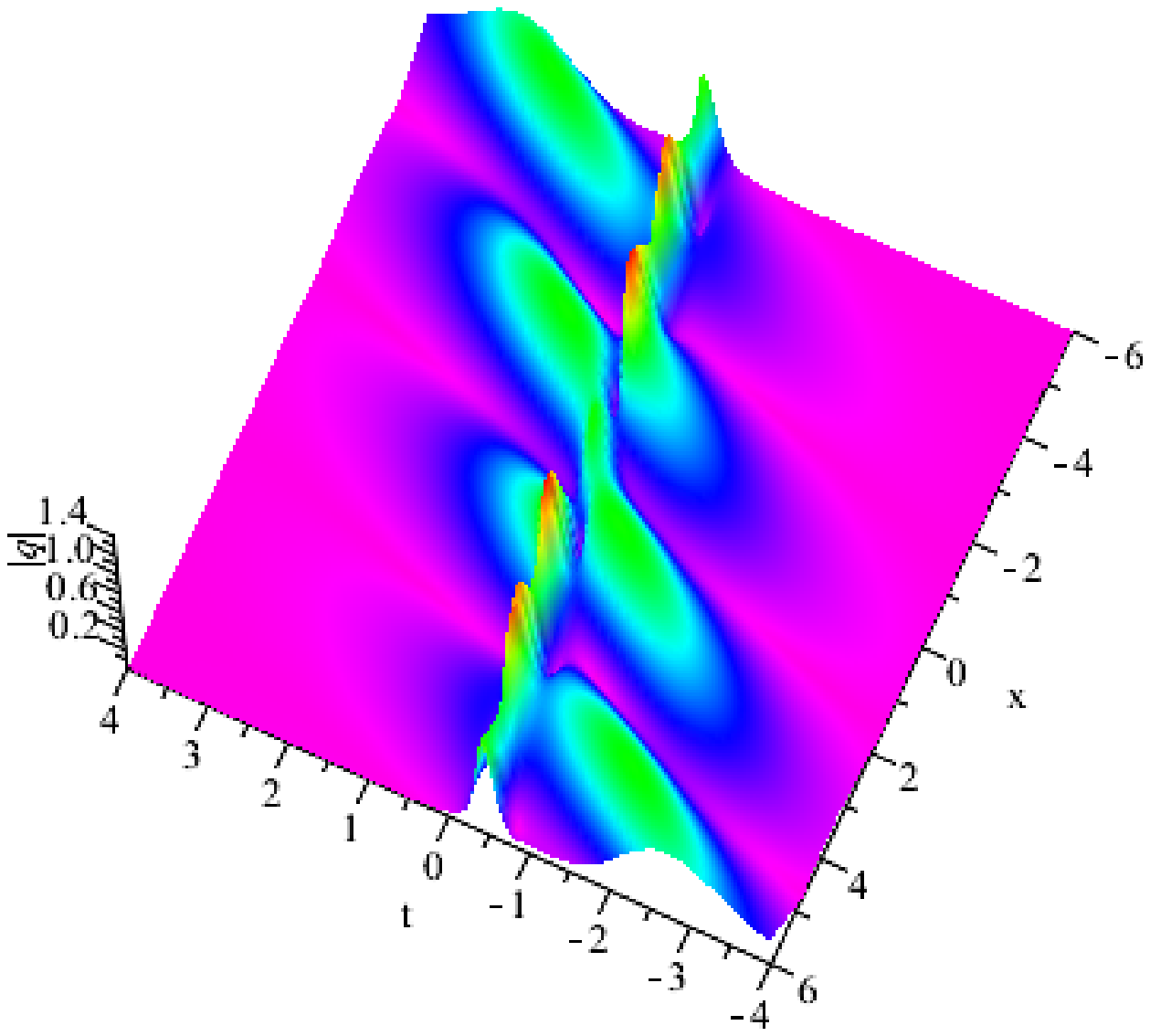}}
~\quad\rotatebox{0}{\includegraphics[width=2.5cm,height=2.4cm,angle=0]{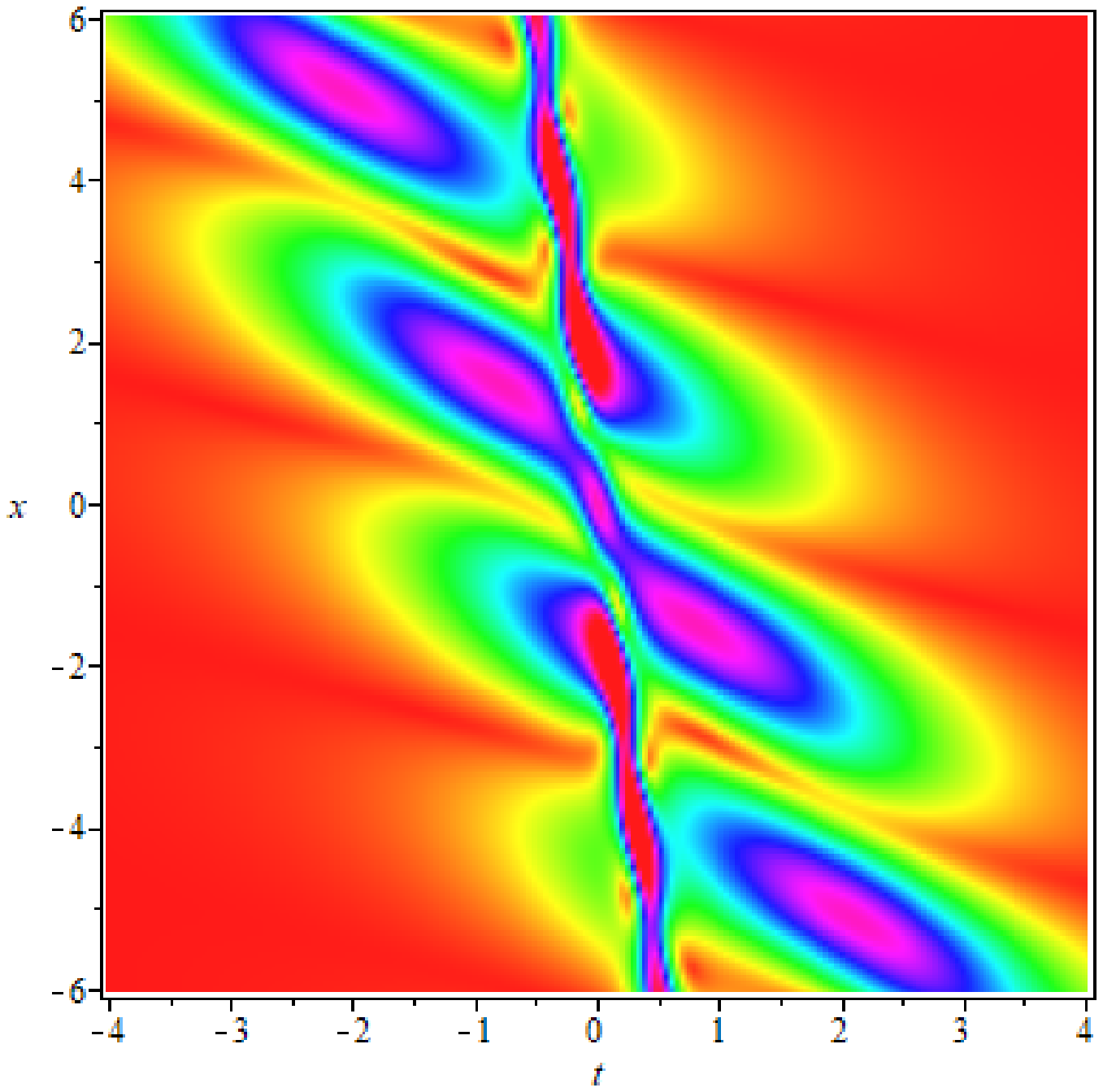}}
~\quad\rotatebox{0}{\includegraphics[width=2.5cm,height=2.4cm,angle=0]{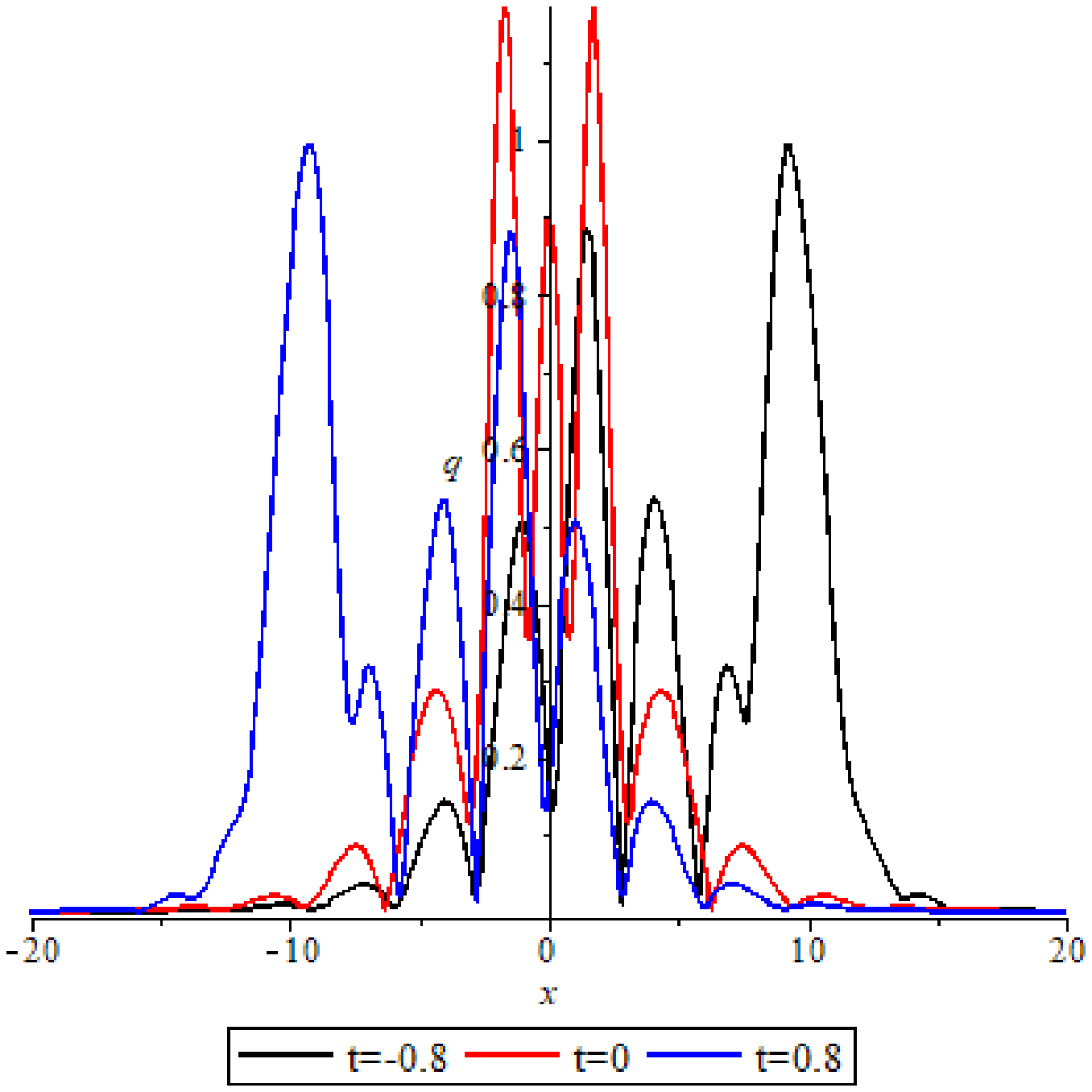}}
~\quad\rotatebox{0}{\includegraphics[width=2.5cm,height=2.4cm,angle=0]{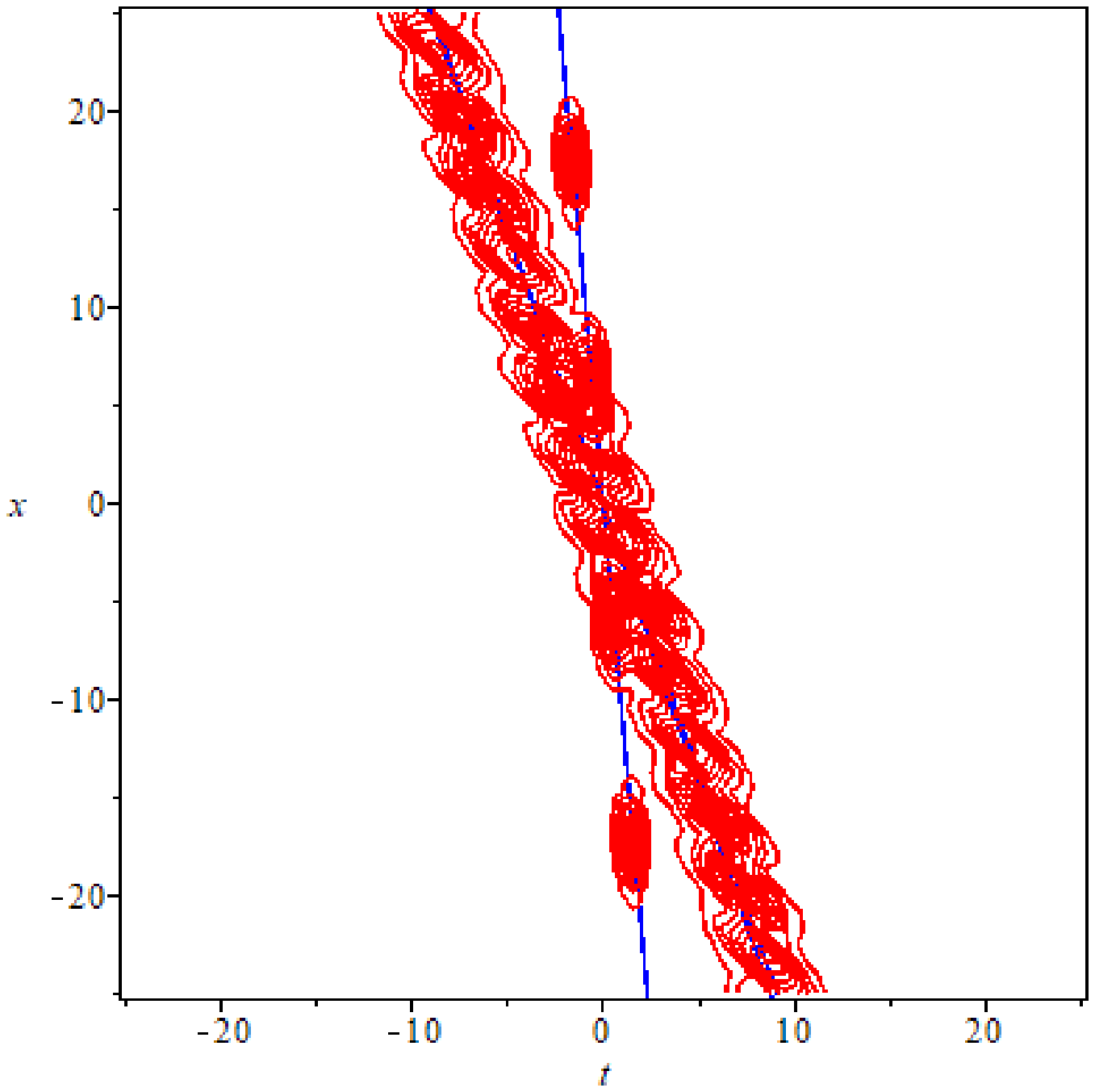}}}

$~~\quad\quad(\textbf{a})\qquad\qquad\qquad\qquad\quad(\textbf{b})
\qquad\qquad\qquad~~~(\textbf{c})\qquad\qquad~~~\qquad\quad(\textbf{d})$\\

\noindent{\rotatebox{0}{\includegraphics[width=3.3cm,height=2.7cm,angle=0]{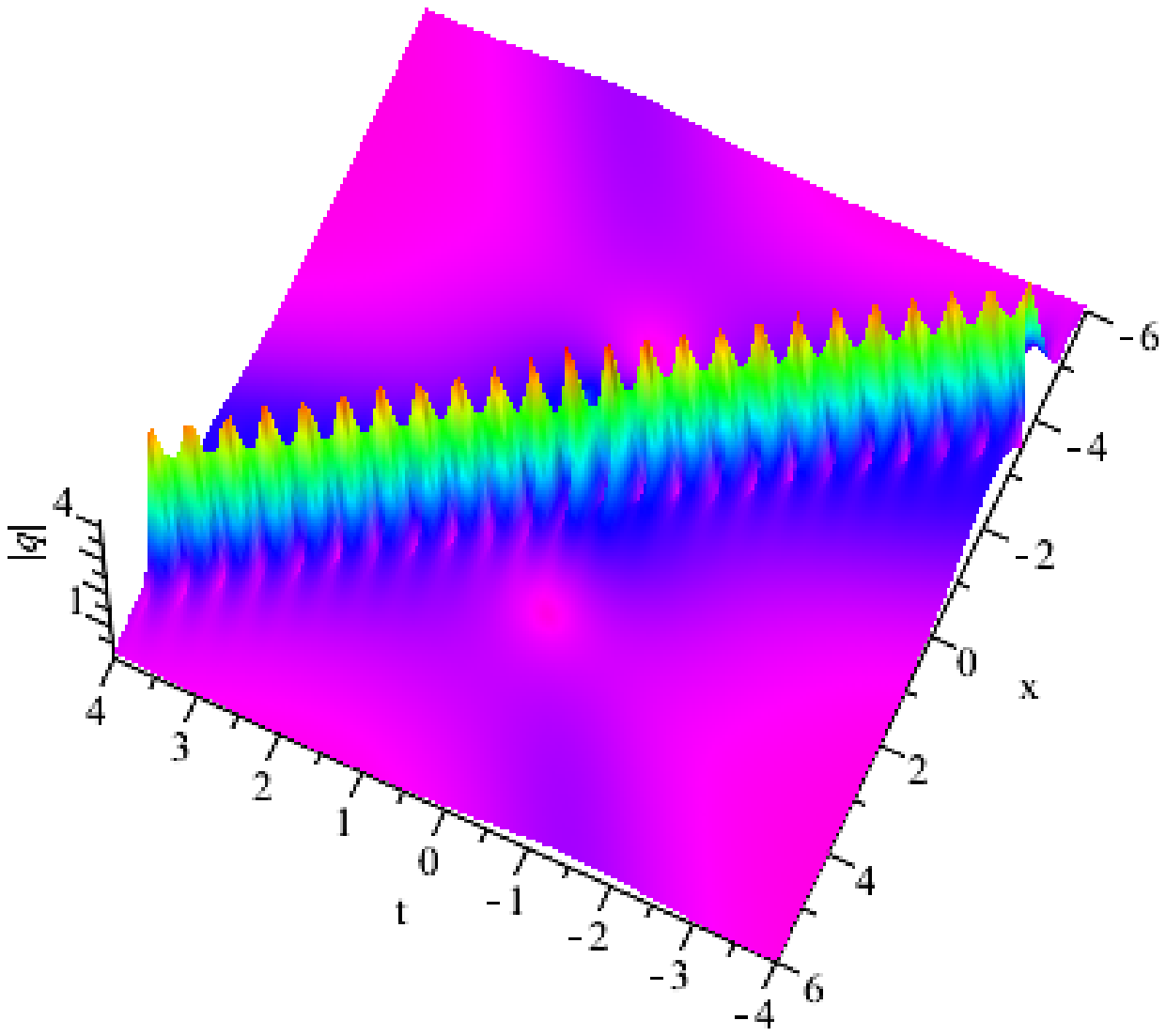}}
~\quad\rotatebox{0}{\includegraphics[width=2.5cm,height=2.4cm,angle=0]{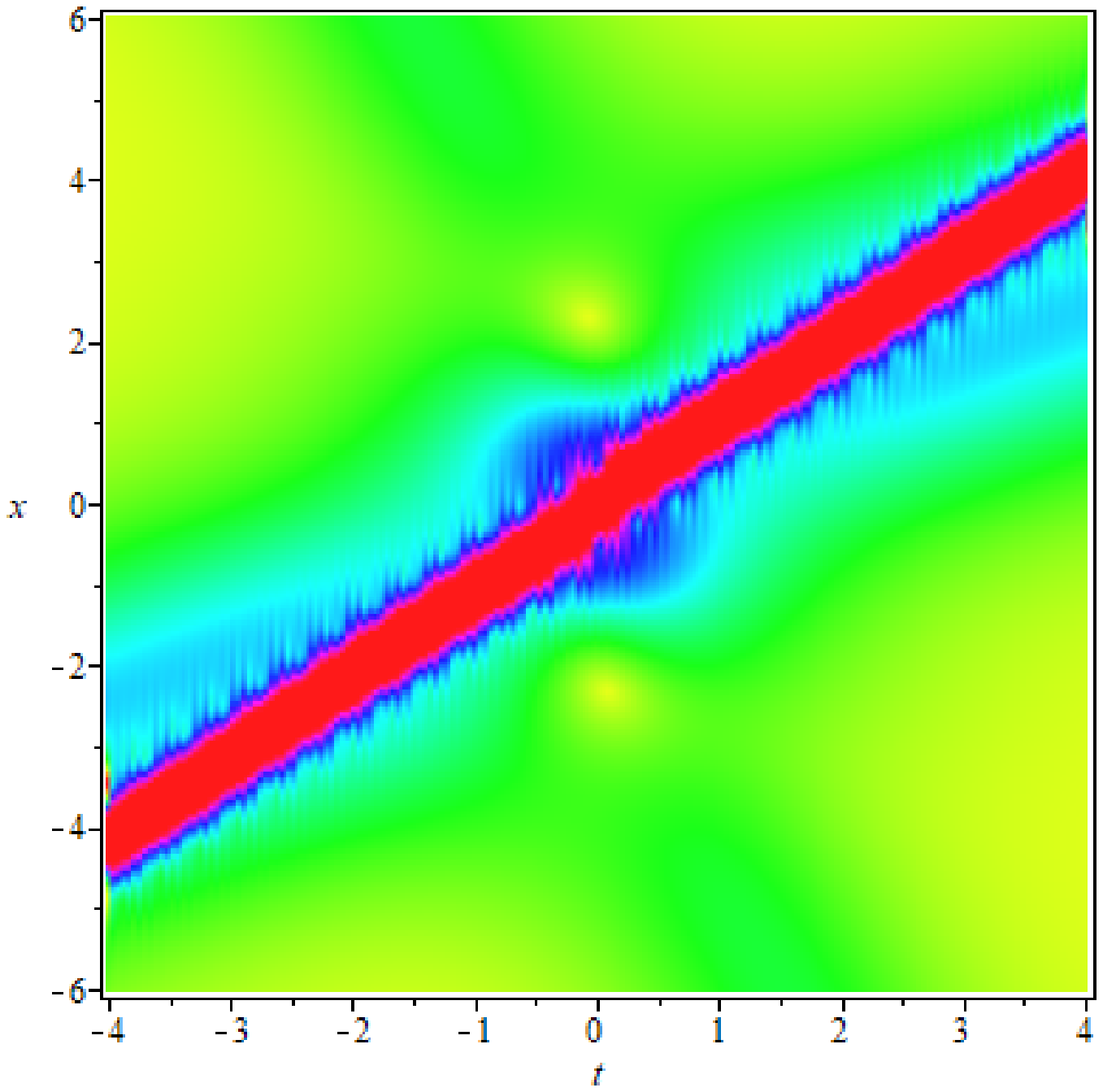}}
~\quad\rotatebox{0}{\includegraphics[width=2.5cm,height=2.4cm,angle=0]{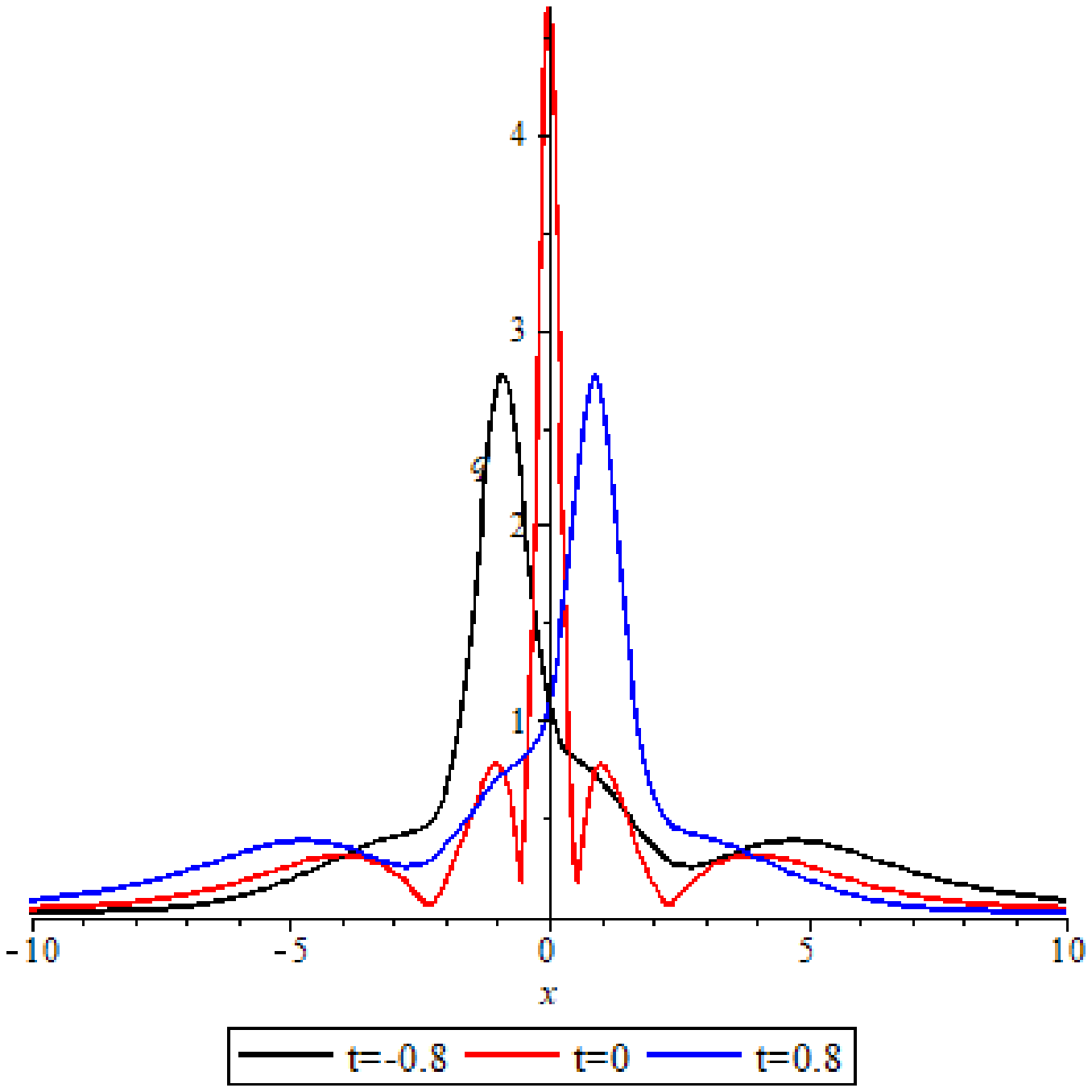}}
~\quad\rotatebox{0}{\includegraphics[width=2.5cm,height=2.4cm,angle=0]{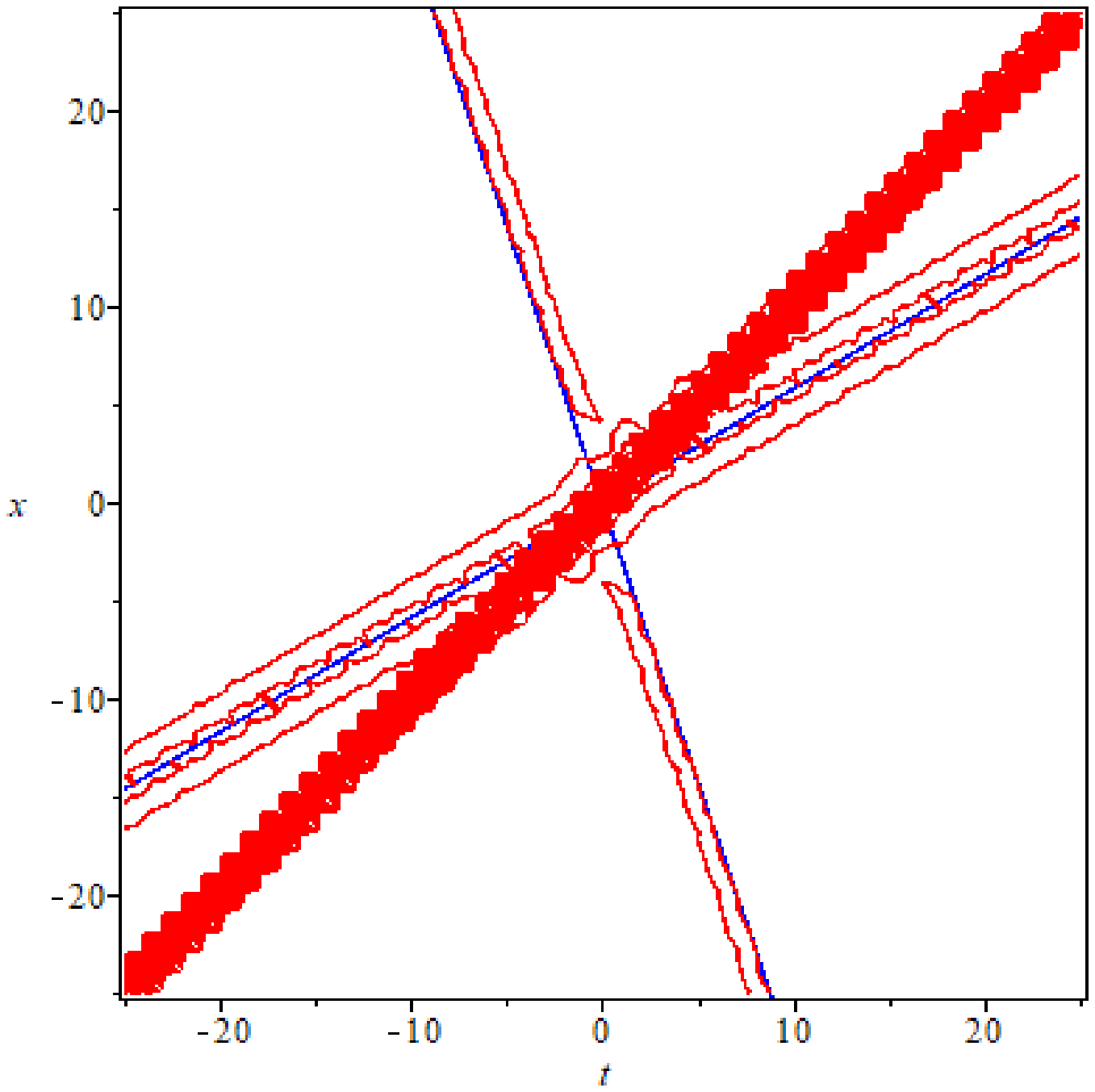}}}

$~~\quad\quad(\textbf{e})\qquad\qquad\qquad\qquad\quad(\textbf{f})
\qquad\qquad\qquad~~~~(\textbf{g})\qquad\qquad~~~\qquad\quad(\textbf{h})$\\

\noindent{\rotatebox{0}{\includegraphics[width=3.3cm,height=2.7cm,angle=0]{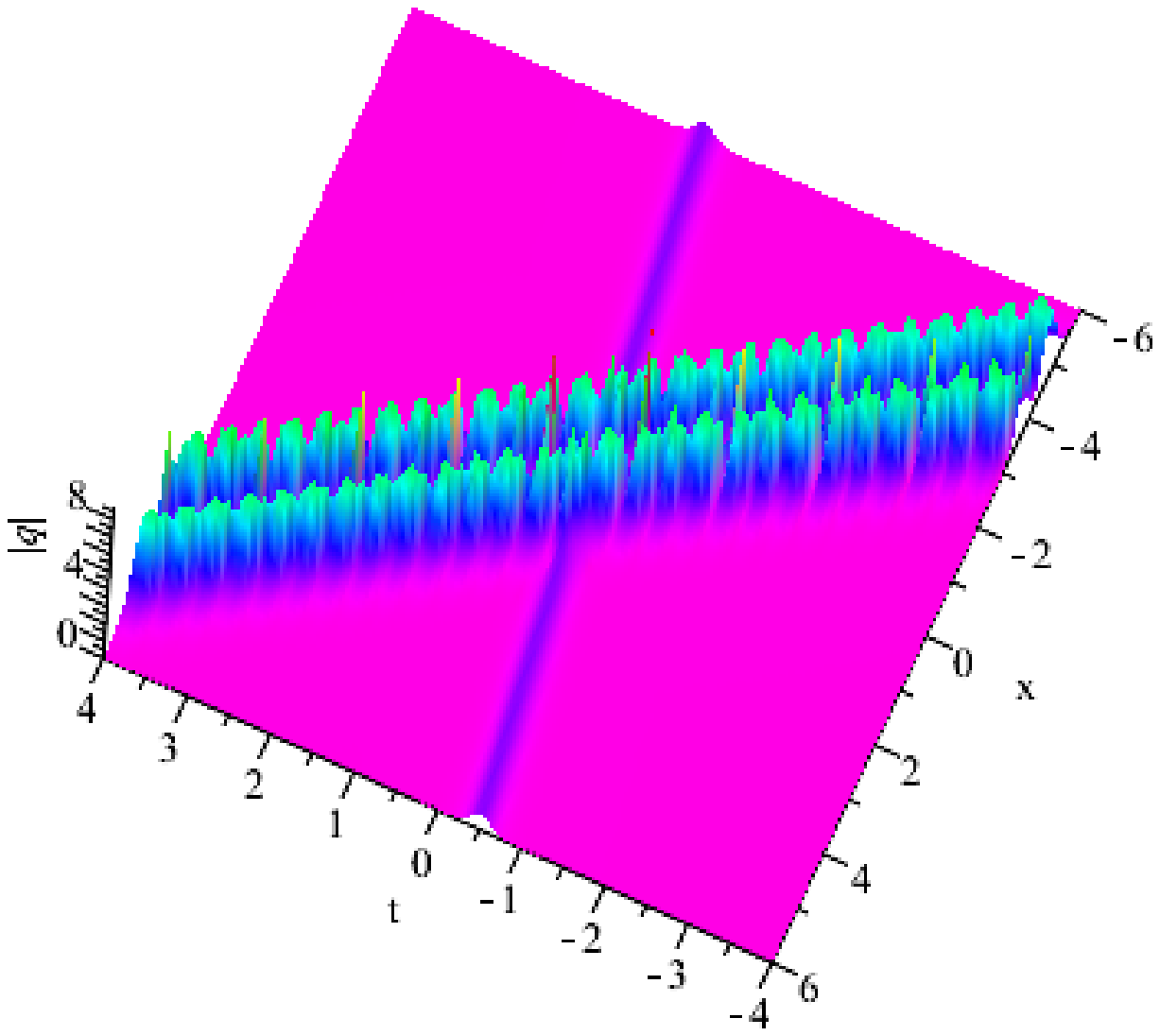}}
~\quad\rotatebox{0}{\includegraphics[width=2.5cm,height=2.4cm,angle=0]{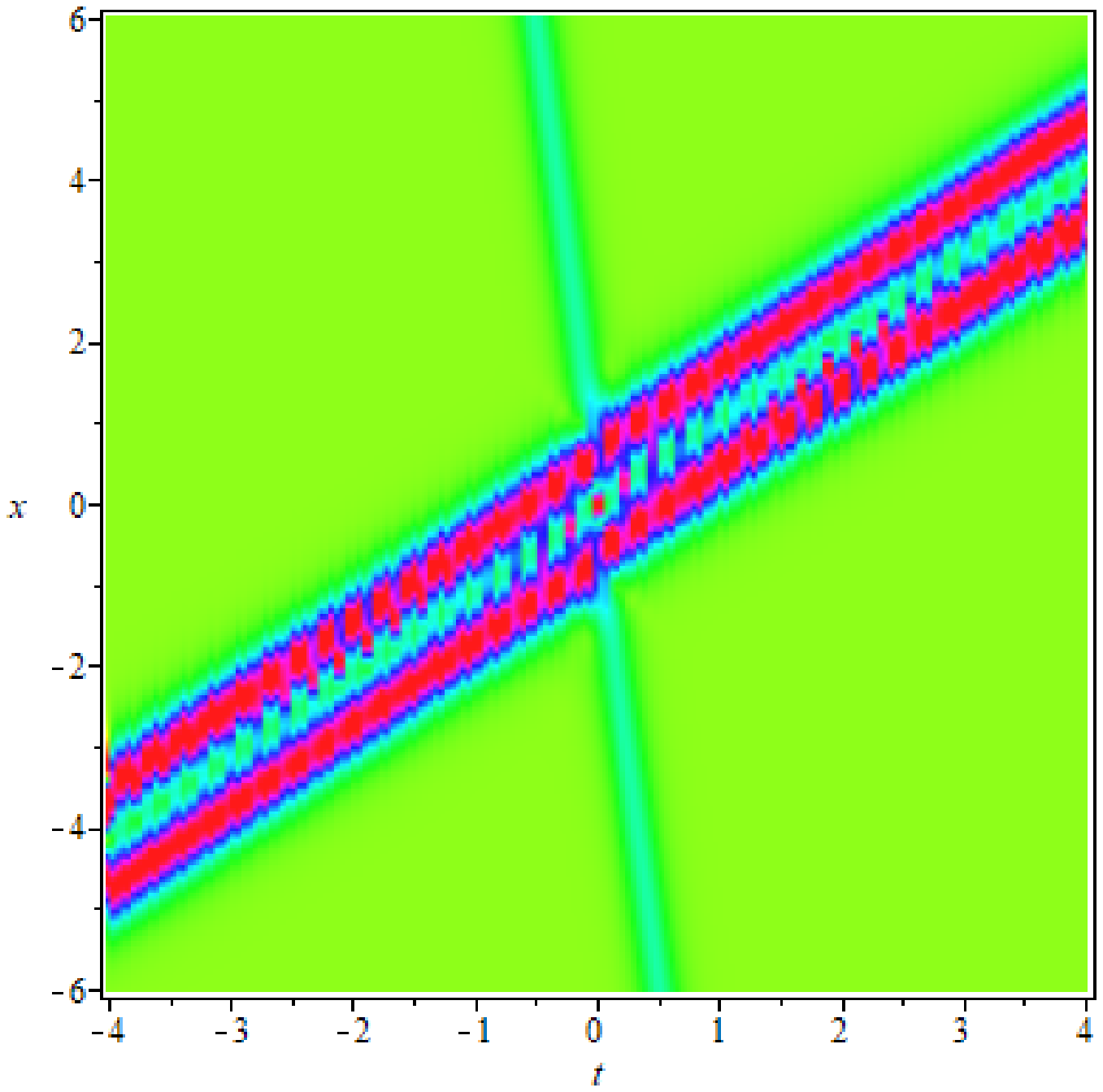}}
~\quad\rotatebox{0}{\includegraphics[width=2.5cm,height=2.4cm,angle=0]{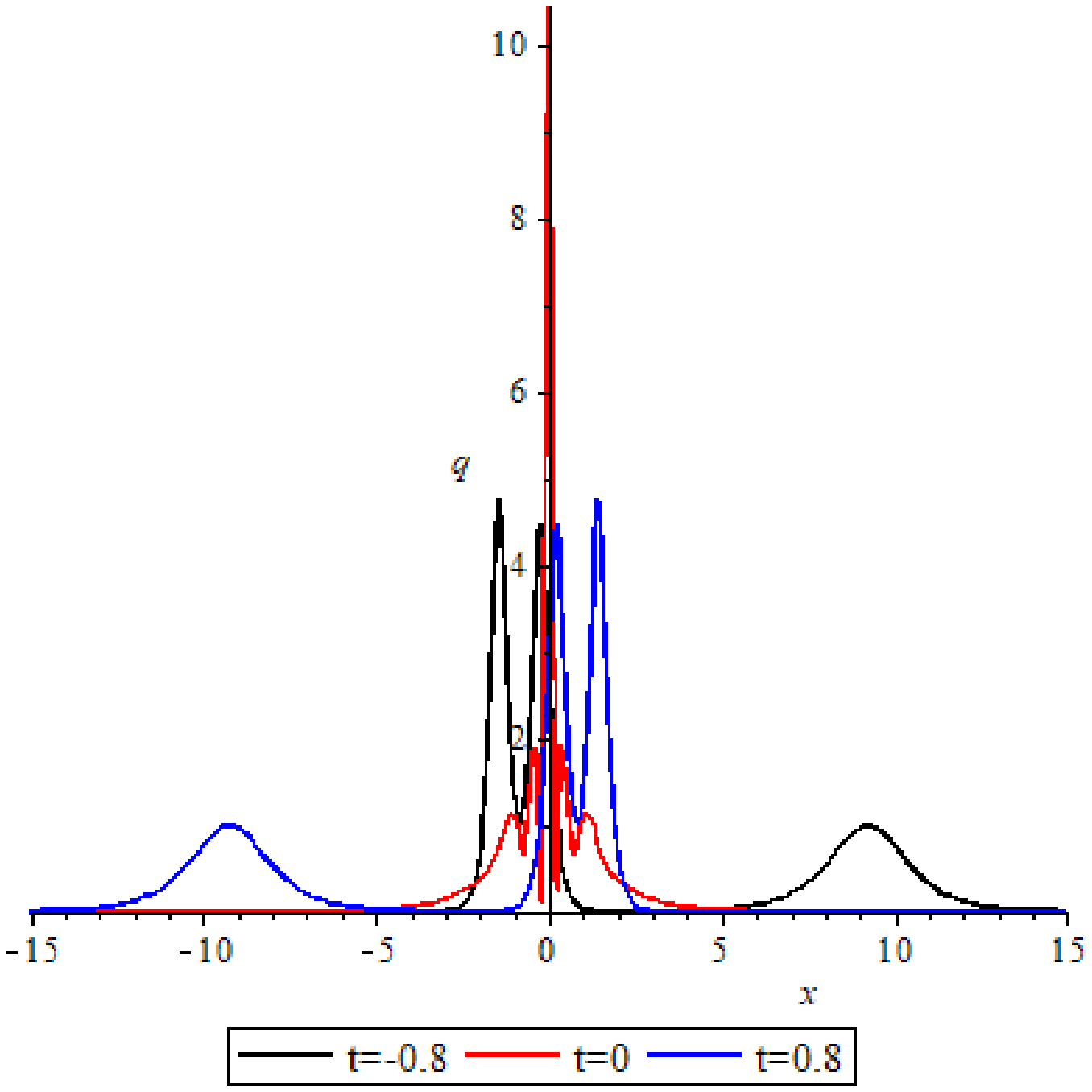}}
~\quad\rotatebox{0}{\includegraphics[width=2.5cm,height=2.4cm,angle=0]{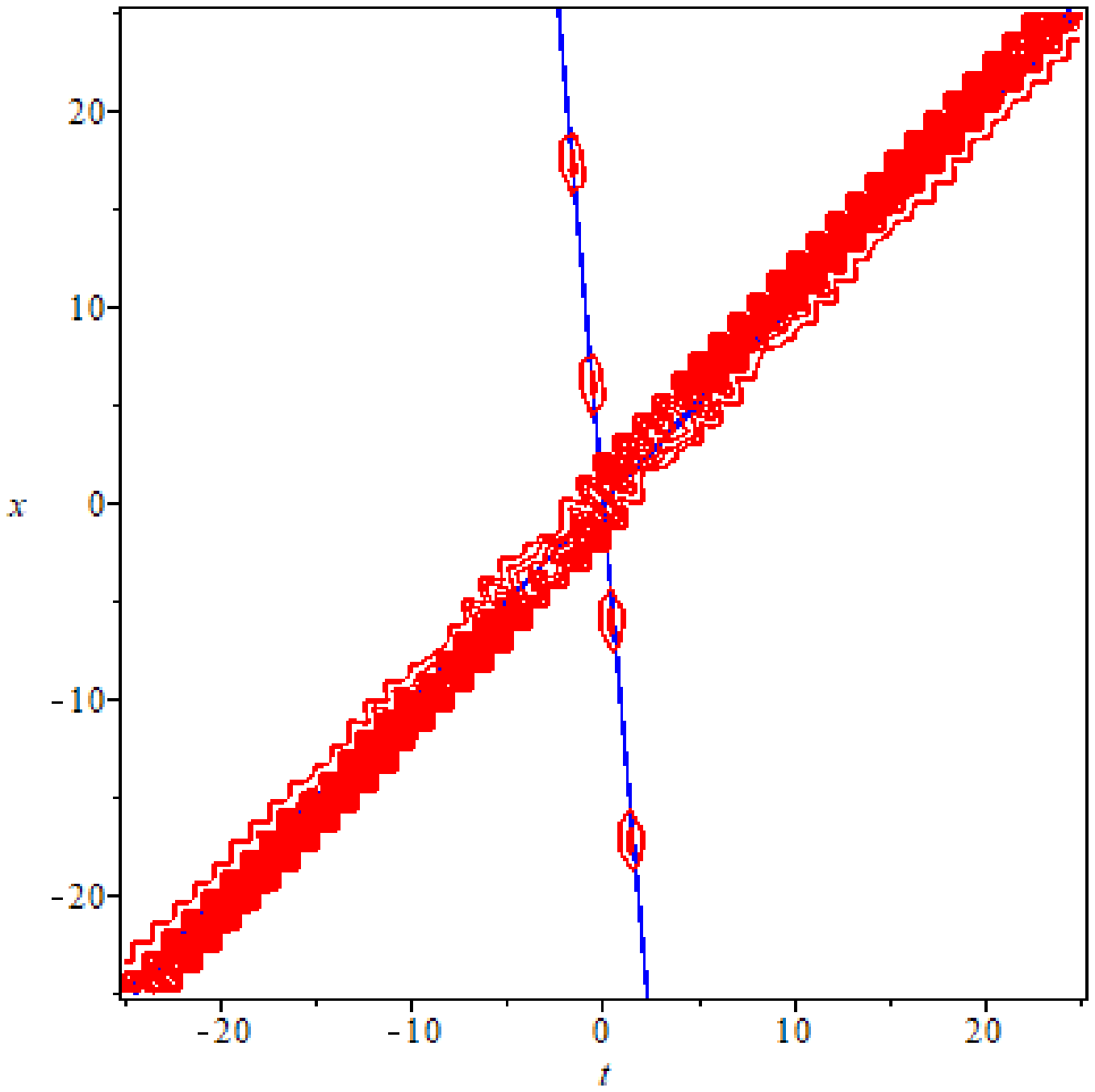}}}

$~~\quad\quad(\textbf{i})\qquad\qquad\qquad\qquad\quad(\textbf{j})
\qquad\qquad\qquad~~~~(\textbf{k})\qquad\qquad~~~\qquad\quad(\textbf{l})$\\

\noindent { \small \textbf{Figure 13.} The solution \eqref{5} of the equation \eqref{Q1} with the parameters $\textbf{(a)}$  $k_{1}=-\frac{1}{2}-\frac{i}{4}$, $\widetilde{k}_{1}=-\frac{1}{2}+\frac{i}{4}$, $k_{2}=-1+\frac{i}{2}$, $\widetilde{k}_{2}=-1-\frac{i}{2}$, $k_{3}=\frac{1}{2}+\frac{i}{5}$, $\widetilde{k}_{3}=\frac{1}{2}-\frac{i}{5}$, $\omega_{1}=\omega_{2}=\omega_{3}=1$, $\widetilde{\omega}_{1}=\widetilde{\omega}_{2}=\widetilde{\omega}_{3}=1$; $\textbf{(b)}$ $k_{1}=\frac{\sqrt{33}}{6}+\frac{\sqrt{3}i}{3}$, $\widetilde{k}_{1}=\frac{\sqrt{33}}{6}-\frac{\sqrt{3}i}{3}$, $k_{2}=\frac{\sqrt{237}}{237}-\frac{5\sqrt{2}i}{2\sqrt{79}}$, $\widetilde{k}_{2}=\frac{\sqrt{237}}{237}-\frac{5\sqrt{2}i}{2\sqrt{79}}$, $k_{3}=\frac{1}{2}+\frac{i}{5}$, $\widetilde{k}_{3}=\frac{1}{2}-\frac{i}{5}$, $\omega_{1}=\omega_{2}=\omega_{3}=1$, $\widetilde{\omega}_{1}=\widetilde{\omega}_{2}=\widetilde{\omega}_{3}=1$;  $\textbf{(c)}$   $k_{1}=\sqrt{2}+\frac{5i}{2}$, $\widetilde{k}_{1}=\sqrt{2}-\frac{5i}{2}$, $k_{2}=-1+\frac{i}{2}$, $\widetilde{k}_{2}=-1-\frac{i}{2}$, $k_{3}=\frac{5}{4}+\frac{\sqrt{2215}i}{2\sqrt{112}}$, $\widetilde{k}_{3}=\frac{5}{4}-\frac{\sqrt{2215}i}{2\sqrt{112}}$, $\omega_{1}=\omega_{2}=\omega_{3}=1$, $\widetilde{\omega}_{1}=\widetilde{\omega}_{2}=\widetilde{\omega}_{3}=1$;
$\textbf{(d)}$  is the characteristic line graph (blue line $L_{1}:x+\frac{11}{4}t=0$, $L_{2}:x+11t=0$, $L_{3}:x+\frac{71}{25}t=0$) and contour map of $\textbf{(a)}$; $\textbf{(h)}$  is the characteristic line graph (blue line $L_{4}:x-t=0$, $L_{5}:x-\frac{46}{79}t=0$, $L_{6}:x+\frac{71}{25}t=0$ and contour map of $\textbf{(e)}$; $\textbf{(l)}$  is the characteristic line graph (blue line $L_{7}:x-t=0$, $L_{8}:x+11t=0$, $L_{9}:x-\frac{115}{112}t=0$ and contour map of $\textbf{(i)}$; }

It can be seen from the images that Fig. 13  produce bounded solutions. We can find that the amplitude of the soliton solution in Fig. 13(\textbf{e}) tends to zero as $x\rightarrow\pm\infty$  after  fluctuating near $x=0$. For the characteristic line of three-soliton interaction, we still find the relationship between $x$ and $t$. For the rotation angle, we make one of the characteristic lines do not move its position, and the other two characteristic lines move their position, so it will lead to some phenomenon changes, which is different from one-soliton and two-solitons. This is also one of the reasons that, in addition to the existence of errors, can lead to different phenomena.

(2) Unbounded three-soliton solutions

\noindent\rotatebox{0}{\includegraphics[width=4.4cm,height=3.5cm,angle=0]{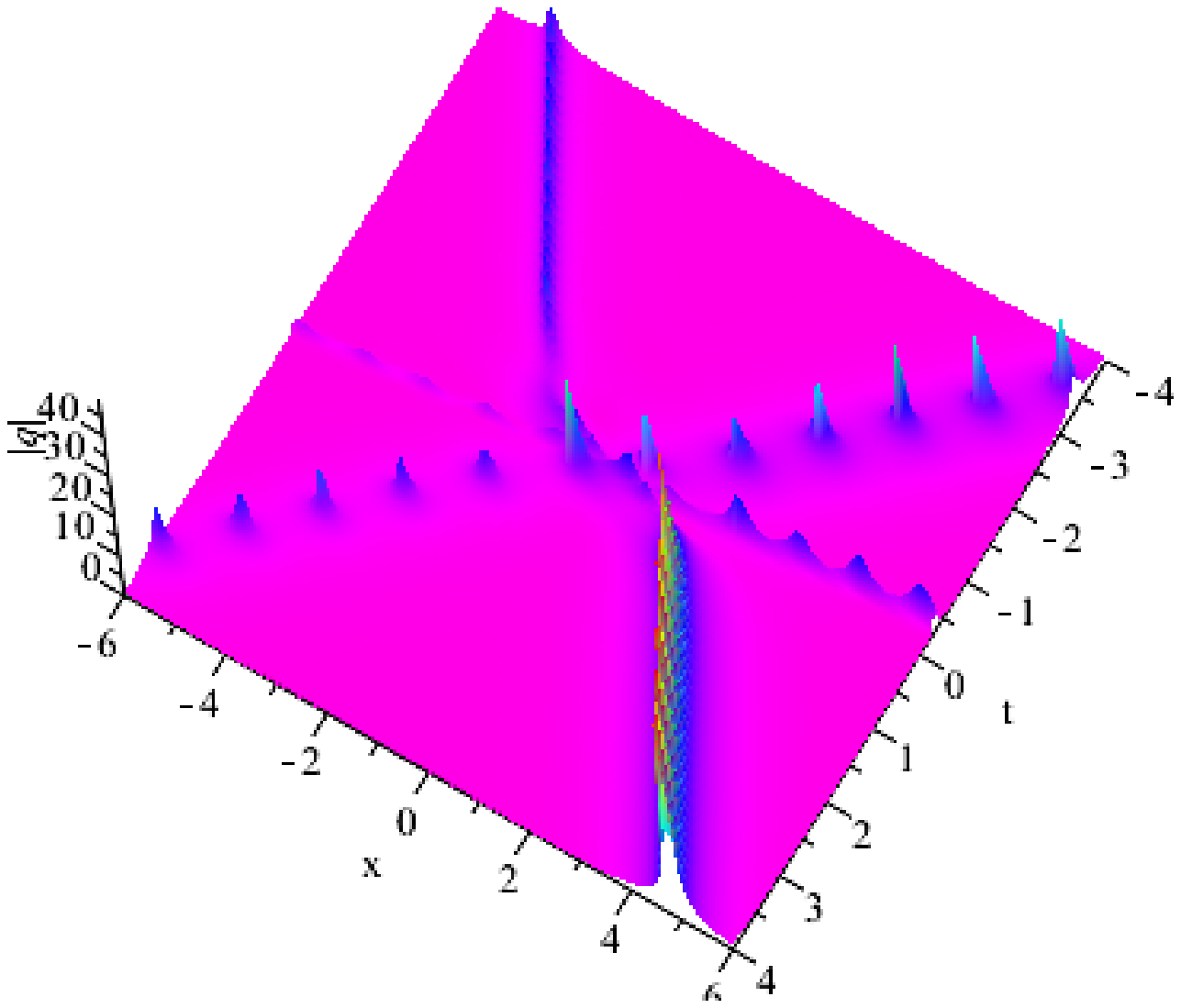}}
\rotatebox{0}{\includegraphics[width=3.3cm,height=2.8cm,angle=0]{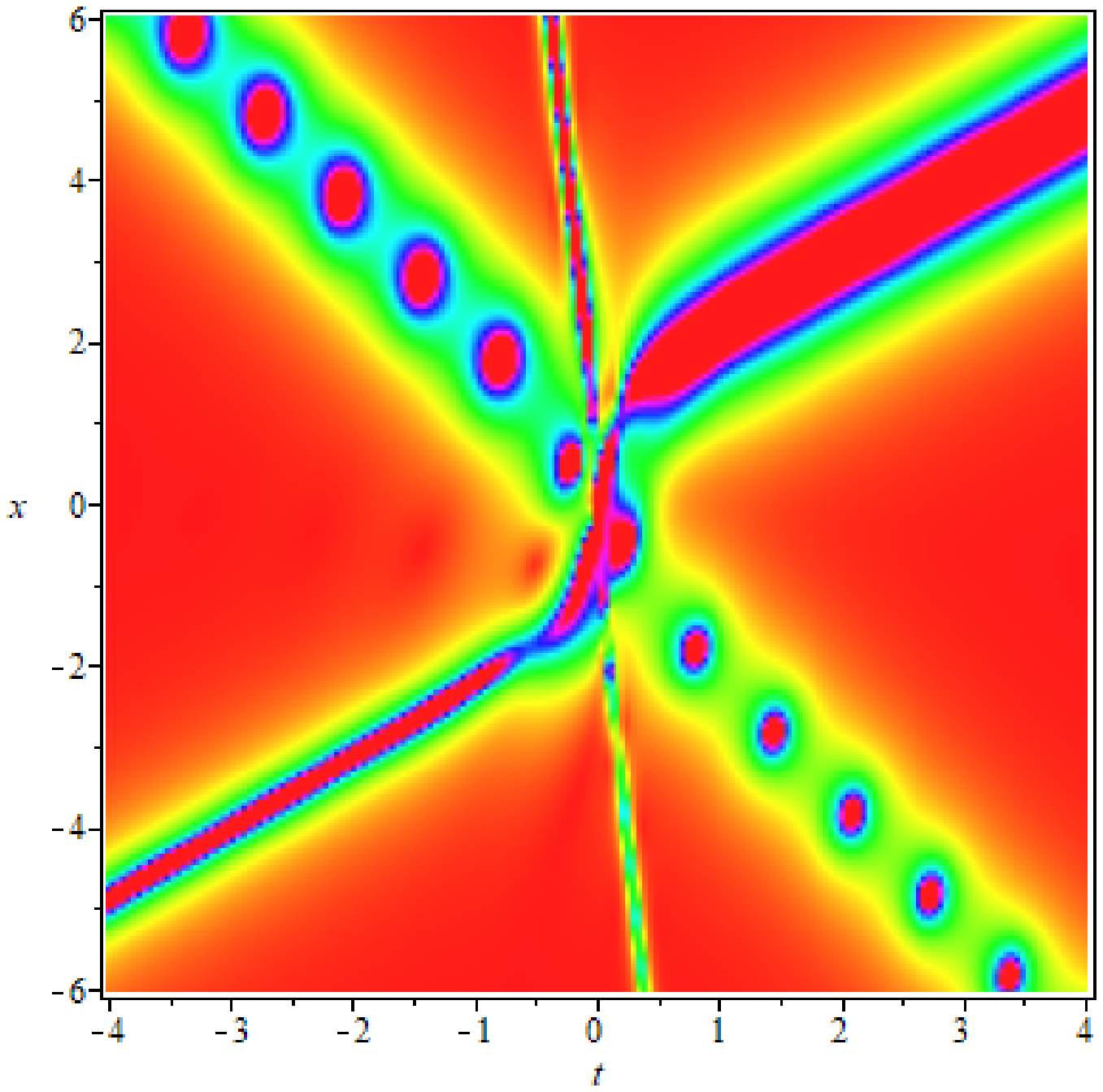}}
\hspace{-3.4cm}
\centerline{\begin{tikzpicture}[scale=0.47]
\draw[-][thick](-3,0)--(-2,0);
\draw[-][thick](-2,0)--(-1,0);
\draw[-][thick](-1,0)--(0,0);
\draw[-][thick](0,0)--(1,0);
\draw[-][thick](1,0)--(2,0);
\draw[-][thick](2,0)--(3,0);
\draw[-][thick](3,0)--(3,2);
\draw[-][thick](3,2)--(3,5.7);
\draw[-][thick](3,5.7)--(0,5.7);
\draw[-][thick](0,5.7)--(-3,5.7);
\draw[-][thick](-3,5.7)--(-3,2);
\draw[-][thick](-3,2)--(-3,0);
\draw[-][thick](-3,0)--(0,0);
\draw[-][thick](0,0)--(3,0);
\draw[-][dashed](-3,3)--(0,3);
\draw[-][dashed](0,3)--(3,3);
\draw[-][dashed](0,5.7)--(0,3);
\draw[-][dashed](0,3)--(0,0);
\draw[fill][red] (-0.3,3.7)circle [radius=0.085];
\draw[fill][red] (-0.3,2.3)circle [radius=0.085];
\draw[fill][blue] (-1.2,3.9)circle [radius=0.085];
\draw[fill][blue] (-1.2,2.1)circle [radius=0.085];
\draw[fill][green] (0.5,2.4)circle [radius=0.085];
\draw[fill][green] (-0.5,3.6)circle [radius=0.085];
\draw[-][dashed](2,3)--(3,3)node[right]{$Re~z$};
\draw[-][dashed](0,5.5)--(0,5.7)node[above]{$Im~z$};
\end{tikzpicture}}

$\qquad\quad\quad(\textbf{a})\qquad\qquad~~\qquad\qquad\qquad(\textbf{b})
\quad~~~~\qquad\qquad\qquad\qquad(\textbf{c})$

\noindent { \small \textbf{Figure 14.} The solution \eqref{5} of the equation \eqref{Q1} with the parameters $\textbf{(a)}$  $k_{1}=-0.3+0.7i$, $\widetilde{k}_{1}=-0.3-0.7i$, $k_{2}=-1.2+0.9i$, $\widetilde{k}_{2}=-1.2-0.9i$, $k_{3}=0.5-0.6i$, $\widetilde{k}_{3}=-0.5+0.6i$. (\textbf{a})(\textbf{b})(\textbf{c}): the local structure, density, distribution pattern of discrete spectral points.}

The corresponding three-solitons are shown in the Fig. 14. The three-solitons move in three directions and collapse repeatedly as they move. The amplitudes of the three moving waves also change with time. We can see that this three-solitons really describes the nonlinear superposition between the three fundamental solitons. As mentioned above, there are also non-singular and bounded solitons under specific parameters.

\section{Conclusions and discussions}

In this work, we have discussed the dynamic behavior of the nonlocal focusing mKdV equation via RH problem. For the scattering problem of the reverse space-time AKNS structure, it is a new nonlocal symmetric reduction, where the symmetric reduction means that it is nonlocal in space and time. Therefore, a new integral reverse space-time nonlocal equation mKdV is derived. Compared with the local equation, the symmetry of the scattering data in the nonlocal equation is different, which directly leads to the difference in the dynamic behavior of the solution of the nonlocal equation. We have started from the initial value problem and got the corresponding eigenfunction and scattering data, from which a Riemann-Hilbert problem is constructed to derived the $N$-soliton solution of the mKdV equation. Then  the $N$-soliton solution of the nonlocal mKdV equation is derived by utilizing the appropriately selected symmetry relationship on the scattering data. Since the general formula of the soliton solution has been successfully obtained, we have combined the characteristic line and selected the eigenvalue with symmetric relationship to analyze the dynamic behavior of one-soliton, two-soliton and three-soliton. We have found that these soliton solutions mainly include bounded solutions, singular solutions, kink solutions and position solutions. The behavior of these solutions mainly has the following types of characteristics. The first case is that they are bounded and stable, and they still retain their original characteristics after collision. The second case is that some solutions will collapse repeatedly, as the parameters of the real or imaginary part of the eigenvalue gradually become larger,  the period gradually decreases. Both of these cases are relatively polarized phenomena. For the third case, which is called semi-elatic collision.

In addition, the real revese space-time mKdV equation  have been also studied in \cite{Ma-2021-JMAA}, but our work is quite different from that of Ma's. In his work, the  $N$-component real revese space-time mKdV equation have been constructed by using a set of nonlocal reductions, then the exact expression of soliton solutions have been obtained via solving the corresponding RH problem. Inspired by other work, the advantages of our work have been reflected in the following points: (1) The symmetry of eigenvalues have been obtained directly from the spectral problem, which is the important factor in the dynamic behaviors of generating more solutions.
(2) We have studied the new dynamic behaviors of one-soliton, two-soliton and three-soliton solutions in detail. By making special reduction  and taking the limit of eigenvalues, we have got the dynamic behaviors of the solutions, including position solution, kink solution, bounded solutions and singular solutions.
(3) In the process of analyzing the behavior of the solution, we have also given the characteristic line expressions and characteristic line graphs of the solutions, as well as the dynamic behaviors of the solution after rotation.

\section*{Acknowledgements}

%The authors would like to thank the editor and the referees for their valuable comments and suggestions.

This work was supported by the National Natural Science Foundation of China under Grant No. 11975306, the Natural Science Foundation of Jiangsu Province under Grant No. BK20181351, the Six Talent Peaks Project in Jiangsu Province under Grant No. JY-059, and the Fundamental Research Fund for the Central Universities under the Grant Nos. 2019ZDPY07 and 2019QNA35.

\section*{References}

\renewcommand{\baselinestretch}{1.2}

\end{document}